\documentclass[floatfix,a4paper,aps,prd,twocolumn,preprintnumbers]{revtex4} 
\usepackage{epsfig}
\usepackage{axodraw}
\usepackage{amsmath}
\usepackage{amssymb}
\usepackage{amsfonts}
\usepackage{graphicx}
\usepackage{graphics,subfigure}
\usepackage{axodraw}
\usepackage{float}
\usepackage{booktabs}
\usepackage{color}
\usepackage{geometry}
\usepackage{rotating}

\newcommand{\newc}{\newcommand}
\newc{\ba}{\begin{array}}
\newc{\ea}{\end{array}}
\newc{\bea}{\begin{eqnarray}}
\newc{\eea}{\end{eqnarray}}
\newc{\beastar}{\begin{eqnarray*}}
\newc{\eeastar}{\end{eqnarray*}}
\newc{\beq}{\begin{equation}}
\newc{\eeq}{\end{equation}}
\newc{\bestar}{\begin{equation*}}
\newc{\eestar}{\end{equation*}}
\newc{\ben}{\begin{enumerate}}
\newc{\een}{\end{enumerate}}
\newc{\bi}{\begin{itemize}}
\newc{\ei}{\end{itemize}}
\newc{\mymed}{\vspace{0.14cm}}
\newc{\met}{p\!\!\!/_T}

\newc{\lam}{\lambda}
\newc{\lamp}{\lam^\prime}
\newc{\lampp}{\lam^{\prime\prime}}
\newc{\Lam}{\lam}
\newc{\BLam}{{\mathbf{\Lambda}}}
\newc{\eps}{\epsilon}
\newc{\kap}{\kappa}

\newc{\ra}{\rightarrow}
\newc{\ovl}{\overline}
\newc{\lsim}{\stackrel{<}{\sim}}
\newc{\Tr}{{~\rm Tr}}
\newc{\itRahmen}[2]{
\begin{center}\fbox{\parbox{#1 cm}{\it #2}}
\end{center}}

\newc{\del}{\partial}
\newc{\veva}{\langle H_1\rangle}
\newc{\vevb}{\langle H_2\rangle}
\newc{\onehalf}{\textstyle \frac{1}{2} \displaystyle}
\newc{\onethird}{\textstyle \frac{1}{3} \displaystyle}
\newc{\mzero}{M_0}
\newc{\mhalf}{{M_{1/2}}}
\newc{\tanb}{\tan\beta}
\newc{\Psix}{{\mathrm{P}_{\!6}}}
\newc{\nPsix}{{\not\!\Psix}}
\newc{\nPsixU}{{\s{\mathrm P}_{\!6}}}

\newc{\muon}{\mu}
\newc{\azero}{A_0}
\newc{\neutralino}{\tilde\chi^0}
\newc{\selectron}{\tilde e}
\newc{\stau}{\tilde\tau}
\newc{\smuon}{\tilde\mu}
\newc{\sneu}{\tilde\nu}
\newc{\higgs}{h^0}
\newc{\sgnmu}{\textrm{sgn}(\mu)}
\newc{\gev}{\mbox{~GeV}}
\newc{\tev}{\mbox{~TeV}}
\newc{\gsim}{\stackrel{>}{\sim}}
\newc{\mgut}{{M_{GUT}}}
\newc{\mweak}{{M_{W}}}

\newbox\charbox
\newbox\slabox
\def\s#1{{      
    \setbox\charbox=\hbox{$#1$}
    \setbox\slabox=\hbox{$/$}
    \dimen\charbox=\ht\slabox
    \advance\dimen\charbox by -\dp\slabox
    \advance\dimen\charbox by -\ht\charbox
    \advance\dimen\charbox by \dp\charbox
    \divide\dimen\charbox by 2
    \raise-\dimen\charbox\hbox to \wd\charbox{\hss/\hss}
    \llap{$#1$}
}}

\newc{\onegraph}[4]{%
  \unitlength=1in
  \begin{picture}(3,2.3)
    \put(-0.6,0){\epsfig{file=#1.eps, width=3.7in}}
    \put(0.33,0.48){\epsfig{file=#12.eps, width=2.509in}}
    \put(2.9,1.1){\rotatebox{90}{$#2$}}
    \put(1.3,0.2){\makebox(0,0){$#3$}}
    \put(0.05,1.2){\rotatebox{90}{$#4$}}
  \end{picture}
}

\newc{\ssup}{\tilde{u}}
\newc{\ssdown}{\tilde{d}}
\newc{\ssstrange}{\tilde{s}}
\newc{\sscharm}{\tilde{c}}
\newc{\sstop}{\tilde{t}}
\newc{\ssbottom}{\tilde{b}}
\newc{\sse}{\tilde{e}}
\newc{\ssmu}{\tilde{\mu}}
\newc{\sstau}{\tilde{\tau}}
\newc{\ssnue}{\tilde{\nu}_{e}}
\newc{\ssnumu}{{\tilde{\nu}_{\mu}}}
\newc{\ssnutau}{{\tilde{\nu}_{\tau}}}
\newc{\ssbnue}{\tilde{\nu}^*_{e}}
\newc{\ssbnumu}{\tilde{\nu}^*_{\mu}}
\newc{\ssbnutau}{\tilde{\nu}^*_{\tau}}
\newc{\neut}{{\tilde{\chi}}^0}
\newc{\charge}{\tilde{{\chi}}}
\newc{\glu}{\tilde{g}}
\newc{\Higgs}{H^0}
\newc{\Azero}{A_0}

\newc{\stext}[1]{{\color{red}  #1}}

\newc{\nue}{\nu_e}
\newc{\numu}{\nu_{\mu}}
\newc{\nutau}{\nu_{\tau}}
\newc{\bnue}{\bar \nu_e }
\newc{\bnumu}{\bar \nu_{\mu}}
\newc{\bnutau}{\bar \nu_{\tau}}
\newc{\Br}{\mathrm{Br}}
\begin{document}
\title{Sneutrino as Lightest Supersymmetric Particle 
in $\text{B}_3$ mSUGRA Models \\ and Signals at the LHC}
\author{M.~A.~Bernhardt\footnote{{\,markus@th.physik.uni-bonn.de}}}
\author{S.~P.~Das\footnote{\,spdas@th.physik.uni-bonn.de}}
\author{H.~K.~Dreiner\footnote{\,dreiner@th.physik.uni-bonn.de}}
\author{S.~Grab\footnote{\,sgrab@th.physik.uni-bonn.de}}
\affiliation{Physikalisches Institut, Nu{\ss}allee 12, 
University of Bonn, 53115 Bonn, Germany}

\begin{abstract}
  We consider $\text{B}_3$ mSUGRA models where we have one lepton
  number violating $L_iQ_j\bar D_k$ operator at the GUT scale.  This
  can alter the supersymmetric mass spectrum leading to a sneutrino as
  the lightest supersymmetric particle in a large region
  of parameter space. We take into account the restrictions from
  neutrino masses, the muon anomalous magnetic moment, $b \rightarrow
  s \gamma$ and other precision measurements. We furthermore
  investigate existing restrictions from direct searches at LEP, the
  Tevatron and the CERN $p\bar p$ collider.  We then give examples for
  characteristic signatures at the LHC.
\end{abstract}

\preprint{BONN-TH-2007-09}

\maketitle
 
\section{Introduction}
Supersymmetry (SUSY) \cite{SUSY_refs} is a promising extension of the
Standard Model of particle physics (SM) \cite{SM_refs}; the simplest
form is denoted the supersymmetric SM (SSM). It should be imminently
testable at the LHC \cite{LHC_refs}. Supersymmetric particles, if they
exist, typically decay instantaneously on collider time scales down to
the lightest supersymmetric particle (LSP). The nature and possible
decay properties of the LSP are thus an essential ingredient for all
SUSY signatures. In the minimal SSM, with conserved proton-hexality,
$\Psix$, \cite{Dreiner:2005rd} (or equivalently conserved R-parity
\cite{Farrar:1978xj}), the LSP is stable. Cosmological constraints as
well as LEP searches then restrict the LSP to be the lightest
neutralino \cite{Ellis:1983ew,Hebbeker:1999pi}.

\mymed

If we allow for violation of proton-hexality, the LSP is no longer
stable and in general any supersymmetric particle can be the LSP
\cite{Dreiner:1997uz}. However, it is impossible to perform a detailed
phenomenological study of the corresponding wide variety of mass
orderings of the sparticle spectrum. We thus must restrict ourselves
to well motivated models. In this paper, we focus on the B$_3$ minimal
supergravity (mSUGRA) model \cite{Allanach:2003eb}. In
Ref.~\cite{Allanach:2003eb,Allanach:2006st} it was shown that in such
models there are three different LSP candidates: the lightest
neutralino, $\neut_1$, the lightest scalar tau (stau),
$\tilde{\tau}_1$ and the sneutrino, $\tilde{\nu}_i$. The lightest
neutralino LSP has been studied extensively in the literature, see for
example
\cite{Dreiner:1991pe,Godbole:1992fb,Dreiner:2000vf,Bartl:2000yh}. 
More recently the stau LSP has also been investigated
\cite{Allanach:2003eb,Allanach:2006st,Allanach:2007vi,Dreiner:2007uj,
Bernhardt:2008mz,Dreiner:2008rv,nonmSUGRA_stauLSP}.

\mymed

In this paper we consider in detail the case of a $\tilde{\nu}_i$
LSP. The $\tilde{\nu}_i$ is special, because unlike the $\neut_1$ and
$\tilde{\tau}_1$, the $\text{B}_3$ contributions to the
renormalization group equations (RGEs) are essential for it to become
the LSP. In Ref.~\cite{Allanach:2006st} only one example $\text{B}_3$
mSUGRA scenario with a $\tilde{\nu}_\tau$ LSP was presented. We go
beyond this work and systematically investigate the $\text{B}_3$
mSUGRA parameter space with a $\tilde{\nu}_i$ LSP. In the first part
of our paper we analyse, which conditions at the grand unification
(GUT) scale lead to a $\tilde{\nu}_i$ LSP. In the second part, we
point out striking collider signatures at the LHC which can lead to a
SUSY discovery and which can distinguish a $\tilde{\nu}_i$ LSP
scenario from a ``standard" mSUGRA scenario with a stable $\neut_1$
LSP.

\mymed

The outline of our paper is as follows. In Sect.~\ref{the_model}, we
briefly review the $\Psix$ and $\text{B}_3$ mSUGRA models and discuss
the RGEs which lead to a $\tilde{\nu}_i$ LSP. We then analyse in
Sect.~\ref{bounds_on_snu_LSP} the experimental bounds, especially on
the $L_i Q_j \bar D_k$ operator, which restrict the $\tilde{\nu}_i$
LSP parameter space. In Sect.~\ref{parameter_space} we investigate
in detail the conditions at the GUT scale leading to a $\tilde{\nu}_i$
LSP. This is the central part of our work. Finally, in
Sect.~\ref{pheno}, we simulate SUSY events at the LHC within one
$\tilde{\nu}_\mu$ LSP scenario. We focus on signatures, which are
special for $\tilde{\nu}_i$ LSP scenarios. We conclude in
Sect.~\ref{conclusion}.

\section{The model}
\label{the_model}

The most general gauge invariant and renormalizable superpotential of
the SSM is \cite{superpot_refs}
\bea
W_{\mathrm{SSM}}&=& W_{\text{P}_6}+W_{\not \text{P}_6}\,,
\label{superpot} 
\\[1.5mm]
W_{\text{P}_6}&=&\eps_{ab}\left[(\mathbf{Y}_E)_{ij}L_i^aH_d^b
  \bar{E}_j + (\mathbf{Y}_D)_{ij}Q_i^{ax}H_d^b\bar{D}_{jx}
\right.  \notag  \\ & & 
\left.+(\mathbf{Y}_U)_{ij}Q_i^{ax}H_u^b\bar{U}_{jx} + \mu
  H_d^aH_u^b\right],
\label{P6-superpot} 
\\[1.5mm]
W_{\not \text{P}_6} & = & \eps_{ab}\left[\frac{1}{2} \lam_{ijk} L_i^aL_j^b
\bar{E}_k + \lam'_{ijk}L_i^aQ_j^{bx}\bar{D}_{kx}\right]\notag 
\\&&
+\epsilon_{ab}\kappa^i  L_i^aH_u^b
+\frac{1}{2}\eps_{xyz} \lam''_{ijk}
\bar{U}_i^{\,x} \bar{D}_j^{\,y} \bar{D}_k^{\,z} \,.
\label{notP6-superpot} 
\eea
where $i,j,k=1,2,3$ are generation indices. We have 
employed the standard notation of Ref.~\cite{Allanach:1999ic}.

\medskip

The superpotential, Eq.~(\ref{superpot}), consists of two different
parts. The second part, $W_{\not \text{P}_6}$, contains lepton and
baryon number violating operators. If simultaneously present, they
lead to rapid proton decay, in disagreement with experimental
observations 
\cite{proton_decay,Dreiner:1997uz,Barbier:2004ez,Shiozawa:1998si}. An
additional discrete symmetry is therefore required to keep the proton
stable \cite{discrete_symmetries,Dreiner:2005rd}. The SSM with
R-parity, which prohibits $W_{\not\text{P}_6}$, is conventionally
denoted the MSSM. In a more general approach, proton-hexality, $\text{P}_6
$, in addition prohibits dangerous dimension-five proton decay
operators \cite{Dreiner:2005rd}. Here, we consider a third possibility,
baryon-triality, $\text{B}_3$, which violates R-parity and $\text{P}_6$
by prohibiting only the $\bar{U}\bar{D}\bar{D}$ operators in
Eq.~(\ref{notP6-superpot}). R-parity, proton-hexality and
baryon-triality are the only discrete gauge anomaly-free symmetries of
the SSM \cite{discrete_symmetries,Dreiner:2005rd}. $\text{B}_3$ models
including also a dark matter candidate have been, for example,
proposed in Refs.~\cite{UMSSM}.

\subsection{$\Psix$ mSUGRA Model}
\label{sec:mSUGRA_param}

The MSSM with conserved $\Psix$ has 124 free parameters
\cite{Haber:1997if}. In the mSUGRA model with conserved $\Psix$ and
radiative electroweak symmetry breaking (REWSB)
\cite{msugramodel,Ibanez:1982fr} this is reduced to five parameters,
which is more manageable for phenomenological studies,
\begin{equation}
M_0,\, M_{1/2},\, A_0,\, \tan \beta,\, \text{sgn}(\mu) \, .
\label{mSUGRA_param}
\end{equation}
$M_0$, $M_{1/2}$ and $A_0$ are the universal scalar mass,
the universal gaugino mass and the universal trilinear scalar 
interaction at the GUT scale ($M_{\rm GUT}$), respectively. $\tan\beta$ 
is the ratio of the vacuum expectation values of the two
Higgs doublets; see Eq.~(\ref{P6-superpot}). Finally, we choose 
with $\text{sgn}(\mu)$ unambiguously one solution of the 
electroweak symmetry breaking scalar potential ($|\mu|$ is
determined by REWSB). 

\mymed   

Using the five parameters at $M_{\rm GUT}$, Eq.~(\ref{mSUGRA_param}),
and the RGEs to evolve the parameters down to the electroweak scale
($M_Z$), the mass spectrum of the sparticles and their interactions is
completely determined. The left- (right-)handed charged
slepton, $\tilde{\ell}_{L(R)}$, and $\tilde{\nu}$ masses of the first
two generations can be approximated at $M_Z$ by \cite{Drees:1995hj}: \bea
m_{\tilde{\ell}_R}^2 &=& M_0^2 + 0.15 M_{1/2}^2 - \sin^2 \theta_W M_Z^2 \cos 2\beta, \nonumber \\
m_{\tilde{\ell}_L}^2 &=& M_0^2 + 0.52 M_{1/2}^2 - (0.5 - \sin^2 \theta_W) M_Z^2 \cos 2\beta, \nonumber \\
m_{\tilde{\nu}}^2 &=& M_0^2 + 0.52 M_{1/2}^2 + 0.5 M_Z^2 \cos 2\beta,
\,
\label{sfermion_masses}
\eea 
where $M_Z$ is the $Z$-boson mass and $\theta_W$ is the weak-mixing
angle; see also the original work of Ref.~\cite{Ibanez:1984vq}.  The
third terms in Eq.~(\ref{sfermion_masses}) originate from the D-term
quartic interactions.

\mymed

For the sleptons of the third generation, the mixing between left and
right chiral states is non-negligible. The stau mass matrix $\mathcal
{M}_{\tilde{\tau}}$ is given by \cite{Gunion:1984yn}
\begin{align}
\mathcal{M}^2_{\tilde\tau} &= \left( {m_{\tau}^2 + A_{LL} 
\qquad m_{\tau} B_{LR}} 
\atop {m_{\tau} B_{LR}\qquad m_{\tau}^2 + C_{RR} } \right) \, ,
\label{eq_staumassmatrix}
\end{align}
with
\begin{align}
\begin{split}
A_{LL}  &=  M^2_{\tilde{\tau}_L} - (0.5 - \sin^2\theta_W) M_Z^2 \cos 2\beta\,, \\
B_{LR}  &= A_{\tau} - \mu \tan{\beta}\,, \\
C_{RR} &=   M^2_{\tilde{\tau}_R} - \sin^2\theta_W M_Z^2 \cos 2\beta \, .
\end{split}
\end{align}
We denote with $m_{\tau}$ the tau lepton mass and with $M_{\tilde{\tau}_L}$ 
and $M_{\tilde{\tau}_R}$ the left- and right-handed third generation
softbreaking stau mass parameters, respectively. At $M_Z$, they can be
approximated by \cite{Drees:1995hj},
\begin{align}
M_{\tilde{\tau}_R}^2 &= M_0^2 + 0.15 M_{1/2}^2 - \frac{2}{3} X_\tau, 
\nonumber \\
M_{\tilde{\tau}_L}^2 &= M_0^2 + 0.52 M_{1/2}^2 - \frac{1}{3} X_\tau,
\label{stau_parameter}\\
X_\tau &\equiv 10^{-4}(1+\tan^2 \beta) \left( M_0^2 + 0.15 M_{1/2}^2 + 
0.33A_0^2 \right).
\nonumber
\end{align}
Here, $X_\tau$ parametrizes the influence of the tau Yukawa coupling on 
the running of the stau masses. 

\mymed

An interesting property of REWSB is that over most of the parameter
space one finds that $\mu^2\gg M_Z^2$. This leads to approximate
relations between neutralino and gaugino masses at $M_Z$. The 
$\tilde{\chi}_1^0$ is dominantly bino-like in mSUGRA
models and its mass can be approximately written as
\cite{Drees:1995hj}
\bea
m_{\tilde{\chi}_1^0} &\simeq & M_1 = 0.41 M_{1/2}, 
\label{neutralino_masses}
\eea
In most of the $\Psix$ mSUGRA parameter space the $\tilde{\chi}_1^0$
or the $\tilde{\tau}_1$ is the LSP
\cite{Ibanez:1984vq,Allanach:2003eb,Allanach:2006st}.

\subsection{$\text{B}_3$ mSUGRA Model}

The SSM allows also for lepton number violating interactions, {\it
cf.} Eq.~(\ref{notP6-superpot}). These additional interactions
increase the number of free parameters from 124 to more than 200. For
detailed phenomenological studies, the ${\text{B}_3}$ mSUGRA model was
proposed in Ref.~\cite{Allanach:2003eb}. Beyond the five mSUGRA
parameters, Eq.~(\ref{mSUGRA_param}), we assume one additional
positive coupling $\BLam' \in\{\lam'_{ijk}\}$ at $M_{\rm GUT}$
\footnote{In general, also $\BLam \in\{\lam_{ijk}\}$ 
at $M_{\rm GUT}$ is allowed in B$_3$ mSUGRA models.
$\kappa_i|_{\rm GUT}$ is rotated away; see 
Ref.~\cite{Allanach:2003eb} for details.}. 
We thus have the six free parameters:
\bea
&&\mzero\,,\, \mhalf\,,\,
\azero\,,\,\tanb\,,\, \sgnmu\,,\,\BLam'\,.
\label{P6V-param} 
\eea
Due to the presence of one $\lam'_{ijk}$ coupling at $M_{\rm GUT}$,
the following changes in collider phenomenology take place compared to
$\Psix$ conserving mSUGRA:
\begin{itemize}
\item The RGEs get additional contributions and consequently the sparticle mass 
spectrum and the couplings at $M_Z$ are altered
\cite{Allanach:2003eb,Allanach:2006st,Jack:2005id}.
\item The LSP can decay into SM particles via the $\lam'_{ijk}$ coupling. 
In principle any sparticle can now be the LSP, because the 
cosmological bound on stable LSPs no longer holds \cite{Ellis:1983ew}. 
\item Sparticles may be produced singly, possibly on resonance
  \cite{Allanach:1997sa,Bernhardt:2008mz,Barbier:2004ez}, \textit{e.g.} single
  slepton production at ha\-dron colliders
  \cite{single_slep,Dreiner:2006sv,Dreiner:2008rv}.
\item The decay patterns of the sparticles can change due  
to changes in the mass spectrum and the additional $\text{B}_3$
interactions; see Refs.~\cite{Allanach:2006st,Dreiner:2008rv} for
explicit examples.
\end{itemize}

\mymed

In this paper we mainly focus on the first aspect and investigate the
effect of a non-vanishing $\lam'_{ijk}|_{\text{GUT}}$ on the
running of the sparticle masses. We show, that a large region in the
$\text{B}_3$ mSUGRA parameter space exists, where a $\tilde{\nu}_i$ is
the LSP. This is consistent with all present experimental
constraints.

\subsection{Sneutrino LSPs in $\text{B}_3$ mSUGRA}
\label{snu_LSP_in_mSUGRA}

In order to understand the dependence of the $\tilde{\nu}_i$ mass at
$M_Z$ on the parameters of Eq.~(\ref{P6V-param}) at the GUT scale, we
must take a closer look at the relevant RGEs. According to
Ref.~\cite{Allanach:2003eb}, the dominant contributions are
\footnote{For $i=3$, we also have terms proportional 
to $(\mathbf{Y}_E)_{33}^2$, {\it i.e.} proportional
to the tau Yukawa coupling squared.}:
\bea
16\pi^2 \frac{d(m^2_{\tilde{\nu}_i})}{dt} &=& 
- \frac{6}{5} g_1^2 M_1^2 - 6 g_2^2 M_2^2
- \frac{3}{5} g_1^2 {\cal S} \nonumber \\
& & + \, 6\lam'^2_{ijk}\left[ (\mathbf{m_{\tilde{L}}})^2_{ii} 
+(\mathbf{m_{\tilde{Q}}})^2_{jj} +(\mathbf{m_{\tilde{D}}})^2_{kk} \right] 
\nonumber \\
& & + \, 6 (\mathbf{h_{D^k}})_{ij}^2 
\label{sneu_RGE}
\eea
with
\bea
(\mathbf{h_{D^k}})_{ij} \equiv \lam'_{ijk} \times A_0 \qquad \text{at} 
\,\,\, M_{\rm GUT}\,, 
\label{hdk_RGE}
\eea
and 
\bea
{\cal S} &\equiv& \Tr[{\bf m_{\tilde{Q}}}^2-
{\bf m_{\tilde{L}}}^2-2{\bf m_{\tilde{U}}}^2 
+ {\bf m_{\tilde{D}}}^2 + {\bf m_{\tilde{E}}}^2] 
\nonumber \\
& & + m_{H_u}^2-m_{H_d}^2  \,.
\label{trace_s}
\eea
Here $g_1$, $g_2$ are the U(1) and SU(2) gauge couplings, respectively.
$t=\ln Q$ with $Q$ the renormalization scale. $({\bf h}_{D^k})_{ij}$
is the soft breaking coupling corresponding to $\lam'_{ijk}$. The
bold-faced soft mass parameters in Eqs.~(\ref{sneu_RGE}) and
(\ref{trace_s}) are $3\times 3$ matrices in flavor space:
$\mathbf{m_{\tilde{Q}}}$ and $\mathbf{m_{\tilde{L}}}$ for the
left-handed doublet squarks and sleptons; $\mathbf{m_{\tilde{U}}}$,
$\mathbf{m_{\tilde{D}}}$ and $\mathbf{m_{\tilde{E}}}$ for the singlet
up-squarks, down-squarks and sleptons, respectively. There is no
summation over repeated indices in Eq.~(\ref{sneu_RGE}).

\mymed

The running of $m_{\tilde{\nu}_i}$ is governed by two different sets
of terms. The first three terms in Eq.~(\ref{sneu_RGE}) are
proportional to the  gauge couplings squared, $g_1^2$ and $g_2^2$. 
We find that the sum of
these three terms is negative at every scale. They therefore
lead to an increase in $m_{\tilde{\nu}_i}$, going from $M_{\rm GUT}$
to $M_Z$.  This effect leads to the contribution proportional to
$M_{1/2}^2$ in the approximate formula, Eq.~(\ref{sfermion_masses}),
for $m_{\tilde{\nu}_i}^2$. Note, that the main contributions come from
the terms proportional to the gaugino masses squared, $M_{1}^2$ and $M_{2}^2$,
because ${\cal S}$ in Eq.~(\ref{sneu_RGE}), which can be negative, is
identical to zero at $M_{\rm GUT}$ for universal scalar
masses. In addition the coefficients of the $M_{1}^2$ and $M_{2}^2$
terms are larger compared to the ${\cal S}$ term.

\mymed

The remaining contributions are proportional to $\lam'^2_{ijk}$ and
$({\bf h}_{D^k})^2_{ij}$; the latter is also proportional to $\lam'^2
_{ijk}$ at $M_{\rm GUT}$, {\it cf.} Eq.~(\ref{hdk_RGE}). These terms
are positive and will therefore reduce $m_{\tilde{\nu}_i}$, going from
$M_{\rm GUT}$ to $M_Z$. They are also new to the B$_3$ mSUGRA model
compared to minimal mSUGRA. The influence of these new contributions
on $m_{\tilde{\nu}_i}$ depends on the magnitude of $\lam'_{ijk}$
and also on the the other mSUGRA parameters, Eq.~(\ref{P6V-param}),
especially on $A_0$, as we will show in Sect.~\ref{parameter_space}.

\begin{figure}[ht!] \centering
\setlength{\unitlength}{1cm}
\includegraphics[scale=0.43, bb = 50 110 500 530, clip=true]{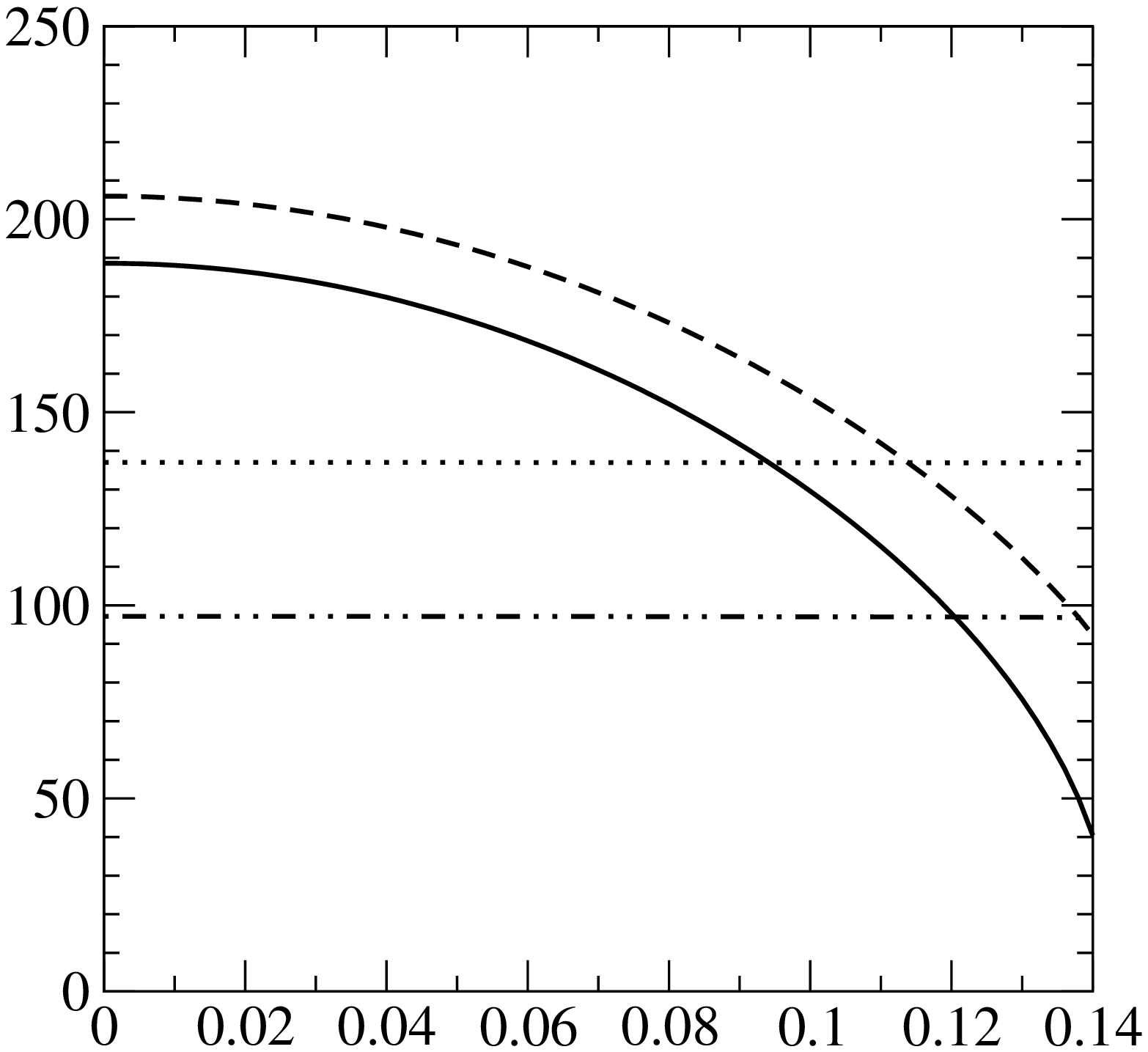}
  \put(-4.3,-0.3){$\lam'_{231}$ at $M_{\rm GUT}$}
  \put(-7.3,2.5){\rotatebox{90}{Mass [GeV]}}
  \put(-5.2,4.3){$\tilde{\nu}_\mu$} \put(-5.2,5.4){$\tilde{\mu}_L$}
  \put(-5.2,3.2){$\sstau_1$} \put(-5.2,2.3){$\neutralino_1$}
  \caption{\label{lambdap231} Masses of $\neutralino_1$, $\sstau_1$,
  $\ssnumu$ and $\tilde{\muon}_L$ at $M_Z$ as a function of
  $\lam'_{231}|_{\text{GUT}}$. The other mSUGRA parameters are that
  of SPS1a \cite{Allanach:2002nj}. We assume up-mixing, {\it cf.}
  Sect.~\ref{quark_mixing}.}
\end{figure}

\mymed

In Fig.~\ref{lambdap231}, we demonstrate the impact of a non-vanishing
$\lam'_{231}|_{\rm GUT}$ on the running of $m_{\tilde{\nu}_i}$. We
have chosen the mSUGRA point SPS1a \cite{Allanach:2002nj}, where in
the $\Psix$ conserving case, the $\neutralino_1$ is the LSP and the
$\sstau_1$ is the next-to-lightest supersymmetric particle (NLSP). See
also Ref.~\cite{Allanach:2006st} for the case of $\lam'_{331}|_{\rm
GUT}$. The mass of the muon sneutrino, $\tilde{\nu}_ {\mu}$, decreases
for increasing $\lam'_{231}|_{\rm GUT}$, as described by
Eq.~(\ref{sneu_RGE}). Furthermore, the mass of the left-handed smuon,
$\tilde{\mu}_L$, decreases, as it belongs to the same SU(2)
doublet. The running of the $\tilde{\mu}_L$ mass squared is also
described by Eq.~(\ref{sneu_RGE}). But note that the mass difference
between $\ssnumu$ and $\tilde{\mu}_L$, is not the same with varying
$\lam'_{231}|_{\rm GUT}$ as can be seen in Fig.~\ref{lambdap231}.
This is due to the different D-term contributions to
$m_{\tilde{\nu}_\mu}$ and $m_{\tilde{\mu}_{L}}$, {\it cf.} 
Eq.~(\ref{sfermion_masses}), for different $\lam'_{231}|_{\rm GUT}$.  
The mass difference is approximately 20
GeV (50 GeV) for $\lam'_ {231}|_{\rm GUT} = 0.0\, (0.14)$.
The $\tilde{\mu}_L$ is also always heavier than the
$\tilde{\nu}_{\mu}$, as long as $\tan \beta>1$.  We calculated the
sparticle masses in Fig.~\ref{lambdap231} with an unpublished
$\text{B}_3$ version of {\tt SOFTSUSY} \cite{rpv_softsusy,footnote_SOFTSUSY}.

\mymed

At one-loop order, the masses of the $\neutralino_1$ and the $\sstau_1
$, are not changed, as can be seen in Fig.~\ref{lambdap231}. They do
not directly couple to the $L_2 Q_3 \bar D_1$ operator, in contrast to
$\tilde{\nu}_{\mu},\,\tilde{\mu}_L$. We therefore obtain for the
parameter set SPS1a with $\lam'_ {231}|_{\rm GUT} > 0.12$ a new
candidate for the LSP, namely the sneutrino! In the following, we
systematically investigate the conditions which lead to a $\tilde{\nu}_i$
LSP in $\text{B}_3$ mSUGRA models. From Eq.~(\ref{sneu_RGE}) it is
clear that we need a coupling $\lam'_ {ijk}|_{\rm GUT}\not=0$. The
smallest $\lam'_{ijk}|_{\rm GUT}$ coupling which we found leading
to a $\tilde{\nu}_i$ LSP is $\lam '_{ijk}|_{\rm GUT}= 0.054$.
Otherwise, the new contributions in the RGE, Eq.~(\ref{sneu_RGE}), are
not large enough to reduce $m_{\tilde{\nu}_i}$ significantly.

\mymed

A non-vanishing $\lam'_{ijk}|_{\rm GUT}$ also reduces the left-handed
squark masses of generation $j$ and the right-handed down-squark
masses of generation $k$, because these squarks couple directly to the
$L_i Q_j \bar D_k$ operator \cite{Allanach:2006st}.  One might worry
that this effect leads to unwanted flavour changing neutral currents
(FCNCs), when we rotate the quarks and squarks from the flavour-basis
to their mass-basis. But, for example for SPS1a with $\lam'_{231}|_{\text{GUT}} =
0.13$, the respective squark masses are reduced by less
than $4\%$, thus avoiding FCNCs which are in contradiction with
experiment \cite{FCNCs_refs}.

\mymed

We investigate in Sect.~\ref{bounds_on_snu_LSP} the experimental
bounds on $\text{B}_3$ mSUGRA models with a $\tilde{\nu}_i$ LSP. For
this purpose, we need to take a closer look at quark-flavour mixing.

\subsection{Quark Mixing}
\label{quark_mixing}

The RGEs of the different $\text{B}_3$ couplings are coupled via the
matrix elements of the lepton- and quark-Yukawa matrices,
Eq.~(\ref{P6-superpot}). Assuming a diagonal lepton Yukawa matrix,
$\mathbf{Y}_E$, a non-vanishing $\lam'_{ijk}$
coupling at $M_{\rm GUT}$ will generate at $M_Z$ all other
$\text{B}_3$ couplings which violate the same lepton number, see also
Ref.~\cite{Dreiner:2008rv}. We thus need to know the up- and
down-quark Yukawa matrices, $\mathbf{Y}_U$ and $\mathbf{Y}_D$,
respectively.

\mymed   

From experiment, we only know the Cabibbo-Kobayashi-Maskawa (CKM)
matrix
\begin{equation}
\mathbf{V}_{\text{CKM}} = \mathbf{V_{uL}} \mathbf{V_{dL}^\dagger}.
\end{equation}
Here, $\mathbf{V_{uL}}$ ($\mathbf{V_{dL}}$) rotates the left-handed
up- (down-) type quarks from the electroweak basis to the mass basis.
For simplicity, we assume that the Yukawa matrices $\mathbf{Y}_U$ and
$\mathbf{Y}_D$ are real and symmetric, thus $\mathbf{V_{uL}}=
\mathbf{V_{uR}}$ and $\mathbf{V_{dL}}=\mathbf{V_{dR}}$. We can imagine
two extreme cases. We refer to ``up-mixing" if
\begin{align}
{\bf V_{u\,L,R}^{}}  = {\bf V_{\rm CKM}}, \quad {\bf V_{d\,L,R}^{}}  
= {\bf 1}_{3\times3}, \label{up_mixing}
\end{align}
at $M_Z$, {\it i.e.} the up-type Yukawa matrix $\mathbf{Y}_U$ is
non-diagonal and $\mathbf{Y}_D$ is diagonal. In the case of
``down-mixing", we have
\begin{align}
{\bf V_{u\,L,R}^{}}={\bf 1}_{3\times3}, \quad {\bf V_{d\,L,R}^{}}=
{\bf V^\dagger_{\rm CKM}},\label{down_mix}
\end{align}
at $M_Z$. Now, the down-type Yukawa matrix $\mathbf{Y}_D$ is
non-diagonal and $\mathbf{Y}_U$ is diagonal. For a more detailed
discussion see for example Refs.~\cite{Agashe:1995qm,
Allanach:2003eb,Dreiner:2008rv}.

\section{Experimental Bounds on $\tilde{\nu}$ LSP Models}
\label{bounds_on_snu_LSP}

We have shown above that a non-vanishing coupling $\lam '_{ijk}$ at
$M_{\rm GUT}$ can affect the spectrum at $M_Z$ such that a $\tilde
{\nu}_i$ is the LSP. This requires $\lam'_{ijk}|_{\rm GUT}\gsim 0.05$,
corresponding to $\lam'_{ijk}\gsim0.15$ at $M_Z$.  In this section, we
investigate for which couplings $\lam'_{ijk}|_{\rm GUT}$ the upper
bounds are sufficiently weak such that a $\tilde{\nu}_i$ LSP can be
generated. For the bounds, we first take into account the generation
of tree level neutrino masses. Then we review other indirect bounds on
these couplings. Finally we discuss the restrictions from direct
searches for supersymmetric particles at LEP, at the Tevatron and the
CERN $p \bar p$ collider.

\subsection{Bounds from Tree Level Neutrino Masses}
\label{tree_neut_bounds}

If $\lam'_{ijk}|_{\rm GUT}\not=0$ and the bilinear coupling
$\kappa_i|_{\rm GUT}=0$, \textit{cf.}  Eq.~(\ref{notP6-superpot}),
$\kappa_i|_{\rm M_Z}\not=0$ will be generated via the RGEs
\cite{Allanach:2003eb,Carlos:1996du,Dreiner:1995hu,Barger:1995qe,Nardi:1996iy}
\begin{equation}
16\pi^2\frac{d\kappa_i}{dt} = - 3 \mu \lam'_{ijk} ({\bf Y}_D)_{jk} + \dots \, .
\label{gen_kappa}
\end{equation}
Furthermore, $\lam'_{ijk}|_{\rm GUT}$ will generate the
corresponding soft breaking term of $\kappa_i$, namely $\tilde{D}_i$,
via
\cite{Allanach:2003eb,Carlos:1996du,Dreiner:1995hu,Barger:1995qe,Nardi:1996iy}
\begin{equation}
16\pi^2\frac{d\tilde{D}_i}{dt} = -3 \left[2 \mu ({\bf h}_{D^k})_{ij} + 
\tilde{B} \lam'_{ijk} \right] 
({\bf Y}_D)_{jk} + \dots \, .
\label{gen_D}
\end{equation}
Here, $\tilde{B}$ is the soft breaking coupling corresponding to $\mu$
and is determined by REWSB \cite{Ibanez:1982fr,Allanach:2003eb}. Since
the RGEs are different for $\kap_i$ and $\tilde D_i$, they are not
aligned at the weak scale and can not be rotated away through a field
redefinition.

\mymed

The neutrino of generation $i$ will develop a vacuum expectation value
$v_i$ due to the non-vanishing couplings $\kappa_i$ and $\tilde{D}_i$.
The vacuum expectation value $v_i$, and the $\kappa_i$ operator will
mix the neutralino fields with the neutrino fields which generates one
massive neutrino, $m_{\nu_i}$, for non-vanishing $\lam'_{ijk}|_{\rm G
UT}$ at tree-level
\cite{Nardi:1996iy,Hall:1983id,Ellis:1984gi,Banks:1995by,Allanach:2003eb}.

\mymed

Demanding that this neutrino mass is smaller than the cosmological
bound on the sum of neutrino masses, determined by the combination of
the WMAP data \cite{Spergel:2003cb} and the 2dFGRS data
\cite{Colless:2003wz},
\begin{equation}
\sum_i m_{\nu_i} < 0.71 \, \text{eV} \, ,
\label{WMAP_bound}
\end{equation}
results in upper bounds on $\lam'_{ijk}|_{\rm GUT}$, which were
calculated in Ref.~\cite{Allanach:2003eb} for the parameter point
SPS1a \cite{Allanach:2002nj}.

\mymed

It was found in Ref.~\cite{Allanach:2003eb}, assuming quark mixing
solely in the down-sector (\ref{down_mix}) and assuming no accidental
cancellations, that the bounds on $\lam'_{ijk}|_{\rm GUT}$ are of the
order of $\mathcal{O}(10^{-3}-10^{-6})$.  However, if quark mixing is
solely in the up-sector (\ref{up_mixing}), than $({\bf Y}_D)_{jk}$ vanishes at $M_Z$ for
$j\not =k$. This suppresses the right hand side of
Eq.~(\ref{gen_kappa}) and Eq.~(\ref{gen_D}).  The neutrino masses and
therefore the bounds on $\lam'_{ijk}|_{\rm GUT}$ are significantly
softened. Taking also two loop effects into account, we summarize in
Table~\ref{RPV_couplings} the $\lam'_{ijk}$ couplings, which are
unrestricted by the neutrino mass bound, Eq.~(\ref{WMAP_bound}), as
long as quark mixing is dominantly in the up-sector, \textit{cf.}
Eq.~(\ref{up_mixing}). We also include the strictest experimental
bound, which we discuss in the following subsection.

\subsection{Indirect Bounds on $\lam'_{ijk}$}
\label{indirect_bounds}

\begin{table}[t!]
\begin{ruledtabular}
\begin{tabular}{ccc}
 coupling & upper bounds at $M_Z$  & LSP \\
 \hline
 $\lam'_{121}$ & $0.03\times (m_{\tilde{c}_L}/100 \, \text{GeV})$ & $\tilde{\nu}_e$  \\ 
 $\lam'_{131}$ & $0.02\times (m_{\tilde{t}_L}/100 \, \text{GeV})$ & $\tilde{\nu}_e$  \\ 
 $\lam'_{112}$ & $0.02\times (m_{\tilde{s}_R}/100 \, \text{GeV})$ & $\tilde{\nu}_e$  \\
 $\lam'_{221}$ & $0.18 \times (m_{\tilde{s}_L}/100 \, \text{GeV})$ & $\tilde{\nu}_\mu$  \\  
 $\lam'_{231}$ & $0.18\times (m_{\tilde{b}_L}/100 \, \text{GeV})$ & $\tilde{\nu}_\mu$  \\ 
 $\lam'_{212}$ & $0.06\times (m_{\tilde{s}_R}/100 \, \text{GeV})$ & $\tilde{\nu}_\mu$  \\   
 $\lam'_{321}$ & $0.52 \times (m_{\tilde{d}_R}/100 \, \text{GeV})$ & $\tilde{\nu}_\tau$  \\
 $\lam'_{331}$ & $0.32 \times (m_{\tilde{d}_R}/100 \, \text{GeV})$ & $\tilde{\nu}_\tau$  \\ 
 $\lam'_{312}$ & $0.11\times  (m_{\tilde{s}_R}/100 \, \text{GeV})$ & $\tilde{\nu}_\tau$     
\end{tabular}
\caption{\label{RPV_couplings} Upper bounds on single couplings
  $\lam'_{ijk}$ from electroweak precision measurements. Only
  couplings are shown, which are consistent with the cosmological
  bound on neutrino masses, Eq.~(\ref{WMAP_bound}); see also
  Ref.~\cite{Allanach:2003eb}. The bounds depend strongly on the
  masses of the relevant squarks, $m_{\tilde{q}}$.  
  The third column shows the $\tilde{\nu}_i$ LSP,
  which can be generated via the respective $\lam'_{ijk}|_{\rm GUT}$
  coupling.}
\end{ruledtabular}
\end{table}

In this section, we review the relevant indirect bounds on the
couplings $\lam'_{ijk}$ from electroweak precision measurements.  In
Table~\ref{RPV_couplings}, we present the strongest bounds on the
single $\lam'_{ijk}$ couplings at the $2\sigma$ level
\cite{Barbier:2004ez,Allanach:1999ic,bounds_rpv,Dreiner:1997uz}.  The
bounds apply to the couplings at $M_Z$. To obtain the respective bound
at $M_{\rm GUT}$ one has to divide the corresponding bound in
Table~\ref{RPV_couplings} by roughly a factor of three. For each
coupling the bound depends linearly on the sfermion mass of the
virtual particle exchanged in the relevant process. In the right
column, we show which sneutrino can become the LSP.  We see that an
electron sneutrino LSP, $\tilde{\nu}_e$, is disfavoured due to the
strong bounds on the couplings $\lam'_{1jk}$. We have found that only
in a small range of mSUGRA parameter space a $\tilde{\nu}_e$ LSP is
found, although large squark masses weaken the bounds.  In the
following we will thus concentrate on muon sneutrinos,
$\tilde{\nu}_\mu$, and tau sneutrinos, $\tilde{\nu}_\tau$ as LSP
candidates.

\mymed

One non-vanishing $\lam'_{ijk}|_{\rm GUT}$ will also generate
additional ($LQ\bar D$ \textit{and} $LL\bar E$) $\text{B}_3$ operators
at $M_Z$, which violate the same lepton number \cite{Dreiner:2008rv}.
For example, for one $\lam'_ {2jk}|_{\rm GUT} \not = 0$, we will
generate all other muon number violating operators at $M_Z$ via one
and two loop effects. Since bounds on products of two different
$\text{B}_3$ couplings are often much stronger than on only one
$\text{B}_3$ coupling
\cite{Barbier:2004ez,Allanach:1999ic,bounds_rpv,Dreiner:1997uz}, we
have also checked that all generated products of the dominant
$\lam'_{ijk}$ coupling with a generated coupling satisfy the bounds. All
products lie at least one order of magnitude below the strongest upper
bounds if $\lambda'_{ijk}|_{\rm GUT}=0.1$.

\mymed

After REWSB, the single coupling scheme, which was assumed in deriving
the bounds in Table~\ref{RPV_couplings}, cannot be realized in the
quark mass eigenbasis \cite{Agashe:1995qm}. In
Sect.~\ref{tree_neut_bounds}, we stated that quark mixing must be
dominantly in the up-sector, Eq.~(\ref{up_mixing}), to fulfill the
cosmological bound on the sum of neutrino masses,
Eq.~(\ref{WMAP_bound}).  Therefore, in the quark mass basis we will
generate the following $\text{B}_3$ couplings
\begin{equation}
\tilde{\lam}'_{imk}=(\mathbf{V}^*_{CKM})_{mj} \lam'_{ijk} \, .
\label{lamp_tilde} 
\end{equation} 
$\tilde{\lam}'_{imk}$ with $m=1,2,3$ couples an up-quark superfield
of generation $m$ (in the mass basis) to a lepton and down-quark
superfield of generation $i$ and $k$, respectively. These effective
couplings can give rise to $D_0$--$\bar D_0$ mixing if $m=1,2$
\cite{Agashe:1995qm,Petrov:2007gp,Golowich:2007ka}. $D_0$ oscillations
were investigated by the BABAR \cite{Aubert:2007aa,Aubert:2007wf},
Belle \cite{Abe:2007rd,Staric:2007dt} and CDF \cite{:2007uc}
collaborations.  The Heavy Flavor Averaging Group combined all
experimental results and obtained windows for the allowed mass
difference and the allowed lifetime difference of the $D_0$--$\bar
D_0$ system \cite{Schwartz:2008wa}.

\mymed

Ref.~\cite{Golowich:2007ka} employed the experimental $2\sigma$ errors
on the $D_0$--$\bar D_0$ mass difference to obtain the following
bounds on $\lam'_{ijk}$ 
\begin{eqnarray}
& &|\tilde{\lam'}_{i21} \, \tilde{\lam'}_{i11}| = |\lam_W \, 
\lam^{\prime 2}_{i21}|
\nonumber \\
& & \leq  0.0029 \left [ \left(\frac{100 \text{GeV} }{m_
{\tilde{\ell}_{Li}}}\right )^2 + \left( \frac{100 \text{GeV} }
{m_{\tilde{d}_R}} \right)^2 \right ]^{-1/2} \, ,
\nonumber \\
\label{bound1_D0D0bar}
\end{eqnarray} 
where $\lam_W=0.23$ is the Wolfenstein parameter \cite{wolf_param} 
and $i=1,2,3$. For the evaluation of
Eq.~(\ref{bound1_D0D0bar}), Ref.~\cite{Golowich:2007ka} assumed that
the mass splitting arises solely from $\text{B}_3$ contributions.
Note that the first equality of Eq.~(\ref{bound1_D0D0bar}) only holds
if quark mixing is solely in the up-sector, Eq.~(\ref{up_mixing}). The
corresponding upper bound on $|\lam_W \, \lam^{\prime 2}_{i12}|$ can be
obtained from Eq.~(\ref{bound1_D0D0bar}) by replacing
$m_{\tilde{d}_R}$ with $m_{\tilde{s}_R}$.

\mymed

The experimentally allowed range for the difference in
lifetime of the $D_0$--$\bar D_0$ system was used in
Ref.~\cite{Petrov:2007gp} to obtain the bounds
\begin{eqnarray}
  |\tilde{\lam'}_{i21} \, \tilde{\lam'}_{i11}| = |\lam_W \, 
\lam^{\prime 2}_{i21}| \leq  0.082  \left(\frac{m_{\tilde{\ell}_
{Li}}}{100 \text{GeV} }\right )^2. 
\label{bound22_D0D0bar}
\end{eqnarray} 
These are valid for $i=1,2$. Unlike Ref.~\cite{Golowich:2007ka},
Ref.~\cite{Petrov:2007gp} also took (destructive) interference between
the $\text{B}_3$ and SM contributions into account. The bound on
$|\lam_W \, \lam^{\prime 2}_{i12}|$ is the same.

\mymed

If we assume a $\tilde{\ell}_{Li}$ with a mass of 200 GeV and squarks
with a mass of 500 GeV, we obtain the upper bounds $\lam'_{i21},
\lam'_{i12} \leq 0.15$ at $M_Z$ from the $D_0$--$\bar D_0$ mass
difference, Eq.~(\ref{bound1_D0D0bar}), and $\lam'_{i21},
\lam'_{i12}\leq 1.2$ at $M_Z$ from the $D_0$--$\bar D_0$ lifetime
difference Eq.~(\ref{bound22_D0D0bar}).  Thus the $\tilde{\nu}_i$ LSP
parameter space is strongly restricted by the $D_0$--$\bar D_0$ mass
difference.  However it was pointed out in Ref.~\cite{Petrov:2007gp}
that destructive interference, for example between $\Psix$ violating
and $\Psix$ conserving contributions, may significantly weaken the
bounds of Eq.~(\ref{bound1_D0D0bar}), as in the case of the
$D_0$--$\bar D_0$ lifetime difference.

\mymed

In the following, we mainly focus on the couplings $\lam'_{231}$ and
$\lam'_{331}$ leading to a $\tilde{\nu}_\mu$ and $\tilde{\nu}_\tau$
LSP, respectively. These couplings are not restricted by $D_0$--$\bar
D_0$ mixing, because the relevant CKM matrix elements to generate
$\tilde{\lam}'_{i21}$ and $\tilde{\lam}'_{i11}$ out of $\lam'_{i31}$
are too small, {\it cf.} Eq.~(\ref{lamp_tilde}).

\subsection{Collider Constraints}

\subsubsection{Constraints from LEP}
\label{LEP_constraints}

We now determine bounds on the $\tilde{\nu}_i$ LSP mass from LEP. For
the case of a non-vanishing $\lam'_{ijk}$ coupling the $\tilde{\nu}_i$
LSP will dominantly decay into two jets:
\bea
\tilde{\nu}_i &\rightarrow& \bar d_j d_k.
\label{sneut_decay}
\eea 
Here, $d_k$ ($\bar d_j$) is a (anti) down quark of generation $k$
($j$). This decay will occur instantaneously in the detector,
\textit{i.e.} with no detached vertex, since in our model $\lam'_
{ijk}$ is bounded from below by the requirement of a $\tilde{\nu}_i$ LSP.
$\tilde{\nu}_i$ pair production followed by the decay,
Eq.~(\ref{sneut_decay}), would lead to four jet events at LEP. 

\mymed

Bounds on the total $\tilde{\nu} _i$ pair production cross section,
with the $\tilde{\nu} _i$ decaying via $\lam'_{ijk}$ were obtained by
the OPAL collaboration \cite{Abbiendi:2003rn} and also by the ALEPH
collaboration \cite{Heister:2002jc}.
\begin{table}[t]
\begin{ruledtabular}
\begin{tabular}{c|ccc}
 & $m_{\tilde{\nu}_e}$ & $m_{\tilde{\nu}_\mu}$ & $m_{\tilde{\nu}_\tau}$  \\
 \hline
 OPAL & $>68-89$ GeV & $>74$ GeV & $>74$ GeV \\
 ALEPH & $>75-95$ GeV & $>79$ GeV & $>79$ GeV  
\end{tabular}
\caption{\label{bounds_LEP} Lower bounds on the $\tilde{\nu}_i$ LSP
  masses from direct $\tilde{\nu}_i$ decay via $\lam'_{ijk}$. The
  bounds were obtained from the OPAL \cite{Abbiendi:2003rn} and ALEPH
  \cite{Heister:2002jc} analyses, respectively.  The $\tilde{\nu}_\mu$
  and $\tilde{\nu}_\tau$ mass bounds are universal. The $\tilde{\nu}
  _e$ mass bound depends on the chargino parameters due to potential
  interference effects.}
\end{ruledtabular}
\end{table} 
{}From these we can obtain lower bounds on the mass of the $\tilde
{\nu}_i$ LSP. We calculated the pair production cross section using
the formulas given in Ref.~\cite{Wendel:1990yc}, with the fine
structure constant equal to its value at $M_Z$, {\it i.e.}  $\alpha =
1/128$. We show in Table~\ref{bounds_LEP} the strongest lower bounds
on the $\tilde{\nu}_i$ LSP masses for different lepton flavours $i$.

\mymed

The $\tilde{\nu}_i$ LSP mass bounds for the second and third
generation ($i=2,3$) are universal. The $\tilde{\nu}_e$ mass bound, in
contrast, depends also on the chargino parameters. The chargino
parameters enter through t-channel diagrams to the sneutrino pair
production cross section. We calculated the different bounds on the
electron sneutrino mass by assuming, that the lightest chargino is
wino-like.  This is the case for most mSUGRA scenarios. We then varied
its mass between 120 GeV and 1000 GeV to obtain the numbers in
Table~\ref{bounds_LEP}.

\mymed

In the following, we investigate the $\tilde{\nu}_\mu$ LSP and $\tilde
{\nu}_\tau$ LSP parameter space in detail.  A $\tilde{\nu}_e$ LSP is
less favoured due to the stronger bounds on the $\lam'_{1jk}$
couplings, {\it cf.} Table~\ref{RPV_couplings}. We employ a lower mass
bound of 78 GeV. This corresponds to the bound obtained by the ALEPH
collaboration, see Table~\ref{bounds_LEP}, reduced by 1 GeV to account
for numerical uncertainties in {\tt SOFTSUSY} \cite{Allanach:2003jw}.

\mymed

Only the mass bounds of the directly decaying $\tilde{\nu}_i$ LSP need
to be considered, because all the other bounds from LEP on direct and
indirect decays of heavier sparticles (compared to the $\tilde{\nu}_i$
LSP) are automatically fulfilled.  In addition, the LEP exclusion
bound on the light Higgs, $h$, is $m_h > 114.4$ GeV at $95 \%$
confidence level \cite{Barate:2003sz}.  Anticipating a numerical error
of 3 GeV of {\tt SOFTSUSY}s prediction of $m_h$
\cite{Allanach:2006st,Allanach:2003jw,Degrassi:2002fi,Allanach:2004rh}, 
we have imposed a lower bound of 111.4 GeV.

\subsubsection{Constraints from the Tevatron}
\label{Tevatron_constraints}

At the Tevatron, a non-vanishing $\lam'_{ijk}$ coupling allows for
resonant single $\tilde{\ell}^-_{Li}$ and $\tilde{\nu}_i$ production
leading to dijet events
\bea \bar u_j d_k & \rightarrow
\tilde{\ell}_{Li}^- & \rightarrow \bar u_j d_k \, ,
\label{res_slep}\\
\bar d_j d_k & \rightarrow  \tilde{\nu}_i & \rightarrow  \bar d_j d_k\, .
\label{res_sneu}
\eea
The expected reach for the slepton resonance search at the Tevatron in
the dijet channel is estimated in Ref.~\cite{Hewett:1998fu} as a
function of the hadronic cross section for the processes in
Eqs.~(\ref{res_slep}),\,(\ref{res_sneu}) and the slepton mass. In
Ref.~\cite{Hewett:1998fu}, the discovery potential for slepton masses
between 200 GeV and 1200 GeV is given assuming an integrated
luminosity of 2 $\text{fb}^{-1}$ and 30 $\text{fb}^{-1}$. We have
checked that all the couplings shown in Table~\ref{RPV_couplings},
assuming $\lam'_{ijk}|_{\rm GUT}=0.1$, lead to production cross
sections which lie at least one order of magnitude below the expected
discovery region for 2 $\text{fb}^{-1}$ given in
Ref.~\cite{Hewett:1998fu}. We have employed the QCD and SUSY-QCD
next-to-leading order (NLO) cross section \cite{Dreiner:2006sv}.

\mymed

Tevatron searches for new resonances in the dijet channel have indeed
been performed by the D0 collaboration \cite{Abazov:2003tj} and the
CDF collaboration \cite{Abe:1995jz,Abe:1997hm,CDFnote}. Although
$\text{B}_3$ models were not considered, bounds on the production
cross section of additional vector bosons, $W^\prime$ and $Z^\prime$,
which decay into two jets, were obtained. These processes are very
similar to the $\text{B}_3$ processes, Eqs.~(\ref{res_slep}) and
(\ref{res_sneu}). $W^\prime$ and $Z^\prime$ masses between 180 GeV
and 1400 GeV were probed.  In this mass region, the production cross
section for a single $\tilde{\ell}_{Li}^-$ and $\tilde{\nu}_i$ with
subsequent decay into two jets, lies at least one order of magnitude
below the experimental limits on $W^\prime$ and $Z^\prime$ production.
We assumed $\lamp_{ijk}|_{\rm GUT}=0.1$ and one coupling of
Table~\ref{RPV_couplings}.

\mymed

\begin{table}
\begin{ruledtabular}
\begin{tabular}{cc|cc}
 process & & \multicolumn{2}{c}{cross section [pb]}   \\
 \hline
 $P \bar P \rightarrow  W (Z) \rightarrow q \bar q$ & & $2.7 \times 10^4$ & ($7.9 \times 10^3$) \\
 $P \bar P \rightarrow  \tilde{\mu}_L \rightarrow q \bar q$ & & $9.2 \times 10^2$ & ($5.7 \times 10^2$) \\
 $P \bar P \rightarrow  \tilde{\nu}_\mu \rightarrow q \bar q$ & & $1.3 \times 10^3$ & ($8.0 \times 10^2$)
\end{tabular}
\caption{\label{WZ_xsection}
        Hadronic cross section for dijet production via an on shell $W$ ($Z$) boson 
		in comparison to $\text{B}_3$ violating dijet production via $\tilde{\mu}_L$, 
        Eq.~(\ref{res_slep}) and $\tilde{\nu}_\mu$, Eq.~(\ref{res_sneu}), with a mass equal to the $W$ ($Z$) mass.
        We assumed $\lam'_{221}|_{\rm GUT}=0.1$. The charge conjugated processes 
		are also taken into account.}
\end{ruledtabular}
\end{table}

We now estimate if the Tevatron has a chance to observe dijet pair
production for $\tilde{\ell}_{Li}^-$ and $\tilde{\nu}_i$ masses {\it
below} 180 GeV. We show in Table~\ref{WZ_xsection} the hadronic cross
sections for dijet production via an on-shell $W$ ($Z$) boson
\cite{D0note,Yao:2006px}. We also give the NLO production cross
section for a $\tilde{\ell}_{Li}^-$ and $\tilde{\nu}_i$ with a mass
equal to the $Z$ and $W$ mass \cite{Dreiner:2006sv}, assuming
$\lamp_{221}|_{\rm GUT}=0.1$. We see that the $\text{B}_3$ cross
sections are roughly one order of magnitude smaller than the SM cross
sections. We conclude that the processes, Eqs.~(\ref{res_slep}) and
(\ref{res_sneu}), for slepton masses below 180 GeV can not be seen at
the Tevatron because the $Z$ and the $W$ have not been observed at the
Tevatron in the dijet channel so far.

\mymed

Singly produced charged sleptons, Eq.~(\ref{res_slep}), may also
cascade decay into a lepton $\ell_i$, two jets and missing energy:
\begin{align}
\tilde{\ell}_{Li}^- \rightarrow  & \tilde{\chi}_1^0 \ell_i^- \nonumber \\
 & \hookrightarrow  \tilde{\nu}_i \bar \nu_i \nonumber \\
 & \qquad \hookrightarrow \bar d_j d_k \, .
 \label{slep_cascade}
\end{align}
In principle, this signature could be more easily distinguished from
the (QCD) background than pure dijet events, due to the additional
isolated lepton in the final state.  However the cascade decay,
Eq.~(\ref{slep_cascade}), is kinematically forbidden in most regions of
the $\tilde{\nu}_i$ LSP parameter space, as we show in
Sect.~\ref{parameter_space}. In that case one might think about the
3-body decay, $\tilde{\ell}_{Li}^- \rightarrow \ell_i^- \bar \nu_i 
\tilde{\nu}_i$, via a virtual neutralino. However, this process can 
only occur at a significant rate, if the 2-body decay mode into two
jets, Eq.~(\ref{res_slep}), is forbidden or kinematically
suppressed. This is the case for $j=3$, {\it i.e.} a top quark in the
final state. But the $\tilde{\ell}_{Li}^-$ can then not be
produced as a single resonance, because we also need a top quark in
the initial state, see Eq.~(\ref{res_slep}). Furthermore the 3-body
decay, $\tilde{\ell}_{Li}^- \rightarrow \ell_i^- \bar \nu_i
\tilde{\nu}_i$, is heavily suppressed compared to the 3-body decay via
a virtual top-quark, as we will see in Sect.~\ref{example_spectrum}.

\mymed

A non-vanishing $\lam'_{i31}$ coupling can lead to $\text{B}_3$
top-quark decay at the Tevatron
\cite{Dreiner:1991dt,Agashe:1995qm,Belyaev:2004qp,Eilam:2001dh,Abraham:2000kx,Ghosh:1996bm}.
For example $t \rightarrow d\,\tilde{\ell}_{Li}$ if $m_{\tilde{\ell
i}} < m_t$.  However, the Tevatron can only
test couplings $\lam'_{i31}$ via top decay, which lie at their upper
bounds \cite{Ghosh:1996bm}, see Table~\ref{RPV_couplings}. We use smaller
$\lam'_{i31}$ couplings in the following.

\mymed

A non-vanishing $\lam'_{i31}$ coupling contributes also to top-pair
production, see Refs.~\cite{Ghosh:1996bm,Hikasa:1999wy,Li:2006he}.
The top quarks in the $t \bar t$ events are polarized, since the
$\text{B}_3$ operator couples only to left-handed top quarks. It is
shown in Refs.~\cite{Ghosh:1996bm,Hikasa:1999wy,Li:2006he}, that the
Tevatron at the end of Run~II can only test couplings $\lam'_{i31}$,
which lie near their current upper bounds, {\it cf.}
Table~\ref{RPV_couplings}. The LHC will be able to probe couplings
$\lam'_{i31}$ down to $\lam'_{i31}=0.2$ via top polarization
\cite{Li:2006he}.

\subsubsection{Constraints from the CERN $p \bar p$ Collider}

Unlike D0 and CDF, the UA2 collaboration at the CERN $p \bar p$
collider was able to measure the hadronic decay mode of the $Z$ and
$W$ \cite{Alitti:1990kw}.  They also searched for a $W^\prime$ and
$Z^\prime$ decaying into two jets.  They found no excess over the SM
background and therefore set exclusion limits for $W^\prime$ and
$Z^\prime$ production with masses between 80 GeV and 320 GeV
\cite{Alitti:1990kw,Alitti:1993pn}.

\mymed

We compared the exclusion limits with our NLO cross section
predictions for single slepton, Eq.~(\ref{res_slep}), and sneutrino,
Eq.~(\ref{res_sneu}), production assuming again $\lamp_{ijk}|_{\rm GUT}=0.1$ 
and one of the couplings shown in Table~\ref{RPV_couplings}
\cite{Dreiner:2006sv}. Our cross section prediction is at least one
order of magnitude smaller than the exclusion limits in the relevant
mass range.

\section{Sneutrino LSP parameter space}
\label{parameter_space}

We have shown in Sect.~\ref{snu_LSP_in_mSUGRA}, that one non-vanishing
coupling $\lam'_{ijk}|_{\rm GUT}=\mathcal{O}(10^{-1})$ may lead to a
$\tilde{\nu}_i$ LSP in $\text{B}_3$ mSUGRA models, {\it cf.}
Fig~{\ref{lambdap231}}. We also presented the $\lam'_{ijk}$ couplings,
which have sufficiently weak upper bounds to allow for a
$\tilde{\nu}_i$ LSP, see Table~\ref{RPV_couplings}.  All lepton
flavours are possible, although a $\tilde{\nu}_e$ LSP is disfavoured
due to the stronger bounds on the $\lam'_{1jk}$. Thus we concentrate
on $\tilde{\nu}_\mu$ and $\tilde{\nu}_\tau$ LSPs in the following.

\mymed

In this section, we investigate in detail the dependence of the
$\tilde{\nu}_i$ LSP parameter space on the mSUGRA parameters $M_0$,
$M_{1/2}$, $A_0$ and $\tan \beta$. This is the central part of our
paper. We explore 2-dimensional parameter spaces, where our scans are
centered around the following points
\begin{align}
\begin{split}
\textnormal{\bf Point I:\,\,}& M_0 = 50 \textnormal{\,GeV},\,  M_{1/2}=500 
\textnormal{\,GeV},
	\\& A_0=-600 \textnormal{\,GeV},\, \tan\beta=10, 
	\\& \textnormal{sgn}(\mu) = +1, \, \lamp_{231}\lvert_{\rm GUT} =0.11,
\\[2ex]
\textnormal{\bf Point II:\,\,}& M_0 = 200 \textnormal{\,GeV},\, M_{1/2}=290 
\textnormal{\,GeV},
	\\& A_0=-550 \textnormal{\,GeV},\, \tan\beta=12, 
	\\& \textnormal{sgn}(\mu) = +1, \, \lamp_{331}\lvert_{\rm GUT} =0.12.
\label{scanpoints}
\end{split}
\end{align}
We perform our parameter scans with an unpublished $\text{B}_3$ version of
{\tt SOFTSUSY} \cite{rpv_softsusy}.

\mymed

Point I results in a $\tilde{\nu}_\mu$ LSP with a mass of 130 GeV. The
NLSP is the left-handed smuon, $\tilde{\muon}_L$, with a mass of 159
GeV. Note that the $\tilde{\muon}_L$ mass is also reduced due to
$\lam'_{231}|_{\rm GUT}\not=0$, and the $\tilde{\muon}_L$ is always
heavier than the $\tilde{\nu}_\mu$ for $\tan \beta > 1$, see
Eq.~(\ref{sfermion_masses}) and Fig.~\ref{lambdap231}.  The masses of
the other LSP candidates, namely the $\tilde{\tau}_1$ and the
$\tilde{\chi}_1^0$, are 186 GeV and 205 GeV, respectively. Due to the
rather large mass difference between the $\tilde{\nu}_\mu$ LSP on the
one side, and $\tilde{\tau}_1$ and $\tilde{\chi}_1^0$ on the other, we
expect an extended $\tilde{\nu}_\mu$ LSP parameter space. This is
indeed the case, as shown in the following.

\mymed
 
Point II results in a $\tilde{\nu}_\tau$ LSP with a mass of 107 GeV.
The NLSP is the $\tilde{\chi}_1^0$ with a mass of 116 GeV.  The NNLSP
is the $\tilde{\tau}_1$, which has a large left-handed component here,
because the soft breaking mass $M_{\tilde{\tau}_L}$,
Eq.~(\ref{stau_parameter}), is also reduced via the non-vanishing
$\lam'_{331}|_{\rm GUT}$ coupling. In contrast, $M_{\tilde{\tau}_R}$
is not affected. The $\tilde{\tau}_1$ mass is 120 GeV.

\mymed

The mass difference between the $\tilde{\nu}_\tau$ LSP and the
$\tilde{\tau}_1$ is smaller for Point II than Point I, because
$\lam'_{331}|_{\rm GUT}$ also reduces the mass of the $\tilde{\tau}_1$,
which is an admixture of $\tilde\tau_L$ and $\tilde\tau_R$. This
competes with the $\tilde{\nu}_\tau$ to be the LSP; \textit{cf.}
Ref.~\cite{Allanach:2006st}. In contrast, $\lam'_ {231}|_{\rm GUT}\not
=0$ reduces the mass of the $\tilde{\muon}_L$.  But the $\tilde{\muon}
_L$ is always heavier than the $\tilde{\nu}_ \mu$. We therefore expect
a smaller $\tilde{\nu}_ \tau$ LSP parameter space around Point II than
the $\tilde{\nu}_\mu$ LSP parameter space around Point I.

\mymed

It is worth mentioning, that Point I leads to a heavier sparticle mass
spectrum than Point II. This stems from the fact, that we have chosen
our central scan points, such that the SUSY contributions to the
anomalous magnetic moment of the muon, $\delta a_\mu^{\rm SUSY}$, can
explain the observed discrepancy, $\delta a_\mu$, between experiment,
$a_\mu^{\rm exp}$, and the SM prediction, $a_\mu^{\rm SM}$,
\begin{equation}
\delta a_\mu= a_\mu^{\rm exp} - a_\mu^{\rm SM} = (29.5 \pm 8.8) \times 10^{-10} \, ,
\label{amu_exp}
\end{equation}
which corresponds to a $3.4 \sigma$ deviation
\cite{Bennett:2006fi,Miller:2007kk,Stockinger:2007pe}. In the
following, we show in our parameter scans in Figs.~\ref{M12-A0:deltaM}
--\ref{fig:M0_M12} contour lines, where the SUSY contributions,
$\delta a_\mu^{\rm SUSY}$, correspond to the
\begin{align}
\begin{split}
\text{central value}:  & \,\, 
\delta a_\mu^{\rm SUSY}=29.5 \times 10^{-10} \\
\Leftrightarrow & \, \text{yellow line},  \, \text{labelled with} ``\, 0 \," \, ,
\\[2ex]
\text{central value} \pm 1 \sigma : &  \,\,
\delta a_\mu^{\rm SUSY}=(29.5 \pm 8.8) \times 10^{-10} \\
\Leftrightarrow & \, \text{blue line}, \,  \text{labelled with} ``\pm 1 " \, ,
\\[2ex]
\text{central value} \pm 2 \sigma : &  \,\,
\delta a_\mu^{\rm SUSY}=(29.5 \pm 17.6) \times 10^{-10} \\
\Leftrightarrow &\, \text{green line} , \,  \text{labelled with} ``\pm 2 " \, ,
\\[2ex]
\text{central value} \pm 3 \sigma : &  \,\,
\delta a_\mu^{\rm SUSY}=(29.5 \pm 26.4) \times 10^{-10} \\
\Leftrightarrow &\, \text{magenta line} , \,  \text{labelled with} ``\pm 3 " \, .
\label{mu_magnetic_moment}
\end{split}
\end{align}
Yellow (labelled with $``\, 0 \,"$), green (labelled with $``\pm1"$),
blue (labelled with $``\pm 2 "$) and magenta (labelled with $`` \pm 3
"$) are the colours of the contour lines in the plots, which we show
in the following sections.

\mymed

The SUSY contributions to the anomalous magnetic moment of the muon,
$\delta a_\mu^{\rm SUSY}$, enter starting at the one loop level, see
for example Refs.~\cite{Grifols:1982vx,Stockinger:2006zn}, and involve
the $\tilde{\muon}_L$ and $\tilde{\nu}_\mu$. Thus, they are enhanced
if the $\tilde{\muon}_L$ and $\tilde{\nu}_\mu$ are light. As a
consequence, $\delta a_\mu^{\rm SUSY}$ increases if we switch on
$\lam'_{231}|_{\rm GUT}$, because the mass of the $\tilde{\muon}_L$ and
$\tilde{\nu}_\mu$ decrease. In contrast, $\lam'_{331}|_{\rm GUT}$ does not affect
$\delta a_\mu^{\rm SUSY}$. Note, that we have not included
$\text{B}_3$ contributions to $\delta a_\mu^{\rm SUSY}$, because they
are at most at the percent level and can therefore be neglected
\cite{Kim:2001se}.  

\mymed

We also consider the constraints from the BR($b \rightarrow s
\gamma$).  The current experimental value is \cite{Barberio:2008fa}
\begin{equation}
\text{BR}(b \rightarrow s \gamma) = (3.52 \pm  0.25) \times 10^{-4}\, .
\label{bsg_exp}
\end{equation}
Here we have added the statistical and systematic errors in quadrature
\cite{Barberio:2008fa}.  If we also include the combined theoretical
error of $0.3 \times 10^{-4}$
\cite{bsg_theo_error} we obtain the $2 \sigma$ window
\begin{equation}
2.74 \times 10^{-4}< \text{BR}(b \rightarrow s \gamma) < 4.30 \times 10^{-4}\, ,
\label{bsg_2sigma}
\end{equation}
where we have now added theoretical and experimental errors in quadrature.

\mymed

The complete $\tilde{\nu}_\mu$ LSP parameter space, which we 
will show in the following, {\it i.e.} Figs.~\ref{M12-A0:deltaM},
\ref{fig:DeltaM_lamp221_a0tanb}, \ref{fig:DeltaM_lamp221_m0m12}, is 
consistent with BR($b \rightarrow s \gamma$) at the $2\sigma$ level
Eq.~(\ref{bsg_2sigma}). The $\tilde{\nu}_\tau$ LSP parameter 
space in the $A_0$--$\tan \beta$ [$M_{1/2}$--$M_{0}$] plane,
Fig.~\ref{fig:DeltaM_lamp331_a0tanb} 
[Fig.~\ref{fig:DeltaM_lamp331_m0m12}] is consistent with BR($b
\rightarrow s \gamma$) at $2\sigma$, Eq.~(\ref{bsg_2sigma}) 
for $\tan \beta\lsim 11$ [$M_{1/2} \gsim 290$ GeV] corresponding to
the dashed black line in Fig.~\ref{fig:DeltaM_lamp331_a0tanb}
[Fig.~\ref{fig:DeltaM_lamp331_m0m12}]. We will show mainly contour
lines for $\delta a_\mu^{\rm SUSY}$ in the following, {\it cf.} 
Eq.~(\ref{mu_magnetic_moment}), because the experimental value of
$a_\mu$ is in general more restrictive on the $\tilde{\nu}_i$ LSP
parameter space than BR($b \rightarrow s \gamma$).

\mymed

We finally want to point out that the complete $\tilde{\nu}_\mu$ and
$\tilde{\nu}_\tau$ LSP parameter space, which we will show in the next
three sections posses a branching ratio for $B_s \rightarrow \mu^+
\mu^-$, which lies at least one order of magnitude below the
current experimental upper bound \cite{Barberio:2008fa},
\begin{equation}
\text{BR}(B_s \rightarrow \mu^+ \mu^-) < 4.7 \times 10^{8}\, .
\end{equation}

\mymed

We have employed {\tt micrOMEGAs1.3.7} \cite{Belanger:2001fz} to
calculate $\delta a_\mu^{\rm SUSY}$, BR($b \rightarrow s \gamma$) and
\text{BR}($B_s \rightarrow \mu^+ \mu^-$). According to Ref.~\cite{
Allanach:2006st}, $\text{B}_3$ contributions to BR($b \rightarrow s
\gamma$) and \text{BR}($B_s \rightarrow \mu^+ \mu^-$) can also be
neglected for only one dominant $\lambda'_{ijk}|_{\rm GUT}$.

\subsection{$A_0$ Dependence}
\label{A0_dependence}

\begin{figure}
  \setlength{\unitlength}{1in} 
  \includegraphics[scale=0.44, bb = 30 40 570 530,
					clip=true]{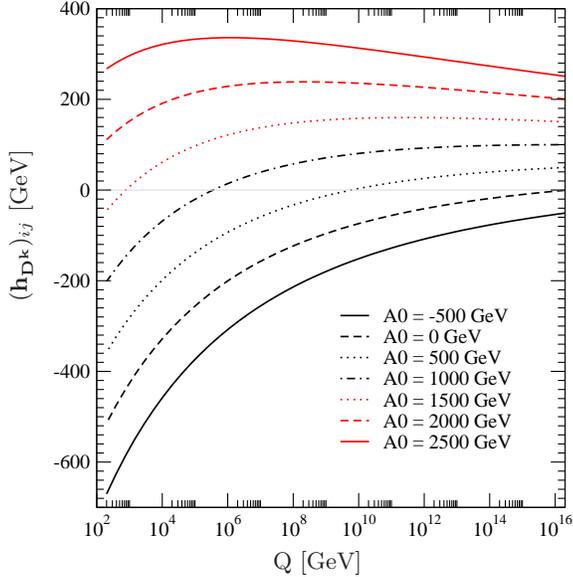}
  \put(-2.0,0.0){Q [GeV]}
  \put(-3.38,1.4){\rotatebox{90}{$(\mathbf{h_{D^k}})_{ij}$ [GeV]}}
  \caption{\label{fig:hDk_M12_500}Running of $(\mathbf{h_{D^k}})_{ij}$
    from $M_{\rm GUT}$ to $M_Z$ for different values of $A_0$.  At
    $M_{\rm GUT}$, we choose $M_{1/2}=500$ GeV and $\lam'_{ijk}=0.1$.}
\end{figure}

\begin{figure}
  \setlength{\unitlength}{1in} 
  \includegraphics[scale=0.44, bb = 30 40 570 530,
					clip=true]{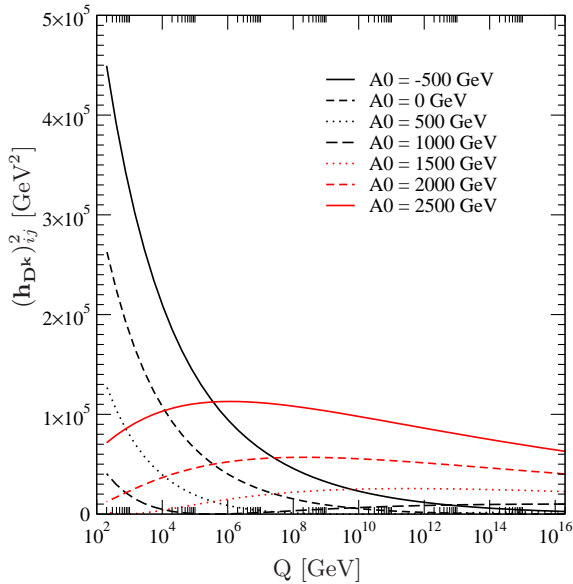}
  \put(-2.0,0.0){Q [GeV]}
  \put(-3.38,1.4){\rotatebox{90}{$(\mathbf{h_{D^k}})^2_{ij}$ [GeV$^2$]}}
  \caption{\label{fig:hDk2_M12_500}Running of
    $(\mathbf{h_{D^k}})^2_{ij}$ from $M_{\rm GUT}$ to $M_Z$ for
    different values of $A_0$.  At $M_{\rm GUT}$, we choose
    $M_{1/2}=500$ GeV and $\lam'_{ijk}=0.1$.}
\end{figure}

We have chosen two scenarios, Point I and Point II,
Eq.~(\ref{scanpoints}), which we use as central values for
2-dimensional mSUGRA parameter scans.  For both points $A_0<0$, with a
magnitude of a few hundred GeV. We now show that this choice of $A_0$
enhances the negative contribution to the $\tilde{\nu}_i$ mass, which
originates from a non-vanishing $\lam'_{ijk}|_{\rm GUT}$ coupling,
{\it cf.} Eq.~(\ref{sneu_RGE}).

\mymed

According to Eq.~(\ref{sneu_RGE}) and (\ref{hdk_RGE}), $A_0$ enters
the running of $m_{\tilde{\nu}_i}$ via the $\text{B}_3$ soft-breaking,
trilinear scalar coupling $(\mathbf{h_{D^k}})_{ij}$ \cite{Allanach:2003eb}. 
Thus $(\mathbf{h_{D^k}})_{ij}$ gives a negative contribution to
$m^2_{\tilde{\nu}_i}$, as $t$ is decreased. It is proportional to the
integral of $(\mathbf{h_{D^k}})_{ij}^2$ over $t$, from $t_{\rm
  min}=\ln(M_{Z})$ to $t_{\rm max}=\ln(M_{\rm GUT})$.

\mymed

We show in Fig.~\ref{fig:hDk_M12_500} the running of the trilinear
scalar coupling $(\mathbf{h_{D^k}})_{ij}$. We assume one non-vanishing
coupling $\lam'_{ijk}\lvert_{\rm GUT}=0.1$ and a universal gaugino
mass $M_{1/2}=500$ GeV. Different lines correspond to different values
of $A_0$.  We have employed the one-loop contributions from gauge
interactions \cite{Allanach:2003eb}, as well as the B$_3$ leading
interaction
\bea 
16\pi^2
\frac{d(\mathbf{h_{D^k}})_{ij}}{dt} &=&
-(\mathbf{h_{D^k}})_{ij} \left( \frac{7}{15}g_1^2 + 3g_2^2 + 
\frac{16}{3}g_3^2 \right) \nonumber \\
&&\hspace{-2cm}+ \lam'_{ijk} \left( \frac{14}{15}g_1^2 M_1 + 6 g_2^2 
M_2 + \frac{32}{3} g_3^2 M_3 \right) \,. \nonumber \\
\label{running_hDK}
\eea 
$M_1$, $M_2$ and $M_3$ are the U(1), SU(2) and SU(3) gaugino masses.
The running of $(\mathbf{h_{D^k}})_{ij}$ is dominated by the strong
interaction, {\it i.e.} by the strong coupling $g_3$ and the gluino
mass $M_3$. The running is governed by two terms with opposite sign in
Eq.~(\ref{running_hDK}), one proportional to $\lam'_{ijk}$ and one
proportional to $(\mathbf{h_{D^k}})_{ij}$.

\mymed

The term proportional to $\lam'_{ijk}$ is always positive and thus
decreases $(\mathbf{h_{D^k}})_{ij}$ when we go from $M_{\rm GUT}$ to
$M_{Z}$. Note, that we assume $\lam'_{ijk}$ is positive.  Furthermore,
the gluino mass $M_3$ will increase by a factor of roughly 2.5 and
also $\lam'_{ijk}$ will increase by roughly a factor of 3 when we run
from $M_{\rm GUT}$ to $M_Z$. Therefore this term gets relatively
more important towards lower scales.

\mymed

The sign of the term proportional to $(\mathbf{h_{D^k}})_{ij}$ depends
on the sign of $A_0$, according to Eq.~(\ref{hdk_RGE}).
At $M_{\rm GUT}$, this term is positive (negative) for negative (positive) $A_0$.
Therefore, for positive $A_0$, the term proportional to
$(\mathbf{h_{D^k}})_{ij}$ increase $(\mathbf{h_{D^k}})_{ij}$ when we
run from $M_{\rm GUT}$ to $M_{Z}$.

\mymed

We can now understand the running of $(\mathbf{h_{D^k}})_{ij}$ in
Fig.~\ref{fig:hDk_M12_500}.  Looking at the solid red line, $A_0=2500$
GeV, we see that $(\mathbf{h_{D^k}})_{ij}$ first increases when we go
from $M_{\rm GUT}$ to smaller scales. Due to the large $A_0$ at
$M_{\rm GUT}$, the negative term proportional to $(\mathbf{h_{D^k}})
_{ij}$ dominates and increases $(\mathbf{h_{D^k}})_{ij}$. Going to
lower scales the positive term proportional to $\lam'_{ijk}$ grows
faster and starts to dominate at $Q\approx 10^{6}$ GeV.  From this
scale on, $(\mathbf{h_{D^k}})_{ij}$ decreases. In contrast, if we
start with negative $A_0$ (solid black line), both terms give negative
contributions to the running of $(\mathbf{h_{D^k}})_ {ij}$. Then,
$(\mathbf{h_{D^k}})_{ij}$ decreases with a large slope.

\mymed

The resulting running of $(\mathbf{h_{D^k}})^2_{ij}$ is shown in
Fig.~\ref{fig:hDk2_M12_500}.  Recall Eq.~(\ref{sneu_RGE}),
$m^2_{\tilde{\nu}_i}$ is reduced proportional to the integral of
$(\mathbf{h_{D^k}})^2_{ij}$ over $t$. A negative value of $A_0$
therefore leads to a smaller $m_{\tilde{\nu}_i}$ compared to a
positive value of $A_0$ with the same magnitude. We expect from
Fig.~\ref{fig:hDk2_M12_500}, that a $\tilde{\nu}_i$ LSP in
$\text{B}_3$ mSUGRA is preferred for negative values of $A_0$ with a
large magnitude. We also expect, that $m_{\tilde{\nu}_i}$ in the $A_0$
direction has a maximum at $A_0=1000$ GeV, if $M_{1/2}=500$ GeV.  In
general, there should be a line in the $M_{1/2}$--$A_{0}$ plane,
where $m_{\tilde{\nu}_i}$ is ``maximal", falling to either side.

\mymed

\begin{figure}
 \setlength{\unitlength}{1in} 
 \includegraphics[scale=0.84, bb = 110 70 570 260,clip=true]{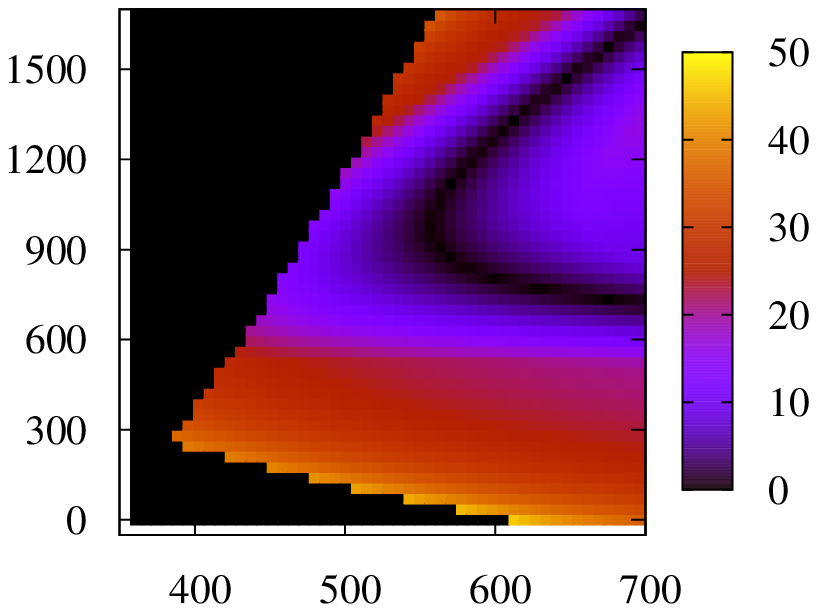}
   \put(-5.39,0.31){\epsfig{file=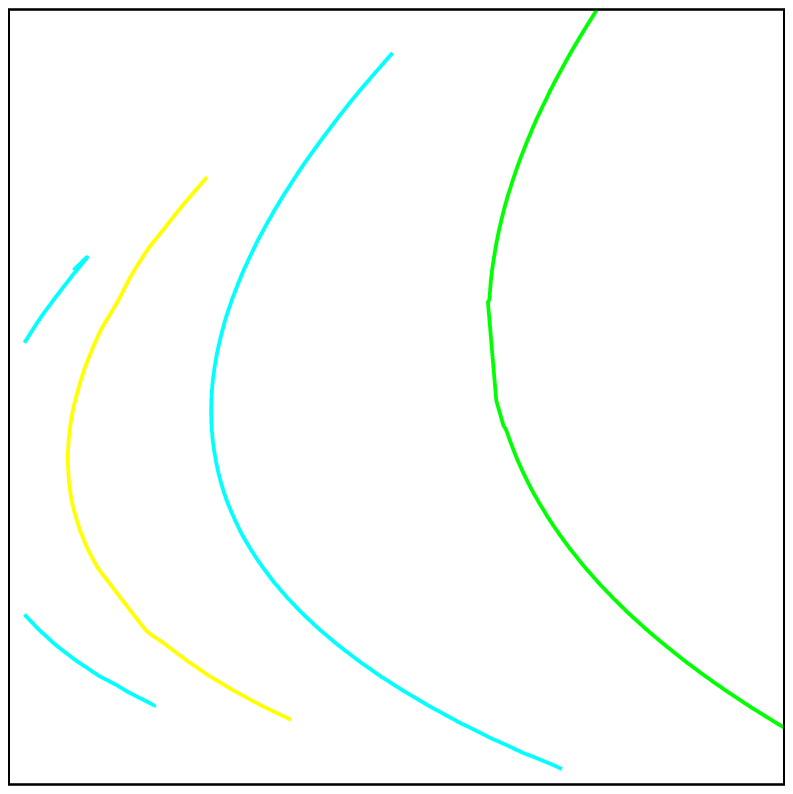,scale=0.5707}}
      \put(-2.4,0.5){\rotatebox{90}{$\Delta_M = M_{\rm NLSP} - M_{\rm LSP}$ [GeV]}}
      \put(-3.9,0.07){\makebox(0,0){$\mhalf$ [GeV]}}
      \put(-5.45,1.1){\rotatebox{90}{$\azero$[GeV]}}
      \put(-3.4,2.1){\makebox(0,0){\color[rgb]{1,1,1}{$\boldsymbol{-2}$}}}
      \put(-3.9,2.02){\makebox(0,0){\color[rgb]{1,1,1}{$\boldsymbol{-1}$}}}
      \put(-4.33,1.88){\makebox(0,0){\color[rgb]{1,1,1}{$\boldsymbol{0}$}}}
      \put(-4.67,1.7){\makebox(0,0){\color[rgb]{1,1,1}{$\boldsymbol{+1}$}}}
\put(-3.4,1.5){\makebox(0,0){\color[rgb]{1,1,1}
{$\boldsymbol{\stau}_1$ \bf{LSP}}}}
   \put(-3.9,1.0){\makebox(0,0){\color[rgb]{1,1,1}{$\boldsymbol{\sneu_\mu}$ \bf{LSP}}}}
   \caption{\label{M12-A0:deltaM} Mass difference in GeV between the
     NLSP and the LSP as a function of $M_{1/2}$ and $A_0$. The other
     mSUGRA parameters are $M_0=0$ GeV, $\tan \beta=10$,
     $\text{sgn}(\mu)=+1$ and $\lam'_{231}\lvert_{\rm GUT}=0.16$.  We
     observe a $\tilde{\nu}_\mu$ LSP and a $\tilde{\tau}_1$ LSP
     region.  The contour lines correspond to different SUSY
     contributions to the anomalous magnetic moment of the muon, {\it
       cf.} Eq.~(\ref{mu_magnetic_moment}). The blackened out region
     is excluded due to tachyons or the LEP $\tilde{\nu}_\mu$, $h$
     mass bounds, see Sect.~\ref{LEP_constraints}.}
\end{figure}

\begin{figure}
      \setlength{\unitlength}{1in} 
      \includegraphics[scale=0.84, bb = 110 70 570 260,
						clip=true]{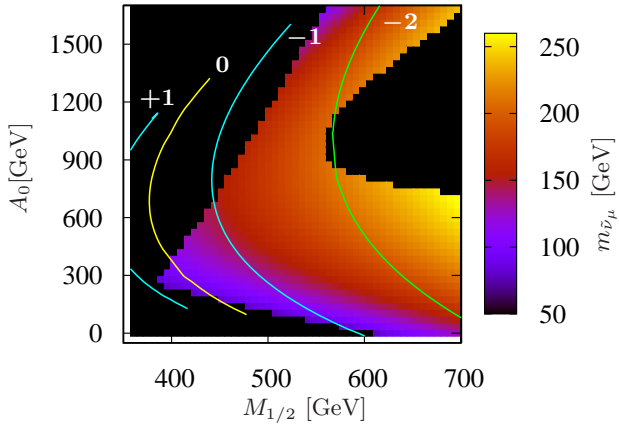}
      \put(-5.39,0.31){\epsfig{file=lamp231_largeA0M12_M00_tanb10_contour.eps,scale=0.5707}}
      \put(-2.4,0.9){\rotatebox{90}{$m_{\sneu_{\mu}}$ [GeV]}}
      \put(-3.9,0.07){\makebox(0,0){$\mhalf$ [GeV]}}
      \put(-5.45,1.1){\rotatebox{90}{$\azero$[GeV]}}
      \put(-3.4,2.1){\makebox(0,0){\color[rgb]{1,1,1}{$\boldsymbol{-2}$}}}
      \put(-3.9,2.02){\makebox(0,0){\color[rgb]{1,1,1}{$\boldsymbol{-1}$}}}
      \put(-4.33,1.88){\makebox(0,0){\color[rgb]{1,1,1}{$\boldsymbol{0}$}}}
      \put(-4.67,1.7){\makebox(0,0){\color[rgb]{1,1,1}{$\boldsymbol{+1}$}}}
  		  \caption{\label{M12-A0:Mnu} Mass of the $\tilde{\nu}_\mu$ in GeV
		  for the $\tilde{\nu}_\mu$ LSP region shown in Fig.~\ref{M12-A0:deltaM}.}
\end{figure} 

We show in Fig.~\ref{M12-A0:deltaM} the mass difference in GeV between
the NLSP and the LSP as a function of $M_{1/2}$ and $A_0$. The other
mSUGRA parameters are $M_0=0$ GeV, $\tan \beta=10$, $\text{sgn}(\mu)
=+1$ and $\lam'_{231}\lvert_{\rm GUT}=0.16$. The yellow (labelled with
$``\, 0 \,"$), blue (labelled with $`` \pm 1 "$) and green (labelled
with $`` \pm 2 "$) line indicate the SUSY contributions to the
anomalous magnetic moment of the muon as described in
Eq.~(\ref{mu_magnetic_moment}).  The blackened out region corresponds
to mSUGRA points, which lead to tachyons or where $m_{\tilde{\nu}_
\mu}$ or $m_h$ lies below the LEP bound, see
Sect.~\ref{LEP_constraints}. In Fig.~\ref{M12-A0:Mnu}, we give the mass
of the $\tilde{\nu}_\mu$ in GeV for the $\tilde{\nu} _\mu$ LSP
region shown in Fig.~\ref{M12-A0:deltaM}.

\mymed

We see in Fig.~\ref{M12-A0:deltaM} a region with a $\tilde{\nu}_\mu$
LSP and a region with a $\tilde{\tau}_1$ LSP. The cross over region is
marked in black. We get a $\tilde{\nu} _\mu$ LSP for small and very
large values of $A_0$, as expected from Fig.~\ref{fig:hDk2_M12_500}.
We also see in Fig.~\ref{M12-A0:Mnu} that $m_{\tilde{\nu}_\mu}$ is
maximal for $M_{1/2}=500$ GeV and $A_{0} \approx 1000$ GeV in the
$A_0$ direction.  The region of negative $A_0$ is not shown in
Figs.~\ref{M12-A0:deltaM}, \ref{M12-A0:Mnu}, because the influence of
$\lam'_{ijk}|_{\rm GUT}$ on $m_{\tilde{\nu} _\mu}$ is so enhanced,
that we violate the mass bound of 78 GeV or even obtain a tachyonic
$\tilde{\nu}_\mu$ in large regions of $A_0<0$ GeV. In the following,
we choose smaller values of $\lam'_{ijk}|_{\rm GUT}$.

\subsection{$A_0$--$\tan \beta$ Plane}
\label{A0tanbplane}

\begin{figure*}[h!]
  \setlength{\unitlength}{1in}
  \subfigure[Mass difference, $\Delta M$, between the NLSP and LSP.
  The LSP candidates in different regions are explicitly mentioned.
  The blackened out region corresponds to parameter points, which
  posses a tachyon or where the $\tilde{\nu}_\mu$ or $h$ mass violate
  the LEP bounds, Sect.~\ref{LEP_constraints}.
	\label{fig:DeltaM_lamp221_a0tanb}]{
    \begin{picture}(3,2.3)
      \put(-0.6,0){\epsfig{file=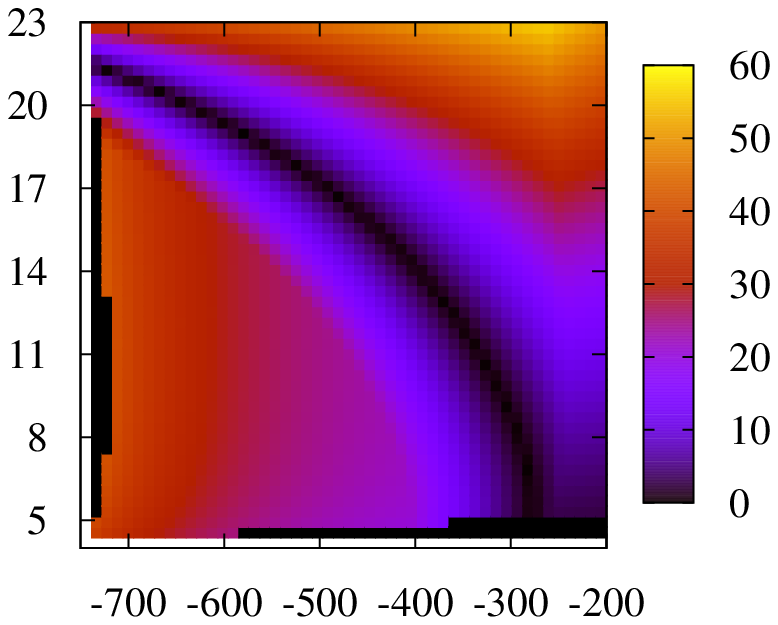,width=3.7in}}
      \put(0.0,0.48){\epsfig{file=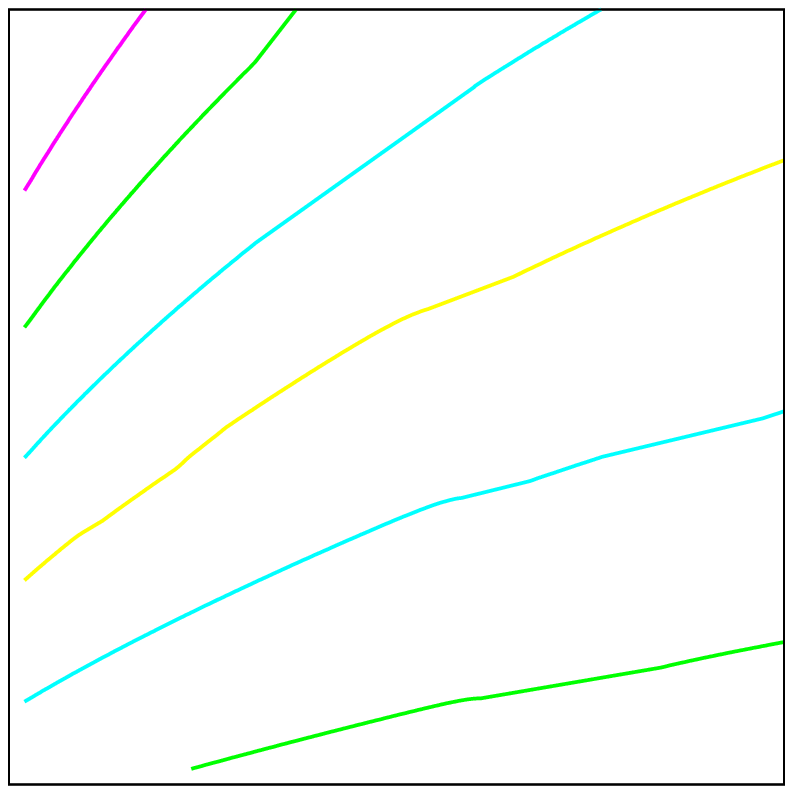,width=2.509in}}
      \put(2.7,0.5){\rotatebox{90}{$\Delta M = M_{\rm NLSP} - M_{\rm LSP}$ [GeV]}}
      \put(1.3,0.2){\makebox(0,0){$\azero$ [GeV]}}
      \put(0.05,1.2){\rotatebox{90}{$\tanb$}}
  \put(1.12,1.2){\makebox(0,0){\color[rgb]{1,1,1}{$\boldsymbol{\sneu_\mu}$ \bf{LSP}}}}
  \put(1.6,1.85){\makebox(0,0){\color[rgb]{1,1,1}{$\boldsymbol{\stau_1}$ \bf{LSP}}}}
    \end{picture}
  }\hfill
  \subfigure[ Mass difference, $\Delta M$, between the NLSP and LSP. 
	The LSP candidates in different regions are explicitly mentioned. 
	The blackened out region corresponds to parameter points, which posses a
	tachyon or where the $\tilde{\nu}_\tau$ or $h$ mass violate the LEP bounds,
	{\it cf.} Sect.~\ref{LEP_constraints}.
	\label{fig:DeltaM_lamp331_a0tanb}]{
    \begin{picture}(3,2.3)
      \put(-0.6,0){\epsfig{file=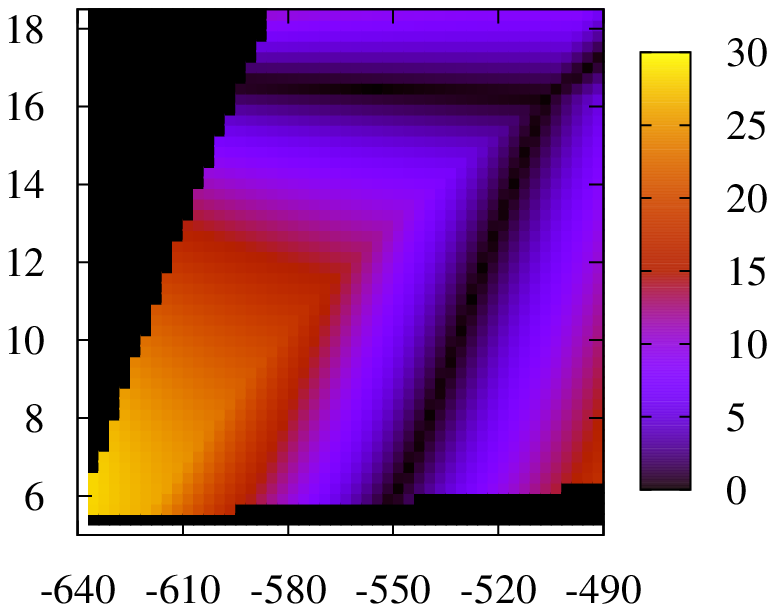,width=3.7in}}
      \put(0.0,0.48){\epsfig{file=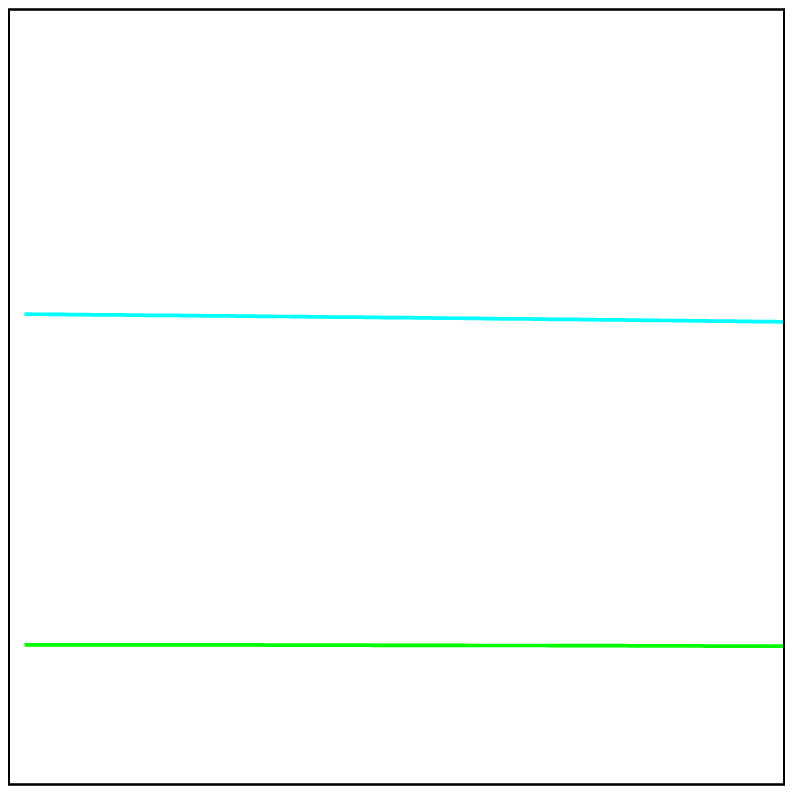,width=2.509in}}
      \put(0.0,0.48){\epsfig{file=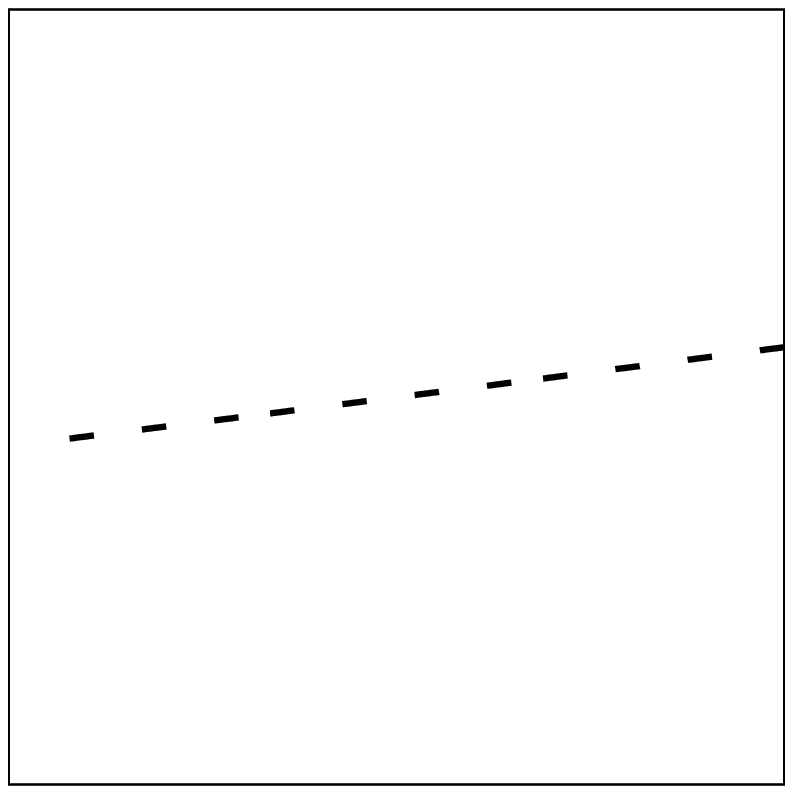,width=2.509in}}
      \put(2.7,0.5){\rotatebox{90}{$\Delta M = M_{\rm NLSP} - M_{\rm LSP}$ [GeV]}}
      \put(1.3,0.2){\makebox(0,0){$\azero$ [GeV]}}
      \put(0.05,1.2){\rotatebox{90}{$\tanb$}}
  \put(1.2,1.15){\makebox(0,0){\color[rgb]{1,1,1}{$\boldsymbol{\sneu_\tau}$ \bf{LSP}}}}
  \put(1.8,1.15){\makebox(0,0){\color[rgb]{1,1,1}{$\boldsymbol{\neutralino_1}$}}}
  \put(1.55,2.04){\makebox(0,0){\color[rgb]{1,1,1}{$\boldsymbol{\stau}_1$ \bf{ LSP}}}}
    \end{picture}
  }

  \subfigure[$\,\tilde{\nu}_\mu$ mass, $m_{\sneu_{\mu}}$, for the
	$\tilde{\nu}_\mu$ LSP region of Fig.~\ref{fig:DeltaM_lamp221_a0tanb}. 
	\label{fig:SnuMass_Lamp221_a0tanb}]{
    \begin{picture}(3,2.3)
      \put(-0.6,0){\epsfig{file=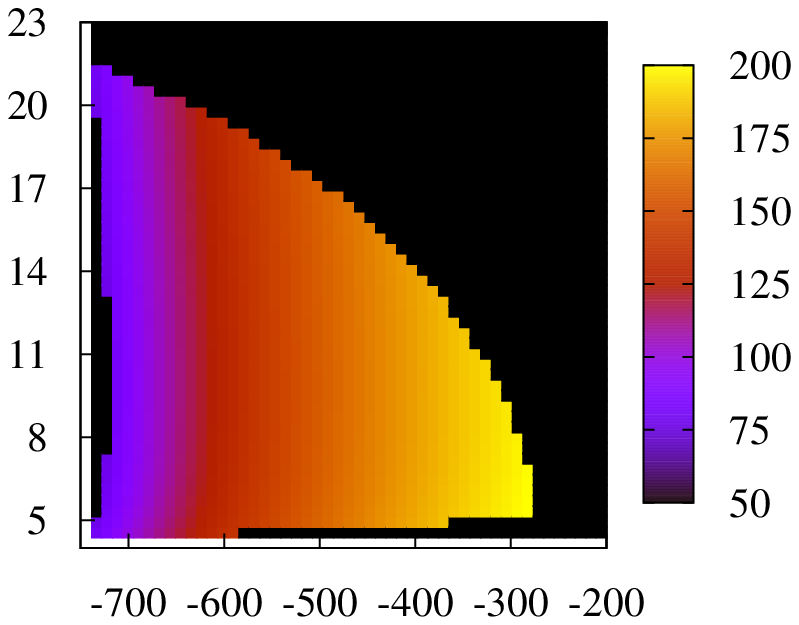, width=3.7in}}
      \put(0.0,0.48){\epsfig{file=lamp231_A0tanb_M12500_M050_contour.eps,width=2.509in}}
      \put(2.7,1.1){\rotatebox{90}{$m_{\sneu_{\mu}}$ [GeV]}}
      \put(1.3,0.2){\makebox(0,0){$\azero$ [GeV]}}
      \put(0.05,1.2){\rotatebox{90}{$\tanb$}}
      \put(1.92,0.74){\makebox(0,0){\color[rgb]{1,1,1}{$\boldsymbol{-2}$}}}
      \put(1.92,1.2){\makebox(0,0){\color[rgb]{1,1,1}{$\boldsymbol{-1}$}}}
      \put(1.96,1.72){\makebox(0,0){\color[rgb]{1,1,1}{$\boldsymbol{0}$}}}
      \put(1.69,2.05){\makebox(0,0){\color[rgb]{1,1,1}{$\boldsymbol{+1}$}}}
      \put(1.12,2.05){\makebox(0,0){\color[rgb]{1,1,1}{$\boldsymbol{+2}$}}}
      \put(0.8,2.05){\makebox(0,0){\color[rgb]{1,1,1}{$\boldsymbol{+3}$}}}
    \end{picture}
  }\hfill
  \subfigure[$\,\tilde{\nu}_\tau$ mass, $m_{\sneu_{\tau}}$, for the
	$\tilde{\nu}_\tau$ LSP region of Fig.~\ref{fig:DeltaM_lamp331_a0tanb}. 
	\label{fig:SnuMass_Lamp331_a0tanb}]{
    \begin{picture}(3,2.3)
      \put(-0.6,0){\epsfig{file=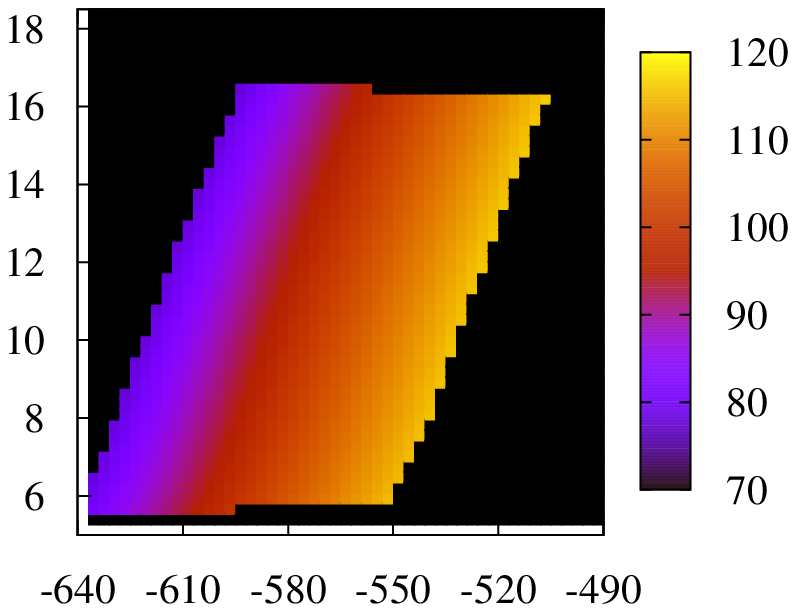, width=3.7in}}
      \put(0.0,0.48){\epsfig{file=lamp331_A0tanb_M0200_M12290_contor.eps,width=2.509in}}
      \put(0.0,0.48){\epsfig{file=lamp331_A0tanb_M0200_M12290_bsgcontor.eps,width=2.509in}}
      \put(2.7,1.1){\rotatebox{90}{$m_{\sneu_{\tau}}$ [GeV]}}
      \put(1.3,0.2){\makebox(0,0){$\azero$ [GeV]}}
      \put(0.05,1.2){\rotatebox{90}{$\tanb$}}
      \put(1.92,0.75){\makebox(0,0){\color[rgb]{1,1,1}{$\boldsymbol{-2}$}}}
      \put(1.92,1.41){\makebox(0,0){\color[rgb]{1,1,1}{$\boldsymbol{-1}$}}}
    \end{picture}
}

  \subfigure[Mass difference of the $\tilde{\mu}_L$ and
  $\tilde{\chi}_1^0$ for the $\tilde{\nu}_\mu$ LSP region of
  Fig.~\ref{fig:DeltaM_lamp221_a0tanb}. We have
  $m_{\tilde{\mu}_L} > m_{\tilde{\chi}_1^0}$ (denoted by
  $\tilde{\mu}_L>\neutralino_1$) and $m_{\tilde{\mu}_L} <
  m_{\tilde{\chi}_1^0}$ (denoted by $\tilde{\mu}_L<\neutralino_1$).
  \label{fig:MOrder_Lamp221_a0tanb}]{ \begin{picture}(3,2.3)
  \put(-0.6,0){\epsfig{file=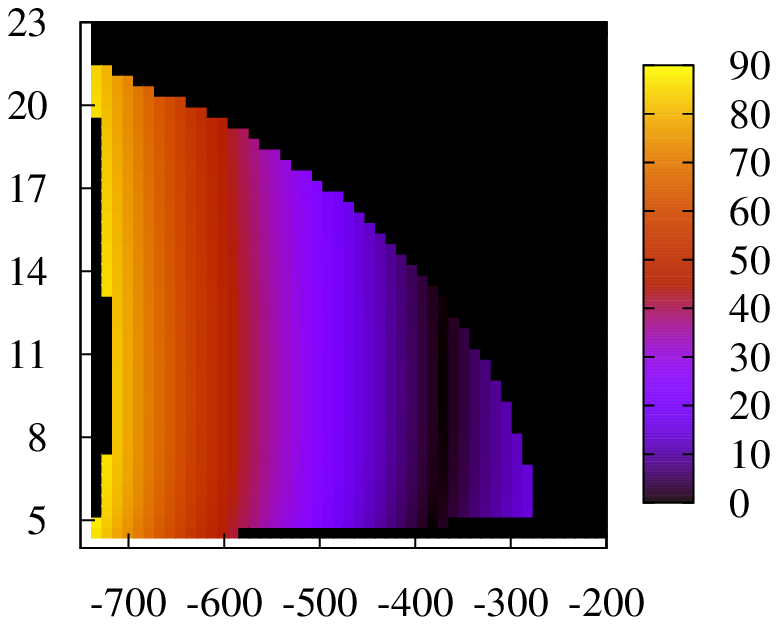,
  width=3.7in}}
  \put(0.0,0.48){\epsfig{file=lamp231_A0tanb_M12500_M050_contour.eps,width=2.509in}}
  \put(2.7,0.8){\rotatebox{90}{$|m_{\tilde{\mu}_L} -
  m_{\tilde{\chi}_1^0}|$ [GeV]}}
  \put(1.1,1.21){\makebox(0,0){\color[rgb]{1,1,1}{$\boldsymbol{\tilde{\mu}}_L<\boldsymbol{\neutralino_1}$}}}
  \put(1.78,1.04){\makebox(0,0){\color[rgb]{1,1,1}{$\boldsymbol{\tilde{\mu}}_L>\boldsymbol{\neutralino_1}$}}}
  \put(1.3,0.2){\makebox(0,0){$\azero$ [GeV]}}
  \put(0.05,1.2){\rotatebox{90}{$\tanb$}} \end{picture} }\hfill
  \subfigure[ Mass difference of the $\tilde{\tau}_1$ and
  $\tilde{\chi}_1^0$ for the $\tilde{\nu}_\tau$ LSP region of
  Fig.~\ref{fig:DeltaM_lamp331_a0tanb}. We have
  $m_{\tilde{\tau}_1} > m_{\tilde{\chi}_1^0}$ (denoted by
  $\tilde{\tau}_1>\neutralino_1$) and $m_{\tilde{\tau}_1} <
  m_{\tilde{\chi}_1^0}$ (denoted by $\tilde{\tau}_1<\neutralino_1$). 
  \label{fig:MOrder_Lamp331_a0tanb}]{ \begin{picture}(3,2.3)
  \put(-0.6,0){\epsfig{file=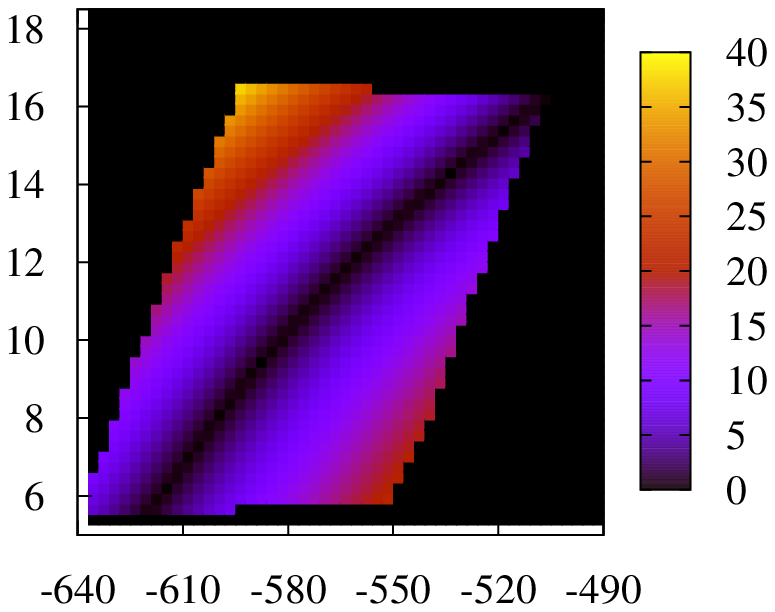,
  width=3.7in}}
  \put(0.0,0.48){\epsfig{file=lamp331_A0tanb_M0200_M12290_contor.eps,width=2.509in}}
  \put(0.0,0.48){\epsfig{file=lamp331_A0tanb_M0200_M12290_bsgcontor.eps,width=2.509in}}
  \put(1.3,1.05){\makebox(0,0){\color[rgb]{1,1,1}{$\boldsymbol{\stau}_1>\boldsymbol{\neutralino_1}$}}}
  \put(1.3,1.75){\makebox(0,0){\color[rgb]{1,1,1}{$\boldsymbol{\stau}_1<\boldsymbol{\neutralino_1}$}}}
  \put(2.7,0.8){\rotatebox{90}{$|m_{\tilde{\tau}_1} -
  m_{\tilde{\chi}_1^0}|$ [GeV]}} \put(1.3,0.2){\makebox(0,0){$\azero$
  [GeV]}} \put(0.05,1.2){\rotatebox{90}{$\tanb$}} \end{picture} }
  \caption{ Sneutrino LSP parameter space in the $\azero$--$\tanb$
  plane. The left panel (right panel) shows the $\tilde{\nu}_\mu$ LSP
  ($\tilde{\nu}_\tau$ LSP) region obtained via
  $\lam'_{231}\lvert_{\rm GUT}=0.11$, $M_0=50$ GeV, $M_{1/2}=500$ GeV
  and $\text{sgn}(\mu)=+1$ ($\lam'_{331}\lvert_{\rm GUT}=0.12$,
  $M_0=200$ GeV, $M_{1/2}=290$ GeV and $\text{sgn}(\mu)=+1$).
  The plots show from top to bottom the mass difference between the
  NLSP and LSP, $\Delta M$, the mass of the sneutrino LSP,
  $m_{\sneu}$, and the mass difference between the $\tilde{\chi}_1^0$
  and the $\tilde{\mu}_L$ (left panel) or between the
  $\tilde{\chi}_1^0$ and $\tilde{\tau}_1$ (right panel). The
  yellow (labelled with $``\, 0 \,"$), blue (labelled
  with $`` \pm 1 "$), green (labelled with $`` \pm 2 "$) and
  magenta (labelled with $`` \pm 3 "$) contours correspond to
  different SUSY contributions to the anomalous magnetic moment of the
  muon, $\delta a_\mu^{\rm SUSY}$, as described in
  Eq.~(\ref{mu_magnetic_moment}). The dashed black line (right
  panel) corresponds to BR($b \rightarrow s \gamma)=2.74 \times
  10^{-4}$, Eq.~(\ref{bsg_2sigma}).  \label{fig:a0_tanb}}
\end{figure*} 

We investigate in this section the sneutrino LSP parameter space in
the $A_0$--$\tan \beta$ plane. As central values for our 2-dimensional
scans, we choose the points given in Eq.~(\ref{scanpoints}).

\mymed

We show in Fig.~\ref{fig:DeltaM_lamp221_a0tanb}
[Fig.~\ref{fig:DeltaM_lamp331_a0tanb}] the $\tilde{\nu}_\mu$ LSP
[$\tilde{\nu}_\tau$ LSP] parameter space in the $A_0$--$\tan \beta$
plane. We have chosen $\lam'_{231}\lvert_{\rm GUT}=0.11$
[$\lam'_{331}\lvert_{\rm GUT}=0.12$]. Both figures show the mass
difference between the NLSP and the LSP in GeV. The solid contour
lines correspond to different SUSY contributions to the anomalous
magnetic moment of the muon $\delta a_\mu^{\rm SUSY}$ as described in
Eq.~(\ref{mu_magnetic_moment}). The dashed black line in
Fig.~\ref{fig:DeltaM_lamp331_a0tanb} corresponds to BR($b\rightarrow
s\gamma)=2.74 \times 10^{-4}$ (\ref{bsg_2sigma}), {\it i.e.} the
parameter space below that line is consistent with $b \rightarrow s
\gamma$ at $2\sigma$.  The blackened out region is excluded due to the
presence of tachyons or by the LEP $\tilde{\nu}_{\mu/\tau}$ and Higgs
mass bound, see Sect.~\ref{LEP_constraints}.

\mymed

We observe that the $\tilde{\nu}_\mu$ LSP lives in an extended region
of $\text{B}_3$ mSUGRA parameter space. For $\tan \beta = 6$, we find
a $\tilde{\nu}_\mu$ LSP between $A_0=-750$ GeV and $A_0=-300$ GeV. For
$A_0 = -700$ GeV, we find a $\tilde{\nu}_\mu$ LSP between $\tan \beta
= 4$ and $\tan \beta =21$. We also observe that most of the
$\tilde{\nu}_\mu$ LSP region is consistent with the observed anomalous
magnetic moment of the muon at the $1 \sigma$ (blue lines) and 
$2 \sigma$ (green lines) level, {\it cf.}  Eq.~(\ref{mu_magnetic_moment}).
Recall, that the complete $\tilde{\nu}_\mu$ LSP region in
Fig.~\ref{fig:DeltaM_lamp221_a0tanb} is also consistent with BR($b
\rightarrow s \gamma$) at $2 \sigma$, Eq.~(\ref{bsg_2sigma}).  The
large region of $\tilde{\nu}_\mu$ LSP parameter space is a
consequence of the choice of our central scan point, {\it i.e.} Point
I of Eq.~(\ref{scanpoints}). Here, the mass difference between the
$\tilde{\nu}_\mu$ LSP and the $\tilde{\tau}_1$ ($\tilde{\chi}_1^0$),
{\it i.e.} the other LSP candidates, is rather large, namely 56 GeV
(75 GeV).

\mymed

We see in Fig.~\ref{fig:DeltaM_lamp221_a0tanb} that we obtain a
$\tilde{\tau}_1$ LSP if we increase $A_0$. We explained this in the
last section. A large magnitude and negative value of $A_0$ enhances the
(negative) effect of $\lam'_{231}|_{\rm GUT}$ on the $\tilde{\nu}_\mu$
mass via the soft breaking trilinear coupling
$(\mathbf{h_{D^1}})_{23}$.  The $\stau_1$ mass on the other hand,
depends only weakly on $A_0$. The dependence is via the tau
Yukawa-coupling, Eq.~(\ref{stau_parameter}), and due to
left-right-mixing, Eq.~(\ref{eq_staumassmatrix}). According to the
last section, there should also be a $\tilde{\nu}_\mu$ LSP for large
values of $A_0$. But in this case the Higgs mass lies below
the LEP bound.

\mymed

We also obtain a $\tilde{\tau}_1$ LSP, when we increase $\tan\beta$.
$\tan \beta$ hardly affects the mass of the $\tilde{\nu}_\mu$
but affects the $\tilde{\tau}_1$ mass in two ways.  First, increasing
$\tan \beta$ increases the tau Yukawa coupling, which reduces the
$\tilde{\tau}_1$ mass going from $M_{\rm GUT}$ to $M_Z$. This is
parametrized by Eq.~(\ref{stau_parameter}). Second, increasing $\tan
\beta$ increases the absolute value of the off diagonal
elements of the stau mass matrix, Eq.~(\ref{eq_staumassmatrix}). This
leads to larger left-right mixing and thus also reduces the
$\tilde{\tau}_1$ mass.

\mymed

Fig.~\ref{fig:DeltaM_lamp221_a0tanb} shows no region with a
$\tilde{\chi}_1^0$ LSP. The entire allowed $A_0$--$\tan\beta$ plane in
Fig.~\ref{fig:DeltaM_lamp221_a0tanb} has a $\tilde{\tau}_1$ LSP for
vanishing $\lam'_{231}$ because $M_{1/2}\gg M_0$.

\mymed

We show in Fig.~\ref{fig:DeltaM_lamp331_a0tanb} the $\tilde{\nu}_\tau$
LSP parameter space. We observe a ``smaller" $\tilde{\nu}_\tau$ LSP
region compared to the $\tilde{\nu}_\mu$ LSP region,
Fig.~\ref{fig:DeltaM_lamp221_a0tanb}. We only find a $\tilde{\nu}_\tau$
LSP between $A_0=-630$ GeV and $A_0=-540$ GeV for $\tan \beta=8$. 
In addition, the experimental $2\sigma$ windows for 
$\delta a_\mu^{\rm SUSY}$, Eq.~(\ref{mu_magnetic_moment}), and 
BR($b \rightarrow s \gamma$), Eq.~(\ref{bsg_2sigma}), restrict the 
allowed $\tilde{\nu}_\tau$ LSP region in Fig.~\ref{fig:DeltaM_lamp331_a0tanb} 
to lie between $\tan\beta=7$ and $\tan\beta=11$.

\mymed

We again obtain in Fig.~\ref{fig:DeltaM_lamp331_a0tanb} the
$\tilde{\tau}_1$ as LSP when we go to larger values of $\tan \beta$
($\tan \beta \approx 17$). Although the $\tilde{\nu}_\tau$ mass will
also be reduced by a larger tau Yukawa coupling, {\it cf.}
Eq.~(\ref{stau_parameter}), the squared mass of the right-handed stau
is reduced twice as much as the $\tilde{\nu}_\tau$ mass. In addition,
$\tan \beta$ increases mixing between the $\tilde{\tau}_R$ and
$\tilde{\tau_L}$, Eq.~(\ref{eq_staumassmatrix}).  But it is not
possible to find a B$_3$ mSUGRA point, where the mass
difference between the $\tilde{\nu}_\tau$ LSP and the $\tilde
{\tau}_1$ is large, because $\lam'_{331}|_{\rm GUT}$ also reduces the
mass of the $\tilde{\tau}_1$.

\mymed

We also obtain in Fig.~\ref{fig:DeltaM_lamp331_a0tanb} a
$\tilde{\chi}_1^0$ LSP instead of a $\tilde{\nu}_\tau$ or
$\tilde{\tau}_1$ LSP if we increase $A_0$ beyond a certain value. The
parameter space shown in Fig.~\ref{fig:DeltaM_lamp331_a0tanb} posses a
$\tilde{\chi}_1^0$ LSP for vanishing $\lam'_{331}|_{\rm GUT}$.
Increasing $A_0$ reduces the effect of $\lam'_{331}|_{\rm GUT}$ on the
$\tilde{\nu}_\tau$ and $\tilde{\tau}_1$ mass, but leaves the
(bino-like) $\tilde{\chi}_1^0$ mass unaffected. Thus, if the influence
of $\lam'_{331}|_{\rm GUT}$ on the $\tilde{\nu}_\tau$ and
$\tilde{\tau}_1$ mass is getting smaller, we re-obtain the
$\tilde{\chi}_1^0$ as the LSP.

\mymed

Finally we want to mention in our discussion of
Fig.~\ref{fig:DeltaM_lamp331_a0tanb} that we have a ``triple-point",
where the $\tilde{\nu}_\tau$, the $\tilde{\tau}_1$ and the
$\tilde{\chi}_1^0$ are degenerate in mass. The existence of this
``triple-point" is a general feature of the sneutrino LSP parameter
space.  This has important consequences for the LHC phenomenology,
because close to a ``triple-point", we effectively have three 
nearly degenerate LSPs at the
same time. There are also large regions in
Fig.~\ref{fig:DeltaM_lamp221_a0tanb} and
Fig.~\ref{fig:DeltaM_lamp331_a0tanb}, where two of the three LSP
candidates are nearly degenerate in mass, {\it i.e.} $\Delta M \leq 5$
GeV.

\mymed

We present in Fig.~\ref{fig:SnuMass_Lamp221_a0tanb}
[Fig.~\ref{fig:SnuMass_Lamp331_a0tanb}] the mass of the $\tilde{\nu}
_\mu$ [$\tilde{\nu}_\tau$] for the corresponding sneutrino LSP regions
of Fig.~\ref{fig:DeltaM_lamp221_a0tanb}
[Fig.~\ref{fig:DeltaM_lamp331_a0tanb}]. The lightest sneutrino LSPs
have a mass of 78 GeV stemming from LEP bounds, {\it cf.}
Sect.~\ref{LEP_constraints}. The heaviest sneutrino LSPs, consistent
with $a_\mu^{\rm exp}$, Eq.~(\ref{amu_exp}), and BR($b \rightarrow s
\gamma$), Eq.~(\ref{bsg_2sigma}), are found in
Fig.~\ref{fig:SnuMass_Lamp221_a0tanb} and posses a mass of roughly 200
GeV. If one wants to have a sneutrino LSP scenario consistent with the
anomalous magnetic moment of the muon, than the sneutrino mass is not
allowed to be much larger than 200 GeV (see also the next section).

\mymed

We show in Fig.~\ref{fig:MOrder_Lamp221_a0tanb}
[Fig.~\ref{fig:MOrder_Lamp331_a0tanb}] the mass difference in GeV
between the $\tilde{\chi}_1^0$ and the $\tilde{\mu}_L$ [mainly
left-handed $\tilde{\tau}_1$]. Whether, $m_{\tilde{\chi}_1^0} >
m_{\tilde{\mu}_L}$ [$m_{\tilde{\tau}_1}$] or $m_{\tilde{\chi}_1^0} <
m_{\tilde{\mu}_L}$ [$m_{\tilde{\tau}_1}$] has important consequences
for collider phenomenology. For example, the $\tilde{\mu}_L$ can not
decay into a $\mu$ and $\tilde{\chi}_1^0$ if $m_{\tilde{\chi}_1^0} >
m_{\mu_L}$. This is the case in most of the $\tilde{\nu}_\mu$ LSP
parameter space. The cascade decay, Eq.~(\ref{slep_cascade}), is then
forbidden and can not be explored at the Tevatron or LHC, as stated in
Sect.~\ref{Tevatron_constraints}. We discuss further phenomenological
implications in Sect.~\ref{pheno}.

\subsection{$M_{1/2}$--$M_{0}$ Plane}

\begin{figure*}[h!]
  \setlength{\unitlength}{1in}

  \subfigure[Mass difference $\Delta M$ between the NLSP and LSP. 
	The LSP candidates in different regions are explicitly mentioned. 
	The blackened out region corresponds to parameter points, which posses a
	tachyon or where the $\tilde{\nu}_\mu$ or $h$ mass violate the LEP bounds,
	{\it cf.} Sect.~\ref{LEP_constraints}.
	\label{fig:DeltaM_lamp221_m0m12}]{
    \begin{picture}(3,2.3)
      \put(-0.6,0){\epsfig{file=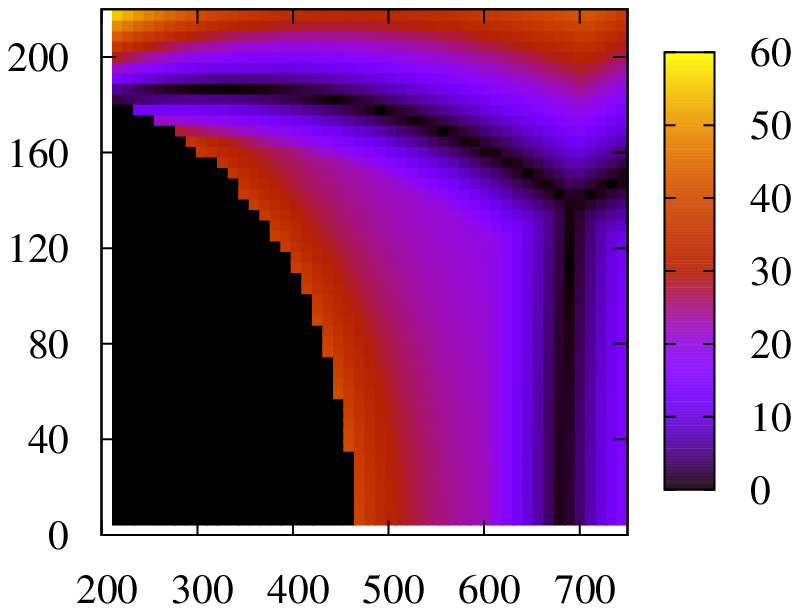, width=3.7in}}
      \put(0.0,0.48){\epsfig{file=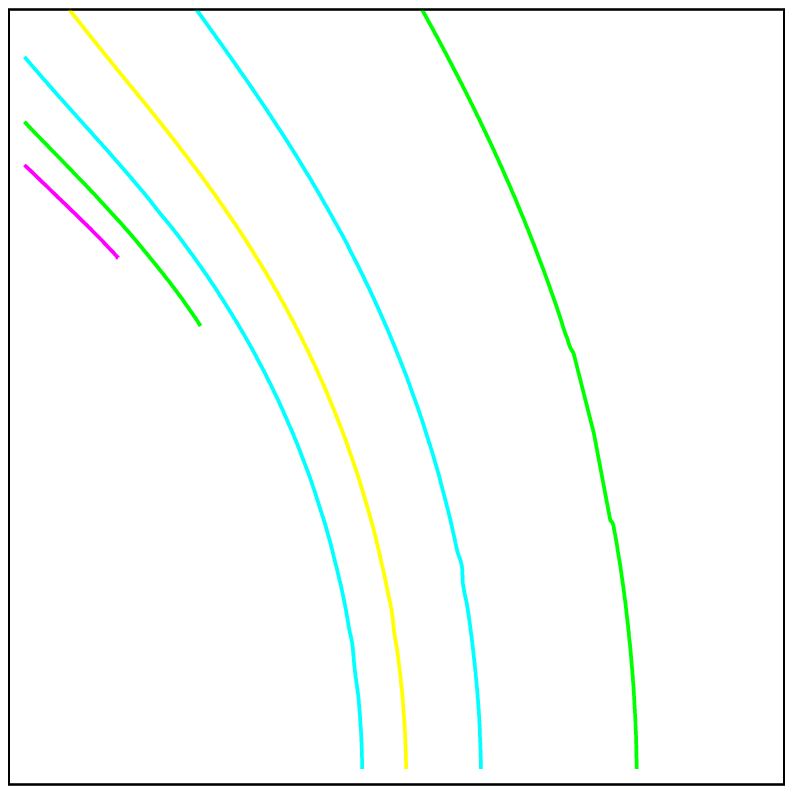,width=2.509in}}
      \put(2.7,0.5){\rotatebox{90}{$\Delta M = M_{\rm NLSP} - M_{\rm LSP}$ [GeV]}}
      \put(1.3,0.2){\makebox(0,0){$\mhalf$ [GeV]}}
      \put(0.03,1.2){\rotatebox{90}{$\mzero$ [GeV] }}
  \put(1.95,1.1){\makebox(0,0){\color[rgb]{1,1,1}{$\boldsymbol{\stau_1}$}}}
  \put(1.455,1.1){\makebox(0,0){\color[rgb]{1,1,1}{$\boldsymbol{\sneu_\mu}$ \bf{LSP}}}}
  \put(1.7,2.0){\makebox(0,0){\color[rgb]{1,1,1}{$\boldsymbol{\neutralino_1}$ \bf{LSP}}}}
    \end{picture}
  }\hfill
  \subfigure[ Mass difference $\Delta M$ between the NLSP and LSP. 
	The LSP candidates in different regions are explicitly mentioned. 
	The blackened out region corresponds to parameter points, which posses a
	tachyon or where the $\tilde{\nu}_\tau$ or $h$ mass violate the LEP bounds,
	Sect.~\ref{LEP_constraints}.
	\label{fig:DeltaM_lamp331_m0m12}]{
    \begin{picture}(3,2.3)
      \put(-0.6,0){\epsfig{file=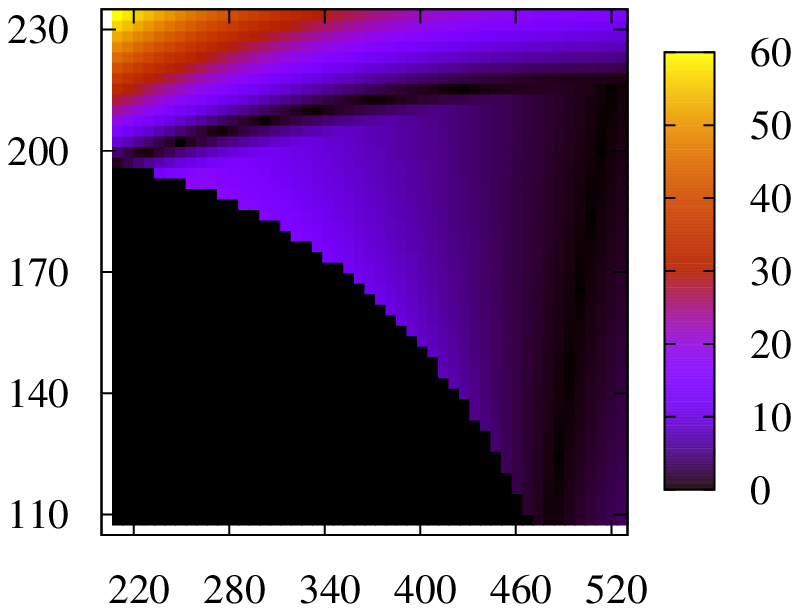, width=3.7in}}
      \put(0.0,0.48){\epsfig{file=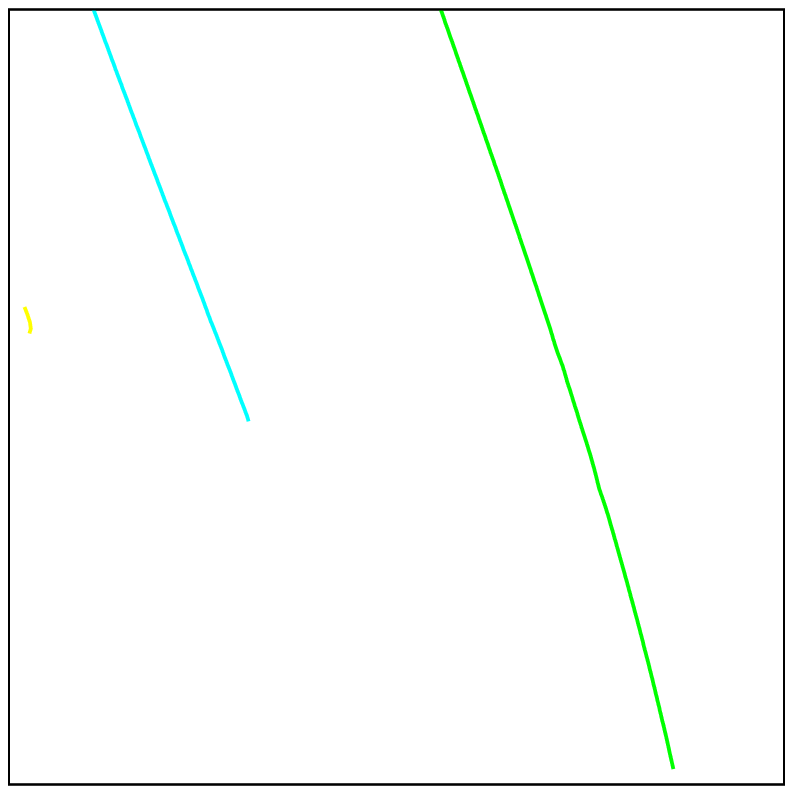,width=2.509in}}
      \put(0.0,0.48){\epsfig{file=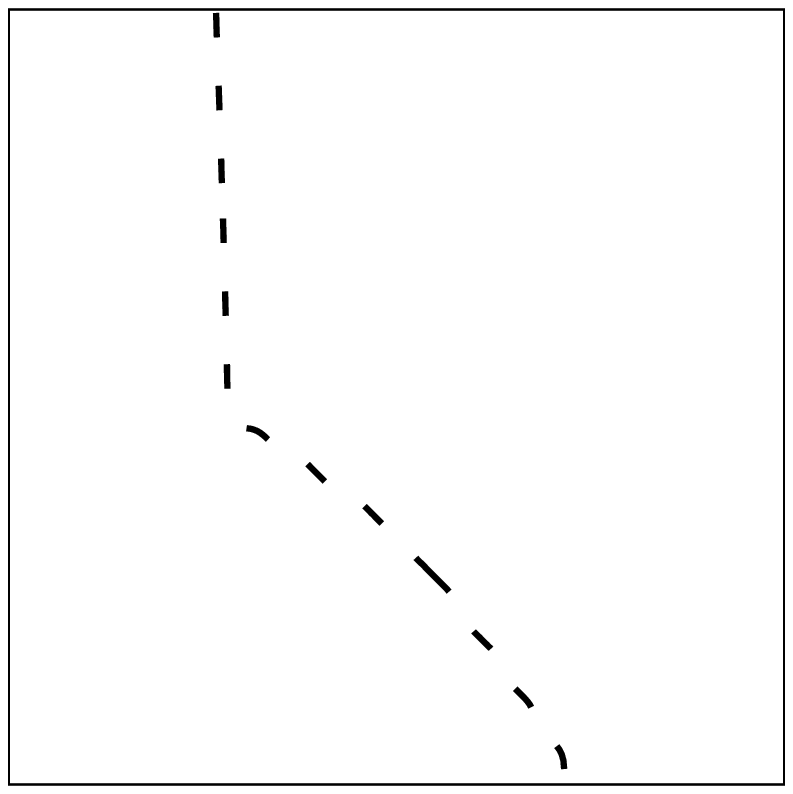,width=2.509in}}
      \put(2.7,0.5){\rotatebox{90}{$\Delta M = M_{\rm NLSP} - M_{\rm LSP}$ [GeV]}}
      \put(1.3,0.2){\makebox(0,0){$\mhalf$ [GeV]}}
      \put(0.03,1.2){\rotatebox{90}{$\mzero$ [GeV] }}
  \put(1.95,1.1){\makebox(0,0){\color[rgb]{1,1,1}{$\boldsymbol{\stau_1}$}}}
  \put(1.25,1.55){\makebox(0,0){\color[rgb]{1,1,1}{$\boldsymbol{\sneu_\tau}$ \bf{LSP}}}}
  \put(1.2,2.){\makebox(0,0){\color[rgb]{1,1,1}{$\boldsymbol{\neutralino_1}$ \bf{LSP}}}}
    \end{picture}
  }

  \subfigure[ $\,\tilde{\nu}_\mu$ mass, $m_{\sneu_{\mu}}$, for the
	$\tilde{\nu}_\mu$ LSP region of Fig.~\ref{fig:DeltaM_lamp221_m0m12}.
	\label{fig:SnuMass_Lamp221_m12m0}]{
    \begin{picture}(3,2.3)
      \put(-0.6,0){\epsfig{file=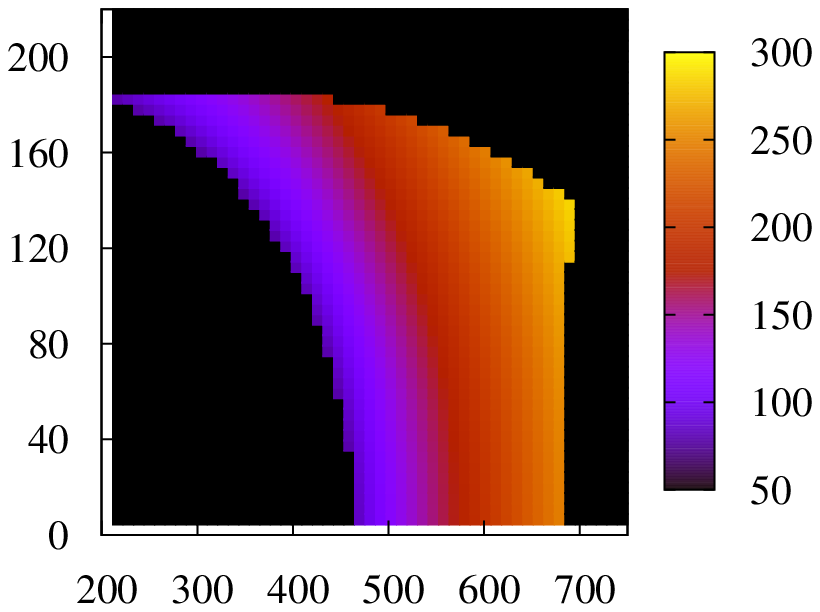,width=3.7in}}
      \put(0.0,0.48){\epsfig{file=lamp231_M0M12_A0m600_tanb10_contour.eps,width=2.509in}}
      \put(0.8,2.0){\makebox(0,0){\color[rgb]{1,1,1}{$\boldsymbol{0}$}}}
	  \put(1.08,2.0){\makebox(0,0){\color[rgb]{1,1,1}{$\boldsymbol{-1}$}}}
      \put(1.5,2.0){\makebox(0,0){\color[rgb]{1,1,1}{$\boldsymbol{-2}$}}}
      \put(0.58,1.6){\makebox(0,0){\color[rgb]{1,1,1}{$\boldsymbol{+3}$}}}
      \put(0.8,1.4){\makebox(0,0){\color[rgb]{1,1,1}{$\boldsymbol{+2}$}}}
      \put(0.95,1.2){\makebox(0,0){\color[rgb]{1,1,1}{$\boldsymbol{+1}$}}}
      \put(2.7,1.1){\rotatebox{90}{$m_{\sneu_{\mu}}$ [GeV]}}
      \put(1.3,0.2){\makebox(0,0){$\mhalf$ [GeV]}}
      \put(0.03,1.2){\rotatebox{90}{$\mzero$ [GeV]  }}
    \end{picture}
  }\hfill
  \subfigure[ $\,\tilde{\nu}_\tau$ mass, $m_{\sneu_{\tau}}$, for the
	$\tilde{\nu}_\tau$ LSP region of Fig.~\ref{fig:DeltaM_lamp331_m0m12}.
	\label{fig:SnuMass_Lamp331_m12m0}]{
    \begin{picture}(3,2.3)
      \put(-0.6,0){\epsfig{file=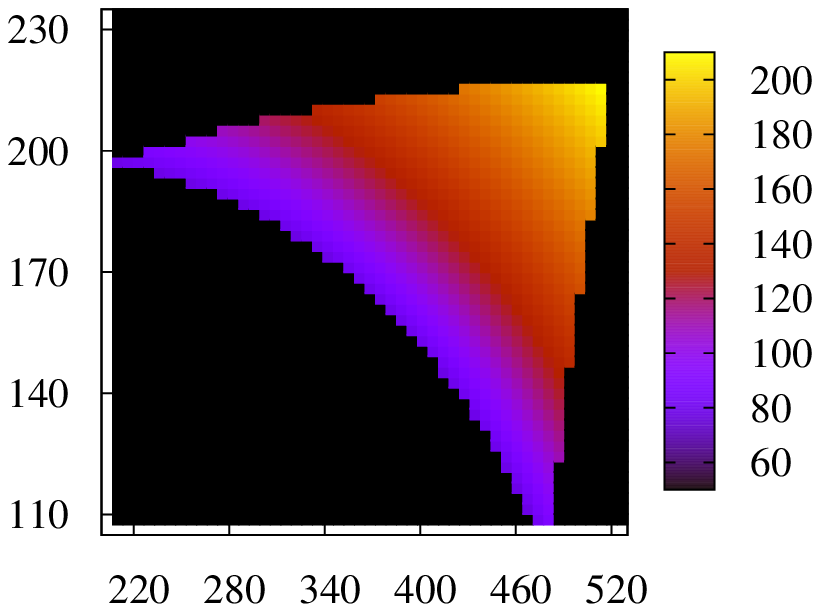,width=3.7in}}
      \put(0.0,0.48){\epsfig{file=lamp331_M0M12_A0m550_tanb12_contour.eps,width=2.509in}}
      \put(0.0,0.48){\epsfig{file=lamp331_M0M12_A0m550_tanb12_bsgcontour.eps,width=2.509in}}
      \put(2.7,1.1){\rotatebox{90}{$m_{\sneu_{\tau}}$ [GeV]}}
      \put(1.3,0.2){\makebox(0,0){$\mhalf$ [GeV]}}
      \put(0.03,1.2){\rotatebox{90}{$\mzero$ [GeV]  }}
      \put(0.84,2.0){\makebox(0,0){\color[rgb]{1,1,1}{$\boldsymbol{-1}$}}}
      \put(1.52,2.0){\makebox(0,0){\color[rgb]{1,1,1}{$\boldsymbol{-2}$}}}
    \end{picture}
}

  \subfigure[Mass difference of the $\tilde{\mu}_L$ and
  $\tilde{\chi}_1^0$ for the $\tilde{\nu}_\mu$ LSP region of
  Fig.~\ref{fig:DeltaM_lamp221_m0m12}. We have $m_{\tilde{\mu}_L} 
  > m_{\tilde{\chi}_1^0}$ (denoted by
  $\tilde{\mu}_L>\neutralino_1$) and $m_{\tilde{\mu}_L} <
  m_{\tilde{\chi}_1^0}$ (denoted by $\tilde{\mu}_L<\neutralino_1$).
  \label{fig:MOrder_Lamp221_m12m0}]{ \begin{picture}(3,2.3)
  \put(-0.6,0){\epsfig{file=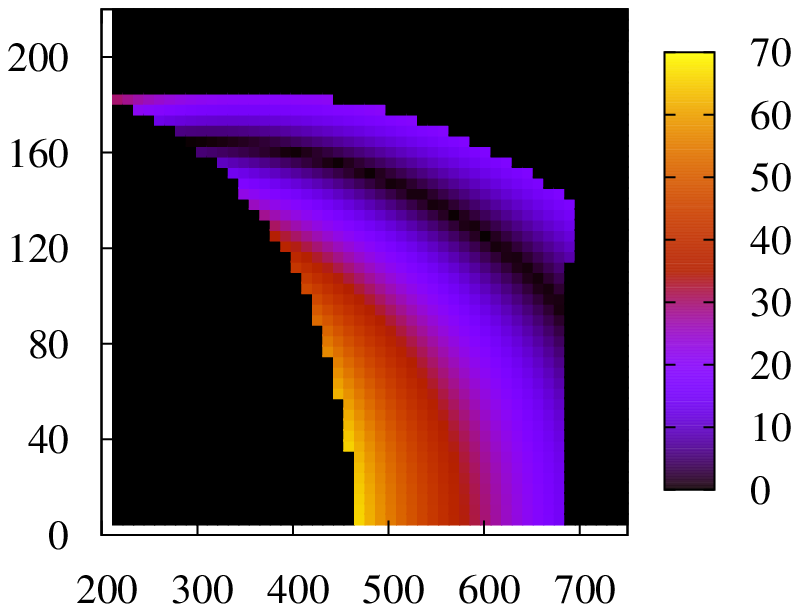,width=3.7in}}
  \put(0.0,0.48){\epsfig{file=lamp231_M0M12_A0m600_tanb10_contour.eps,width=2.509in}}
  \put(2.7,0.8){\rotatebox{90}{$|m_{\tilde{\mu}_L} -
  m_{\tilde{\chi}_1^0}|$ [GeV]}}
  \put(1.43,1.2){\makebox(0,0){\color[rgb]{1,1,1}{$\boldsymbol{\tilde{\mu}}_L<\boldsymbol{\neutralino_1}$}}}
  \put(1.58,1.73){\makebox(0,0){\color[rgb]{1,1,1}{$\boldsymbol{\tilde{\mu}}_L>\boldsymbol{\neutralino_1}$}}}
  \put(1.3,0.2){\makebox(0,0){$\mhalf$ [GeV]}}
  \put(0.03,1.2){\rotatebox{90}{$\mzero$ [GeV] }} \end{picture}
  }\hfill \subfigure[ Mass difference of the $\tilde{\tau}_1$ and
  $\tilde{\chi}_1^0$ for the $\tilde{\nu}_\tau$ LSP region of
  Fig.~\ref{fig:DeltaM_lamp331_m0m12}.  We have
  $m_{\tilde{\tau}_1} > m_{\tilde{\chi}_1^0}$ (denoted by
  $\tilde{\tau}_1>\neutralino_1$) and $m_{\tilde{\tau}_1} <
  m_{\tilde{\chi}_1^0}$ (denoted by $\tilde{\tau}_1<\neutralino_1$).
  \label{fig:MOrder_Lamp331_m12m0}]{ \begin{picture}(3,2.3)
  \put(-0.6,0){\epsfig{file=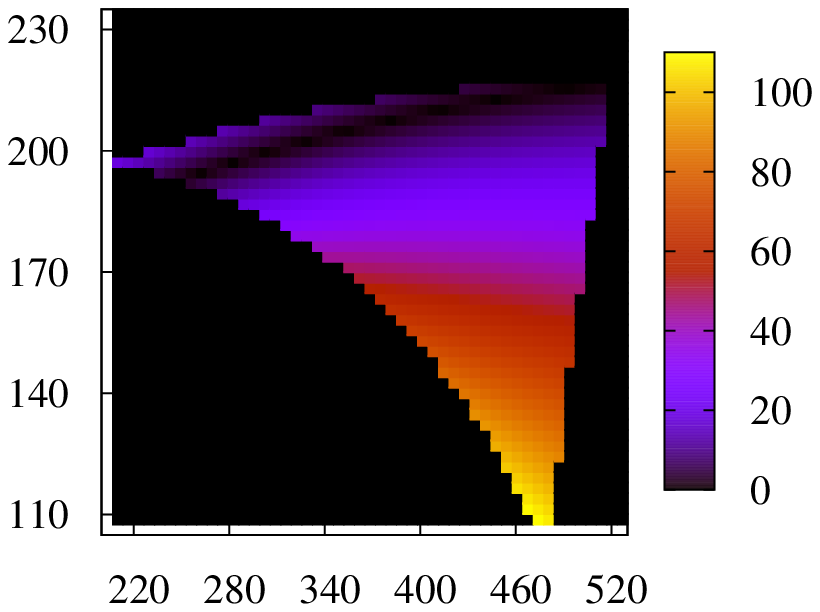,width=3.7in}}
  \put(0.0,0.48){\epsfig{file=lamp331_M0M12_A0m550_tanb12_contour.eps,width=2.509in}}
  \put(0.0,0.48){\epsfig{file=lamp331_M0M12_A0m550_tanb12_bsgcontour.eps,width=2.509in}}
  \put(1.68,1.4){\makebox(0,0){\color[rgb]{1,1,1}{$\boldsymbol{\stau}_1<\boldsymbol{\neutralino_1}$}}}
  \put(0.86,1.8){\makebox(0,0){\color[rgb]{1,1,1}{$\boldsymbol{\stau}_1>\boldsymbol{\neutralino_1}$}}}
  \put(2.7,0.8){\rotatebox{90}{$|m_{\tilde{\tau}_1} -
  m_{\tilde{\chi}_1^0}|$ [GeV]}} \put(1.3,0.2){\makebox(0,0){$\mhalf$
  [GeV]}} \put(0.03,1.2){\rotatebox{90}{$\mzero$ [GeV] }}
  \end{picture} } \caption{ Sneutrino LSP parameter space in the
  $M_{1/2}$--$M_{0}$ plane. The left panel (right panel) shows the
  $\tilde{\nu}_\mu$ LSP ($\tilde{\nu}_\tau$ LSP) region obtained via
  $\lam'_{231}\lvert_{\rm GUT}=0.11$, $A_0=-600$ GeV, $\tanb=10$ and
  $\text{sgn}(\mu)=+1$ ($\lam'_{331}\lvert_{\rm GUT}=0.12$, $A_0=-550$
  GeV, $\tanb=12$ and $\text{sgn}(\mu)=+1$).  The plots show
  from top to bottom the mass difference between the NLSP and LSP,
  $\Delta M$, the mass of the sneutrino LSP, $m_{\sneu}$, and the mass
  difference between the $\tilde{\chi}_1^0$ and the $\tilde{\mu}_L$
  (left panel) or between the $\tilde{\chi}_1^0$ and $\tilde{\tau}_1$
  (right panel). The yellow (labelled with $``\, 0 \,"$), blue
  (labelled with $`` \pm 1 "$), green (labelled with $`` \pm 2 "$) and
  magenta (labelled with $`` \pm 3 "$) contours correspond to
  different SUSY contributions to the anomalous magnetic moment of the
  muon, $\delta a_\mu^{\rm SUSY}$, as described in
  Eq.~(\ref{mu_magnetic_moment}). The dashed black line in
  Fig.~\ref{fig:DeltaM_lamp331_m0m12} corresponds to BR($b \rightarrow
  s \gamma)=2.74 \times 10^{-4}$, Eq.~(\ref{bsg_2sigma}).
  \label{fig:M0_M12}}
\end{figure*}


We present in Fig.~\ref{fig:DeltaM_lamp221_m0m12}
[Fig.~\ref{fig:DeltaM_lamp331_m0m12}] the $\tilde{\nu}_\mu$ LSP
[$\tilde{\nu}_\tau$ LSP] region in the $M_{1/2}$--$M_{0}$ plane.  We
have chosen $\lam'_{231}\lvert_{\rm GUT}=0.11$
[$\lam'_{331}\lvert_{\rm GUT}$=0.12].  The figures show the mass
difference in GeV between the NLSP and the LSP.  The solid contour lines
correspond again to SUSY scenarios, which contribute to $a_{\mu}$ the
amount described in Eq.~(\ref{mu_magnetic_moment}) and the 
dashed black line in Fig.~\ref{fig:DeltaM_lamp331_m0m12} 
corresponds to BR($b \rightarrow s \gamma)=2.74 \times 10^{-4}$ 
Eq.~(\ref{bsg_2sigma}).

\mymed

The $\tilde{\nu}_\mu$ LSP lives in an extended region of $\text{B}_3$
mSUGRA parameter space. This 
stems from the fact, that we were able to choose a central scan point,
Point I of Eq.~(\ref{scanpoints}), where the mass difference between
the $\tilde{\nu}_\mu$ LSP and the other LSP candidates,
$\tilde{\tau}_1$ and $\tilde{\chi}_1^0$, is large, namely 56 GeV and
75 GeV, respectively. We find a $\tilde{\nu}_\mu$ LSP between
$M_{1/2}=350$ GeV and $M_{1/2}=600$ GeV for $M_0=140$ GeV, which is
consistent with $a_\mu^{\rm exp}$, Eq.~(\ref{amu_exp}), and
BR($b \rightarrow s \gamma$), Eq.~(\ref{bsg_2sigma}), at $2
\sigma$. For $M_{1/2}=500$ GeV, we obtain a consistent $\tilde
{\nu}_\mu$ LSP for $M_0<170$ GeV. 

\mymed

Nearly the entire $\tilde{\nu}_\mu$ LSP region of
Fig.~\ref{fig:DeltaM_lamp221_m0m12} is consistent with the observed
value of $a_\mu$ at the $1\sigma$ (blue lines) and $2\sigma$ (green
lines) level, {\it cf.} Eq.~(\ref{mu_magnetic_moment}).  It is also
consistent with BR($b \rightarrow s \gamma$) at $2\sigma$,
Eq.~(\ref{bsg_2sigma}).

\mymed

We see in Fig.~\ref{fig:DeltaM_lamp221_m0m12}, all three LSP
candidates, the $\tilde{\nu}_\mu$, the $\tilde{\tau}_1$ and the
$\tilde{\chi}_1^0$.  If we increase $M_0$, we re-obtain at $M_0\approx
150$ GeV the $\tilde{\chi}_1^0$ LSP instead of the $\tilde{\nu}_\mu$
or the $\tilde{\tau}_1$ LSP.  This is easy to understand. $M_0$
increases the mass of all the sfermions, see
Eq.~(\ref{sfermion_masses}), but leaves the mass of the (bino-like)
$\tilde{\chi}_1^0$ unaffected, {\it cf.}
Eq.~(\ref{neutralino_masses}).

\mymed

We get a $\tilde{\tau}_1$ LSP instead of a $\tilde{\nu}_\mu$ LSP 
for $M_{1/2} > 650$ GeV and $M_0<140$ GeV. 
Remember that the $\tilde{\tau}_1$ is mainly right-handed for non-vanishing 
$\lam'_{231}|_{\rm GUT}$ (not for large $\lam'_{331}|_{\rm GUT}$). 
According to Eq.~(\ref{stau_parameter}), the right-handed stau mass 
increases more slowly with $M_{1/2}$ than the left-handed $\tilde{\nu}_\mu$ 
mass, Eq.~(\ref{sfermion_masses}), because the right-handed sfermions couple 
only to the U(1) gaugino, whereas the left-handed sfermions couple also 
to the SU(2) gauginos.
  
\mymed
  
For $M_0$ between 140 GeV and 180 GeV, we obtain a $\tilde{\chi}_1^0$
LSP instead of a $\tilde{\nu}_\mu$ LSP if we increase $M_{1/2}$. In
this region of parameter space, {\it i.e.} $M_0$ between 140 GeV and
180 GeV and $M_{1/2}<700$ GeV, we have a $\tilde{\chi}_1^0$ LSP for vanishing
$\lam'_{231}|_{\rm GUT}$. With $\lam'_{231}\lvert_{\rm GUT}=0.11$, 
we must retrieve the $\tilde{\chi}_1^0$ LSP for increasing
$M_{1/2}$, because the (left-handed) $\tilde{\nu}_\mu$ couples
stronger via the gauge interactions than the (bino-like) $\tilde{\chi}_1^0$; see
Eq.~(\ref{sfermion_masses}) and Eq.~(\ref{neutralino_masses})
respectively.

\mymed

The $M_{1/2}$--$M_{0}$ plane showing the $\tilde{\nu}_\tau$ LSP region,
Fig.~\ref{fig:DeltaM_lamp331_m0m12}, looks similar to the
$\tilde{\nu}_\mu$ LSP region, Fig.~\ref{fig:DeltaM_lamp221_m0m12}: We
again get a $\tilde{\chi}_1^0$ LSP when we increase $M_0$, and a
$\tilde{\tau}_1$ LSP for larger values of $M_{1/2}$. Most of the
$\tilde{\nu}_\tau$ LSP region is also consistent with the observed
value of $a_\mu$ at the $1\sigma$ (blue line) or $2\sigma$ (green
line) level, Eq.~(\ref{mu_magnetic_moment}). But we must have $M_{1/2}
\gsim 290$ GeV [dashed black line in
Fig.~\ref{fig:DeltaM_lamp331_m0m12}] to be consistent with BR($b
\rightarrow s \gamma$) at $2 \sigma$, {\it cf.} 
Eq.~(\ref{bsg_2sigma}). The allowed $\tilde{\nu}_\tau$ LSP region in
the $M_{1/2}$--$M_{0}$ plane is therefore ``smaller" compared to the
$\tilde{\nu}_\mu$ LSP region. It is worth mentioning, that one can
also obtain a $\tilde{\nu}_\tau$ LSP via $\lambda'_{331}|_{\rm GUT}$
consistent with $a_\mu^{\rm exp}$, Eq.~(\ref{amu_exp}), and BR($b
\rightarrow s \gamma$), Eq.~(\ref{bsg_2sigma}), within $1 \sigma$; see
an example in Ref.~\cite{Allanach:2006st}. However the allowed
$\tilde{\nu}_\tau$ LSP region in the $M_{1/2}$--$M_{0}$ [$A_0$--$\tan
\beta$] plane is smaller in that case compared to
Fig.~\ref{fig:DeltaM_lamp331_m0m12}
[Fig.~\ref{fig:DeltaM_lamp331_a0tanb}].

\mymed

As explained
before, $\lam'_{331}|_{\rm GUT}$ reduces also the mass of the
$\tilde{\tau}_1$, which is also a candidate for the LSP. We can see
this in Fig.~\ref{fig:DeltaM_lamp331_m0m12} by noting that the mass
difference between the $\tilde{\nu}_\tau$ LSP and the $\tilde{\tau}_1$
NLSP is rather small, {\it i.e.} $\Delta M \lsim 15$ GeV. A way to
increase this mass difference is to decrease $\tan \beta$; see the
discussion in Sect.~\ref{A0tanbplane}.

\mymed

Another difference between the $\tilde{\nu}_\tau$ LSP region,
Fig.~\ref{fig:DeltaM_lamp331_m0m12}, and the $\tilde{\nu}_\mu$ LSP
region, Fig.~\ref{fig:DeltaM_lamp221_m0m12}, is that the corresponding
SUSY mass spectra for a $\tilde{\nu}_\mu$ LSP scenario are in average
heavier than the SUSY mass spectra for a $\tilde{\nu}_\tau$ LSP
scenario. For example, $M_0=100$ GeV (200 GeV) and $M_{1/2}=500$ GeV
(320 GeV) lead to squark masses of roughly 1000 GeV (700 GeV) in the
$\tilde{\nu}_\mu$ LSP ($\tilde{\nu}_\tau$ LSP) parameter space. The
reason is, that we have chosen our scenarios consistent with the
measured value of $a_\mu$; see discussion after
Eq.~(\ref{mu_magnetic_moment}).

\mymed

We have again in Fig.~\ref{fig:DeltaM_lamp221_m0m12} as well as in
Fig.~\ref{fig:DeltaM_lamp331_m0m12} a ``triple-point", where the three
LSP candidates are degenerate in mass.

\mymed

We give in Fig.~\ref{fig:SnuMass_Lamp221_m12m0}
[Fig.~\ref{fig:SnuMass_Lamp331_m12m0}] the mass of the $\tilde{\nu}
_\mu$ LSP [$\tilde{\nu}_\tau$ LSP] for the sneutrino LSP region of
Fig.~\ref{fig:DeltaM_lamp221_m0m12}
[Fig.~\ref{fig:DeltaM_lamp331_m0m12}].  The sneutrino LSP masses,
which lead to SUSY scenarios in agreement with $a_\mu^{\rm exp}$ (and
$b \rightarrow s \gamma$), range from 78 GeV (LEP bound,
Sect.~\ref{LEP_constraints}) up to roughly 250 GeV. Relaxing this
bound, we claim that $a_\mu^{\rm exp}$ puts an upper bound of roughly
300 GeV at the $2\sigma$ level on the mass of a sneutrino LSP within
$\text{B}_3$ mSUGRA. Note that BR($b \rightarrow s \gamma$) increases
if we increase $M_{1/2}$, whereas $\delta a_\mu^{\rm SUSY}$
decreases, {\it cf.} for example Fig.~4 and Fig.~5 in
Ref.~\cite{Allanach:2006st}. The upper bound on the
sneutrino LSP mass is thus due to $a_\mu^{\rm exp}$.
 
\mymed
 
We finally show in Fig.~\ref{fig:MOrder_Lamp221_m12m0}
[Fig.~\ref{fig:MOrder_Lamp331_m12m0}] the mass difference in GeV
between the $\tilde{\chi}_1^0$ and the $\mu_L$
[mainly left-handed $\tilde{\tau}_1$].  We again observe that the
$\tilde{\chi}_1^0$ is heavier than the $\tilde{\mu}_L$ in most regions
of the $\tilde{\nu}_\mu$ LSP parameter space. The cascade decay,
Eq.~(\ref{slep_cascade}), is therefore not observable at the Tevatron.
Further phenomenological consequences at hadron colliders will be
discussed in Sect.~\ref{pheno}.

\subsection{Sneutrino LSPs with $\lam'_{ijk}|_{\rm GUT} \not = 
\lam'_{231}$ or $\lam'_{331}$}

We investigated in the last three sections in detail the $\tilde{\nu}
_\mu$ LSP ($\tilde{\nu}_\tau$ LSP) parameter space with $\lam'_
{231}\lvert_{\rm GUT}=0.11$ ($\lam'_{331}\lvert_{\rm GUT}=0.12$).
We briefly consider the other couplings of Table~\ref{RPV_couplings}.

\mymed

For $\lam'_{131}|_{\rm GUT}$, we obtain nearly the same parameter
space as in Fig.~\ref{fig:DeltaM_lamp221_a0tanb} and
Fig.~\ref{fig:DeltaM_lamp221_m0m12}, where $\lam'_{231}\lvert_{\rm GUT}=0.11$.
We now have a $\tilde{\nu}_e$ LSP instead of a
$\tilde{\nu}_\mu$ LSP.  Also the mass of the left-handed selectron,
$\tilde{e}_L$, (for $\lam'_{131}\lvert_{\rm GUT}=0.11$) equals the
mass of the $\tilde{\mu}_L$ (for $\lam'_{231}\lvert_{\rm GUT}=0.11$) 
and vice versa. But note, that the $\tilde{\nu}_e$ LSP
parameter space is much more restricted than the $\tilde{\nu}_\mu$ LSP
parameter space due to the stronger bounds on $\lam'_{131}$, {\it
 cf.}  Table~\ref{RPV_couplings}. Also the LEP bound on $m_{\tilde
{\nu}_e}$ is more model dependent, see Table~\ref{bounds_LEP}.

\mymed

We also obtain a $\tilde{\nu}_\mu$ LSP scenario via $\lam'_{221}
|_{\rm GUT}$ and $\lam'_{212}|_{\rm GUT}$. If we choose $\lam
'_{221}\lvert_{\rm GUT}$ or $\lam'_{212}\lvert_{\rm GUT}= 0.097$, we find similar
regions to Fig.~\ref{fig:DeltaM_lamp221_a0tanb} and
Fig.~\ref{fig:DeltaM_lamp221_m0m12}, where the $\tilde{\nu}_\mu$ is
the LSP. The effect of $\lam'_{221}|_{\rm GUT}$ and
$\lam'_{212}|_{\rm GUT}$ on $m_{\tilde{\nu}_\mu}$ is stronger,
because the running of both couplings involves no loops containing the
large top Yukawa coupling. In contrast, the top Yukawa coupling
weakens the running of $\lam'_{231}$ ($j$=3!) when we go from $M_{\rm
  GUT}$ to $M_Z$ \cite{Allanach:2003eb,Dreiner:2008rv}.

\mymed

Analogously, similar to Fig.~\ref{fig:DeltaM_lamp331_a0tanb} and
Fig.~\ref{fig:DeltaM_lamp331_m0m12}, we find parameter regions, where
the $\tilde{\nu}_\tau$ is the LSP. We now have to choose $\lam'_{321}|_{\rm GUT}$
or $\lam'_{312}\lvert_{\rm GUT}=0.104$ instead of $\lam'_{331}\lvert_{\rm GUT}=0.12$.

\mymed

Note however, that different couplings $\lam'_{ijk}$ lead to a
different collider phenomenology, because the $L_i Q_j \bar D_k$
operator couples to different generations of lepton and quark
superfields.  We discuss this topic in the next section.

\section{Hadron Collider Phenomenology}
\label{pheno}

We have shown in the last section, that a sneutrino LSP exists in an
extended region of $\text{B}_3$ mSUGRA parameter space. We now
investigate the corresponding phenomenology at hadron colliders,
especially at the LHC. The main phenomenological differences between a
$\Psix$ mSUGRA scenario with a stable $\tilde{\chi}_1^0$ LSP
and a $\text{B}_3$ mSUGRA scenario with an unstable sneutrino LSP are:

\begin{itemize}

\item The mass spectrum is changed. We now have a sneutrino LSP.
Also some of the sleptons might be lighter than the $\tilde{\chi}_
1^0$, for example the $\tilde{\mu}_L$ in the presence of $\lam'_
{231}|_{\rm GUT}$; see Figs.~\ref{fig:MOrder_Lamp221_a0tanb},
\ref{fig:MOrder_Lamp221_m12m0}. Thus the decay chains and final state
topologies are different.

\item The LSP is not stable anymore and directly decays to SM
particles via the $\text{B}_3$ coupling. In the following analysis,
with $\lam'_{231}|_{\rm GUT}\not=0$, we have two extra jets from each
$\tilde{\nu}_{\mu}$ LSP decay. This also results in less missing
transverse momentum, $\met$.

\item We have shown, that $\lam'_{ijk}|_{\rm GUT}=\mathcal{O} (10^{-1})$ 
is needed to obtain a $\tilde{\nu}_i$ LSP.  This large coupling
can lead to direct and dominating $\text{B}_3$ decays of heavy
sparticles; namely of left-handed charged sleptons of generation $i$,
of left-handed squarks of generation $j$ and of right-handed down-type
squarks of generation $k$. The SM decay products naturally have large
momenta.
\end{itemize} 

In the following, we investigate these aspects in detail. We perform a
Monte Carlo simulation at the parton level using the {\tt HERWIG}
event generator \cite{Herwig,4bodyHERWIG}.

\subsection{Example Spectrum and Branching Ratios}
\label{example_spectrum}

To investigate the sneutrino LSP phenomenology at the LHC,
we choose as an example a scenario with a $\tilde{\nu}_\mu$ LSP:
\begin{eqnarray}
&& \lamp_{231}\lvert_{\text{\rm GUT}} = 0.11, \, M_0 = 100 \textnormal{\,GeV},\,  
M_{1/2}=450 \textnormal{\,GeV},
\nonumber \\ && A_0 = -600 \textnormal{\,GeV},\, \tan\beta=10, \,  
\textnormal{sgn}(\mu) = +1 \, . 
\label{example_point}
\end{eqnarray}
This benchmark point can be found in
Fig.~\ref{fig:DeltaM_lamp221_m0m12} and is consistent with
$a_{\mu}^{\rm exp}$, Eq.~(\ref{amu_exp}), and BR($b \rightarrow s \gamma$),
Eq.~(\ref{bsg_2sigma}), at 1$\sigma$. See also
Ref.~\cite{Allanach:2006st} for a benchmark scenario with a
$\tilde{\nu}_\tau$ LSP.

\mymed 

The resulting sparticle masses and branching ratios (BRs) are given in
Table~\ref{BRs_point_lp231a}. The $\text{B}_3$ decays are shown in
bold-face. Sparticle masses which are significantly affected by
$\lam'_{231}|_{\rm GUT}$ are also bold-face. We calculate the decay
rates by piping the output of {\tt SOFTSUSY} through {\tt
ISAWIG1.200}. This is linked to {\tt ISAJET7.75} \cite{Paige:2003mg}
in order to calculate the decay widths of the SUSY particles.  This
output is later fed into {\tt HERWIG} to simulate events at the LHC.

\mymed

\begin{table*}[ht!]
  \centering
\begin{tabular}{cc}
  \begin{tabular}{|lc|ll|ll|}
    \hline
     & mass [GeV] &channel &BR &channel &BR \\
    \hline
$\ssnumu$ & {\bf 124} &$\bar b d$
    & {\bf 100}$\%$ &&\\ \hline
$\ssmu^-_L$& {\bf 147} &$ W^- \bar b d$
     &{\bf 79.0}$\%$ &$\bar c d$&{\bf 21.0}$\%$\\ \hline
$\neut_1$ & 184 &$\ssnumu^* \nu_\mu$ & $36.0 \%$ 
&$\ssnumu \bar \nu_\mu$ & $36.0 \%$ \\
 & & $\ssmu^+_L \mu^-$ & $14.0 \%$ 
&$\ssmu^-_L \mu^+$ & $14.0 \%$ \\ \hline
$\sstau_1^-$ & 188 &$\neut_1 \tau^-$ &$100 \%$&& \\ \hline
$\sse^-_R$ ($\ssmu^-_R$)& 206 &$\neut_1 e^- (\mu^-)$
     &$100\%$ &&\\ \hline
$\ssnutau$ & 316 & $\neut_1\nutau$ & $67.3\%$ 
&$W^+\sstau_1^-$ & $32.7\%$ \\ \hline
$\ssnue$ & 319 & $\neut_1 \nue$
    & $100 \%$ &&\\ \hline
$\sse^-_L$& 329 &$\neut_1 e^-$
     &$100\%$ &&\\ \hline
$\sstau_2^-$& 329 &$\neut_1 \tau^-$ & $65.1\%$
&$h^0\sstau_1^-$ & $18.2\%$ \\ 
& & $Z^0\sstau_1^-$ & $16.7\%$
& & \\ \hline 
$\neut_2$& 350 &$\ssnumu \bnumu$ & $23.7\%$ 
&$\ssnumu^* \numu$ & $23.7\%$ \\
& &$\ssmu_L^- \mu^+$ & $22.4\%$ 
&$\ssmu_L^+ \mu^-$ & $22.4\%$ \\ 
& &$\ssnutau \bnutau$ & $1.1\%$ 
&$\ssbnutau \nutau$ & $1.1\%$ \\\hline
$\charge_1^-$& 350 &$\ssnumu^* \mu^-$ & $49.7\%$  
&$\ssmu_L^- \bnumu$ & $42.6\%$ \\
& &$\ssnutau^* \tau^-$ & $2.3\%$
&$\ssnue^* e^-$ & $1.8\%$ \\
& & $\sstau_1^- \bnutau$ & $1.6\%$ & & \\ \hline
$\neut_3$ & 691 &$\charge_1^- W^+$ & $29.7\%$ 
&$\charge_1^+ W^-$ & $29.7\%$ \\
& &$\neut_2 Z^0$ & $26.1\%$ 
&$\neut_1 Z^0$ & $8.3\%$ \\
& &$\neut_1 \higgs$ & $1.7\%$
&$\neut_2 \higgs$ & $1.7\%$  \\
\hline
$\sstop_1$& 650 &$\charge^+_1 b$ & $42.1\%$
&$\neut_1 t$ & $33.5\%$ \\
& &$\neut_2 t$ & $13.8\%$ & $\mu^+ d$ & {\bf 10.6}$\%$ \\ 
\hline
$\charge_2^-$& 702 &$\neut_2 W^-$ & $28.0\%$ 
&$\charge_1^- Z^0$ & $26.6\%$ \\
& &$\charge_1^- \higgs$ & $23.8\%$ 
&$\neut_1 W^-$ & $7.9\%$ \\
& &$\sstop_1^* b$ & $4.1\%$ 
&$\ssmu_L^- \bnumu$ & $2.5\%$ \\
& &$\sstau_2^- \bnutau$ & $2.0\%$ 
&$\sse_L^- \bnue$ & $1.7\%$ \\
& &$\ssbnutau \tau^-$ & $1.3\%$ & & \\
\hline
$\neut_4$& 702 &$\charge_1^- W^+$ & $28.3\%$
&$\charge_1^+ W^-$ & $28.3\%$ \\
& &$\neut_2 \higgs$ & $22.3\%$
&$\neut_1 \higgs$ & $7.0\%$ \\
& &$\neut_2 Z^0$ & $2.0\%$ 
&$\neut_1 Z^0$ & $1.8\%$ \\
& &$\ssnumu \bnumu$ &$1.2\%$ 
&$\ssbnumu \numu$ &$1.2\%$ \\
\hline
  \end{tabular} &
\hspace{1cm}
  \begin{tabular}{|lc|ll|ll|} 
    \hline
     &mass [GeV] &channel &BR &channel &BR \\
    \hline
$\ssbottom_1$& {\bf 842} &$W^- \sstop_1$ & $35.8\%$ 
& $\charge^-_1 t$ & $31.3\%$  \\
& &$\neut_2 b$ & $18.8\%$ 
& $\bar \nu_\mu d$ &  {\bf 12.4}$\%$\\
& &$\neut_1 b$ & $1.2\%$ & &  \\
\hline
$\tilde d_R$& {\bf 897} &$\nu_\mu b$ & {\bf 45.3}$\%$ 
& $\mu^- t$ & {\bf 42.1}$\%$ \\
& &$\neut_1 d$ & $12.6\%$ & & \\ 
\hline
$\sstop_2$& {\bf 906} &$Z^0 \sstop_1$ & $28.2\%$ 
&$\charge^+_1 b$ & $23.7\%$ \\    
& &$\higgs \sstop_1$ & $11.7\%$
& $\neut_2 t$ & $10.2\%$\\
& &$\mu^+ d$ & {\bf 9.0}$\%$
&$\neut_4 t$ & $7.5\%$ \\
& &$\charge^+_2 b$ & $5.4\%$
&$\neut_1 t$ & $2.6\%$ \\
& & $\neut_3 t$ & $1.7\%$
& &  \\
\hline
$\ssbottom_2$& 919 &$\neut_1 b$ & $41.3\%$ 
&$W^- \sstop_1$ & $25.3\%$ \\
& &$\charge^-_2 t$ & $14.4\%$
&$\neut_4 b$ & $5.3\%$ \\ 
& &$\neut_3 b$ & $5.0\%$ 
& $\bar \nu_\mu d$ & {\bf 3.4}$\%$\\
& & $\charge^-_1 t$ & $3.2\%$ 
& $\neut_2 b$ & $1.9\%$\\
\hline
$\tilde s_R$& 928 &$\neut_1 s$ & $99.8\%$ & &  \\ 
\hline
$\tilde u_R$ ($\tilde c_R$)& 932 &$\neut_1 u (c)$ & $99.8\%$ & &  \\ 
\hline
$\tilde u_L$ ($\tilde c_L$)& 963 &$\charge^+_1 d (s)$ & $65.6\%$ & $\neut_2 u (c)$ & $32.6\%$ \\
& &$\neut_1 u (c)$ & $1.2\%$ & & \\ 
\hline
$\tilde d_L$ ($\tilde s_L$)& 966 &$\charge^-_1 u (c)$ & $64.5\%$ & $\neut_2 d (s)$ & $32.5\%$ \\
& &$\neut_1 d (s) $ & $1.6\%$ & $\charge^-_2 u (c)$ & $1.0\%$ \\  
\hline
$\glu$ & 1046 &$\sstop_1 \bar{t}$ & $15.0\%$ 
&$\sstop^*_1 t$ & $15.0\%$ \\
& &$\ssbottom_1 \bar{b}$ & $9.2\%$ 
&$\ssbottom_1^* b$ & $9.2\%$ \\
& &$\tilde d_R \bar d$ & $5.2\%$ 
&$\tilde d_R^* d$ & $5.2\%$ \\
& &$\ssbottom_2 \bar{b}$ & $3.9\%$ 
&$\ssbottom_2^* b$ & $3.9\%$ \\
& &$\tilde s_R \bar s$ & $3.4\%$ 
&$\tilde s_R^* s$ & $3.4\%$ \\
& &$\tilde u_R \bar u$ ($\tilde c_R \bar c$) & $3.2\%$ 
&$\tilde u_R^* u$ ($\tilde c_R^* c$) & $3.2\%$ \\
& &$\tilde u_L \bar u$ ($\tilde c_L \bar c$)& $1.7\%$ 
&$\tilde u_L^* u$ ($\tilde c_L^* c$) & $1.7\%$ \\
& &$\tilde d_L \bar d$ ($\tilde s_L \bar s$)& $1.6\%$ 
&$\tilde d_L^* d$ ($\tilde s_L^* s$)& $1.6\%$ \\
\hline
\end{tabular} \\
\end{tabular}
\caption{\label{BRs_point_lp231a} Branching ratios (BRs) and sparticle
masses for the example scenario defined in Eq.~(\ref{example_point}).
BRs smaller than $1\%$ are neglected. $\text{B}_3$ decays are shown in 
bold-face. Masses which are reduced by more than 5 GeV 
(compared to the $\Psix$ spectrum) due to $\lam'_{231}|_{\rm GUT}=0.11$ 
are also shown in bold-face.}
\end{table*}

We find that the decay of the $\tilde{\nu}_\mu$ LSP with a mass of 124
GeV is completely dominated by the $\lam'_{231}$ coupling.  Each LSP
decay leads to a bottom and a down quark and no $\met$
\cite{footnote_4body,footnote_2body}.  However, $\met$ can be obtained
from cascade decays of heavy sparticles. In principle, reconstruction
of the $\tilde{\nu}_\mu$ mass should be possible, although
combinatorial backgrounds might complicate this task.

\mymed

The $\ssmu_L$ with a mass of 147 GeV is the NLSP.  This is the case in
most of the $\tilde{\nu}_\mu$ LSP parameter space, {\it cf.}
Figs.~\ref{fig:MOrder_Lamp221_a0tanb}, \ref{fig:MOrder_Lamp221_m12m0}.
The $\ssmu_L$ decays mainly via the $L_2 Q_3 \bar D_1$ operator into
SM fermions, in principle to $\bar t d$. If this decay mode is not
kinematically allowed, like for the benchmark point under study, we
obtain a dominant 3-body decay into $W^-\bar bd$
\cite{Dreiner:2008rv}.  We thus have at least two jets, where one of
the jets is a $b$-jet.  As mentioned in
Sect.~\ref{Tevatron_constraints}, another possible 3-body decay is
$\tilde{\mu}^-_L \ra \mu^- \bar\nu_\mu \tilde{\nu}_\mu$ via a virtual
neutralino. But this decay is suppressed by four orders of magnitude
compared to the 3-body decay via a virtual top quark. The reasons are:
small couplings (left-handed sleptons couple to a bino-like
$\neut_1$), less phase space ($m_{\tilde{\mu}_L} -
m_{\tilde{\nu}_\mu}=23$ GeV), destructive interferences between
diagrams with a virtual $\neut_1$ and $\neut_2$, and the decay via the
virtual top is enhanced by a colour factor of 3
\cite{Dreiner:2008rv}. However, there is an additional 2-body decay
mode, $\tilde{\mu}_L\ra\bar c d$, in Table~\ref{BRs_point_lp231a}.
This decay proceeds via a non-vanishing $\lam'_{221}$ coupling, which
is generated out of $\lam'_{231}|_{\rm GUT}$ via RGE running
\cite{Carlos:1996du,Allanach:2003eb,Dreiner:2008rv}.

\mymed

The electroweak gauginos decay dominantly via $\Psix$ conserving gauge
interactions to 2-body final states. The lightest gaugino is the
$\neut_1$, which is only the NNLSP within our benchmark scenario;
$m_{\tilde{\chi}_1^0}=184$ GeV. It decays into either the LSP or NLSP.
These then undergo direct $\text{B}_3$ decays, as discussed before.
So, the $\neut_1$ decays lead to dijet events with $\met$ or a muon.
Due to the Majorana nature of the $\neut_1$, negatively and positively
charged muons are possible. Cascade decays of pair produced sparticles
can therefore lead to like sign-muon events via $\neut_1$ decays; see
Sect.~\ref{spart_pair_prod}. Note, that $\tilde{\nu}_\mu$ LSP
scenarios exist where the $\neut_1$ is also heavier than the
$\tilde{\tau}_1$ or even the right-handed smuon, $\tilde{\mu}_R$, and
selectron, $\tilde{e}_R$. These scenarios can lead to multi-lepton
final states. We will not consider these scenarios here, because the
relevant $\tilde{\mu}_R$ and $\tilde{e}_R$ decays into the
$\tilde{\nu}_\mu$ LSP and the $\tilde{\mu}_L$ NLSP are not implemented
in {\tt HERWIG}.

\mymed     

The $\neut_2$ also has a significant BR to $\ssmu_L^\pm \mu^\mp$ and
$\ssnumu \nu_\mu$. Similarly, the lightest chargino, $\charge_1^-$,
decays either predominantly into $\ssnumu^* \mu^-$ or $\ssmu^-_L\bar
\nu_{\mu}$, leading to either a muon or missing energy in the final
state.  The $\neut_2$ and $\charge^-_1$ are wino-like in mSUGRA
models.  They thus decay predominantly to the left-handed $\ssmu_L$
and $\tilde{\nu}_\mu$.  The decays of the heavier chargino, $\charge_
2^-$, and neutralinos, $\neut_{3/4}$ are similar to $\Psix$ mSUGRA
scenarios.

\mymed

The $\sstau_1$ in Table~\ref{BRs_point_lp231a} is the next-to-NNLSP
(NNNLSP) with a mass of 188 GeV and almost degenerate with the
$\neut_1$. The $\sstau_1$ can in general be the NLSP, the NNLSP or
NNNLSP in $\text{B}_3$ mSUGRA scenarios with a sneutrino LSP.  Here we
have $\sstau_1^-$ $\ra$ $\neut_1$ $\tau^-$.

\mymed 

The $\tilde{\mu}_R$, $\tilde{e}_R$, $\tilde{e}_L$, $\tilde{\nu}_e$,
$\tilde{\nu}_\tau$ and $\sstau_2$ in Table~\ref{BRs_point_lp231a}
decay into the $\neut_1$ or, in the case of the $\sstau_2$ and
$\tilde{\nu}_\tau$, also into the $\sstau_1$ similar to $\Psix$ mSUGRA
scenarios. But as mentioned above, the $\sstau_1$, the $\tilde{\mu}_R$
and the $\tilde{e}_R$ can in general be lighter than the $\neut_1$ in
$\tilde{\nu}_\mu$ LSP scenarios. These particles then decay
preferentially into the $\tilde{\nu}_\mu$ LSP via a 3-body decay.

\mymed

The masses of the top-squarks, $\tilde{t}_{1,2}$, and the
bottom-squarks, $\tilde{b}_{1,2}$, are slightly reduced due to the
presence of $\lam'_{231}$ in the corresponding RGEs. The $\tilde
{t}_{1}$ is the lightest squark with a mass of 650 GeV and has four
2-body decay modes with appreciable BRs.  Three decays are via gauge
interactions and one via $\lam'_{231}$.  Since the electroweak gauge
couplings and $\lam'_{231}$ have the same order of magnitude, we also
expect $\Psix$ conserving and violating decays at a similar rate.  The
situation for the $\tilde{t}_{2}$, $\tilde{b}_{1}$ and $\tilde{b}_{2}$
is similar to $\tilde{t}_{1}$. All of these particles couple via their
left-handed component to the $L_2 Q_3 \bar D_1$ operator and can
therefore decay into two SM particles.

\mymed

The masses of the left-handed and right-handed squarks of the 1st and
2nd generation are around 900 GeV. The right-handed down-squark
($m_{\tilde d_R}$= 897 GeV) is lighter than the right-handed
strange-squark ($m_{\tilde s_R}$= 928 GeV). In contrast both squarks
are degenerate in mass in $\Psix$ mSUGRA. However, they are
so heavy, that no problems should occur with flavour changing
neutral currents. $\lam'_{231}|_{\rm GUT}$
couples only to the right-handed down squarks and not to the
right-handed strange squarks.  So, $m_{\tilde d_R}$ is reduced,
keeping $m_{\tilde s_R}$ unchanged.  For the same reason, there exist
no $\text{B}_3$ decays of $\tilde s_R$ via $\lambda'_{231}$ at tree-level. 
In contrast, $\tilde d_R$ has
dominant direct $\text{B}_3$ decays to SM particles, which than have large
momenta, see Sect.~\ref{spart_pair_prod}.

\mymed

The heaviest sparticle is the gluino, $\tilde{g}$, with a mass of 1046
GeV.  It decays only via the strong interaction. The allowed decay
modes and their relative BRs depend upon the sum of the final state
masses.  For example, $\tilde{g} \ra \tilde{t}_{1} t$ has the largest
BR, since the $\tilde{t}_{1}$ is the lightest squark.

\mymed

We conclude that the heavy part of the mass spectrum looks very
similar to $\Psix$ mSUGRA scenarios with a stable $\tilde{\chi}_1^0$
LSP. However, a non-vanishing $\lam'_{ijk}$ coupling, which has the
same order of magnitude as the gauge couplings, allows for additional
2-body $\text{B}_3$ decays of some of the squarks. Which squarks are
allowed to decay via $\lam'_{ijk}$ depend on the indices $j$, $k$.
The masses and compositions of the electroweak gauginos are also very
similar to $\Psix$ mSUGRA. However, the $\tilde
{\chi}_1^0$ is no longer the LSP. Depending on the specific $\tilde
{\nu}_i$ LSP scenario, the $\tilde{\chi}_1^0$ can decay into charged
sleptons and sneutrinos of different generations.  Therefore, the main
difference can be found in the light part of the mass spectrum where
we have the $\tilde{\nu}_i$ LSP. The $\tilde{\nu}_i$ LSP decays
preferentially into two jets via $\lam'_{ijk}$.

\subsection{Sparticle Pair Production}
\label{spart_pair_prod}

We have investigated in the last section the mass spectrum and the BRs
of SUSY particles for one representative $\text{B}_3$ mSUGRA scenario
with a $\tilde{\nu}_\mu$ LSP, described by Eq.~(\ref{example_point}).
We have pointed out the general differences compared to mSUGRA
scenarios with a stable $\neut_1$ LSP. We now explore signatures at
the LHC which arise from pair production of sparticles via the gauge
interactions, {\it i.e.} mainly squark and gluino production via the
strong interaction. For this purpose we use the {\tt HERWIG} event
generator. We investigate single sparticle production in
Sect.~\ref{single_spart_prod}.

\mymed 

The masses of the strongly interacting sparticles are roughly 1
TeV. We therefore obtain from {\tt HERWIG} a total sparticle pair
production (leading order) cross section at the LHC of
\begin{equation}
\sigma_{\text{total}} = 3.0 \, \text{pb} \, .
\label{total_xsection}
\end{equation} 
So, one can expect approximately $300\,000$ SUSY pair production
events for an integrated luminosity of 100 $\text{fb}^{-1}$. The
sparticle decays follow those in Table~\ref{BRs_point_lp231a}. The
different decay chains lead to different final states. Moreover, the
$p_T$ distributions of the final state particles and the $\met$ can be
very distinctive compared to $\Psix$ mSUGRA with a stable $\neut_1$ LSP.

\mymed

\begin{figure}
	\includegraphics[scale=0.40, bb = 30 70
    530 530, clip=true]{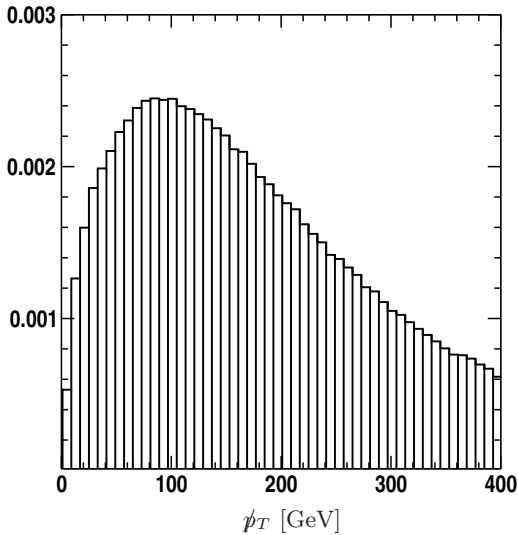}
    \put(-107.0,-13.0){$\met$ [GeV]}
 	\caption{\label{missing_pT_from_neutrinos} 
	$\met$ distribution due to neutrinos in the final state
    for the example scenario Eq.~(\ref{example_point}). 
	The distribution is normalized to one. Note that 
	events with no $\met$ in the final state are not
	shown.}
\end{figure}

We show in Fig.~\ref{missing_pT_from_neutrinos} the $\met$
distribution due to neutrinos in the final state. Note, that here
roughly $20 \%$ of all SUSY events posses no $\met$ in contrast to
$\Psix$ mSUGRA scenarios. For example, if the decay chains of the pair
produced sparticles into the $\tilde{\nu}_\mu$ LSP contain no neutrino
than there is no $\met$. The $\met$ distribution in
Fig.~\ref{missing_pT_from_neutrinos} peaks at roughly 90 GeV. Thus,
$\met$ might still be used to distinguish the SUSY signal from its SM
background. Large amounts of $\met$, {\it i.e.} $\met$ of a few
hundred GeV, can arise if a squark decays directly via $\lam'_{231}$
into a quark and a neutrino. For example $\tilde{d}_R\ra\nu_\mu b$,
{\it cf.} Table~\ref{BRs_point_lp231a}. This decay also leads to a 
high-$p_T$ $b$-jet,  {\it i.e.} $p_T$ of $\mathcal{O}(100\,\text{GeV})$.
 
\mymed

\begin{figure}
	\includegraphics[scale=0.40, bb = 30 70 530 530,
	clip=true]{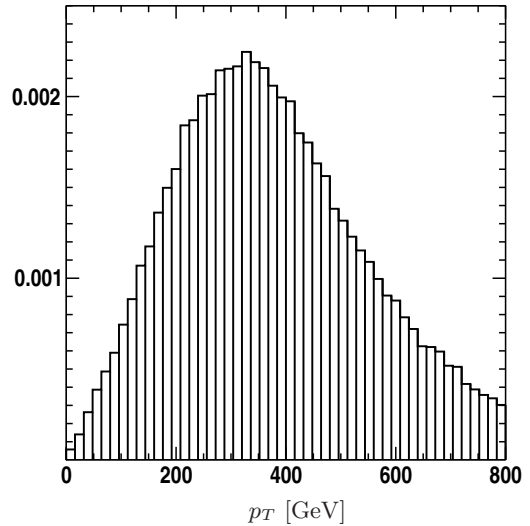}
	\put(-107.0,-13.0){$p_T$ [GeV]}
	\caption{\label{mu_pT_from_sdR_or_stop} $p_T$ distribution of
	the muon from the decays $\tilde{d}_R \rightarrow \mu t $ and
	$\tilde{t}_{1/2} \rightarrow \mu d$ ({\it cf}. Table
	\ref{BRs_point_lp231a}) at the LHC.  The distribution is
	normalized to one.}
\end{figure}

\begin{figure}
	\includegraphics[scale=0.40, bb = 30 70
    530 530, clip=true]{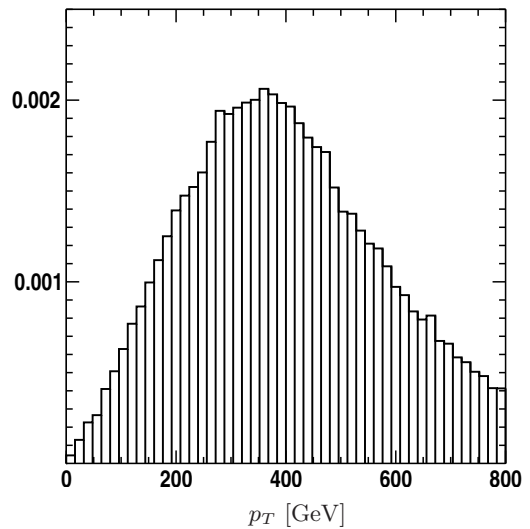}
    \put(-107.0,-13.0){$p_T$ [GeV]}
 	\caption{\label{top_pT_from_sdR} 
	$p_T$ distribution of the top quark from the decay 
	$\tilde{d}_R \rightarrow \mu t $ ({\it cf}. Table 
	\ref{BRs_point_lp231a}) at the LHC.  
	The distribution is normalized to one.}
\end{figure}

Instead of high-$p_T$ neutrinos, we can also have high-$p_T$ muons
from the direct decays of $\tilde{d}_R$ and $\tilde{t}_{1/2}$ via
$\lam'_{231}$, see Table~\ref{BRs_point_lp231a}.  We show in
Fig.~\ref{mu_pT_from_sdR_or_stop} the $p_T$ distribution of these
muons. The distribution peaks at 340 GeV. The large momenta are a
consequence of the large squark masses. Nearly the entire mass of the
squarks is transformed into the momenta of two SM particles. These
high-$p_T$ SM particles might also be used to reconstruct the squark
mass. The muon $p_T$-distribution will peak at smaller values, if the
squarks are lighter than in our benchmark scenario. But at the same
time we will produce more squarks and muons compared to the cross
section, Eq.~(\ref{total_xsection}). If the mass spectrum is heavier
compared to our example point, the cross section will be smaller. But
the muon $p_T $-distribution will now peak at larger values. Thus
stronger cuts on the muon $p_T$ can be applied.  We conclude that the
high-$p_T$ muons might be used on the one hand to distinguish the SUSY
signal from the SM background and on the other hand to distinguish the
$\text{B}_3$ mSUGRA model with a $\tilde{\nu}_\mu$ LSP from mSUGRA
with a stable $\neut_1$ LSP.  For our benchmark scenario
Eq.~(\ref{example_point}), we find that $11 \%$ of all sparticle pair
production events lead to at least one high-$p_T$ muon from a squark
decay. A fraction of roughly $10 \%$ is a general feature of our
$\tilde{\nu}_\mu$ LSP scenarios.

\mymed

The neutrino or muon from the squark decay will be accompanied by a
quark with roughly the opposite $p_T$. These quarks lead to high-$p_T$
jets, which might be $b$-jets depending on the flavour indices of
$\lam'$. For our benchmark point, we obtain high-$p_T$ $b$-jets from
the $\text{B}_3$ decay $\tilde{d}_R \ra \nu_\mu b$. We also can get a
top-quark, $t$, from the decay $\tilde{d}_R \ra \mu^- t$. We show in
Fig.~\ref{top_pT_from_sdR} the $p_T$-distribution of this
top-quark. The distribution peaks at 360 GeV. The top decay will also
produce a $b$-jet and a $W$. The $W$ might produce additional jets or
leptons with $\met$. These decay products will be boosted due
to the large top momentum. Thus isolated leptons can most likely 
not be used to reconstruct the top quark.

\mymed    

\begin{figure}
	\includegraphics[scale=0.40, bb = 30 70
    530 530, clip=true]{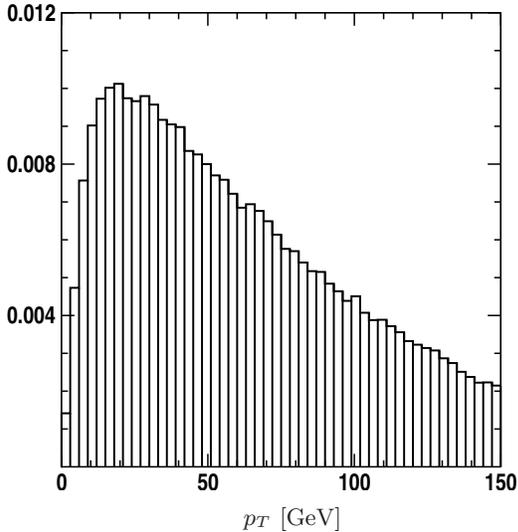}
    \put(-107.0,-13.0){$p_T$ [GeV]}
 	\caption{\label{mu_pT_from_chi} 
	$p_T$ distribution of the muon from the decay 
	$\tilde{\chi}_1^0 \rightarrow \tilde{\mu}_L \mu$ 
	({\it cf}. Table 
	\ref{BRs_point_lp231a}) at the LHC .  
	The distribution is normalized to one.}
\end{figure}

Finally we want to mention an effect arising from the mass ordering in
the light part of the spectrum. We have shown in
Figs.~\ref{fig:MOrder_Lamp221_a0tanb}, \ref{fig:MOrder_Lamp221_m12m0}
that the $\tilde{\mu}_L$ is lighter than the $\neut_1$ in most regions
of $\tilde{\nu}_\mu$ LSP parameter space allowing for the decay
$\neut_1 \ra \tilde{\mu}_L^\pm \mu^\mp$. Since many decay chains in
Table~\ref{BRs_point_lp231a} involve the $\neut_1$, we expect more
muons in the final state than in mSUGRA with a stable $\neut_1$ LSP
\footnote{Note, that also the $\neut_2$ and $\tilde{\chi}_1^-$ decay
to a muon with a BR of roughly $50\%$, see
Table~\ref{BRs_point_lp231a}.}. For example, all right-handed squarks,
which do not directly couple to the $L_2Q_3\bar D_1$ operator will
predominantly decay into the $\neut_1$. Thus pair production of
right-handed squarks, $\tilde{q}_R$, has a large fraction of the
signature
\begin{equation}
\tilde{q}_R \tilde{q}_R \ra \mu^\pm \mu^\pm \, jjjjjj \,(W W) \,.
\label{squark_pair_mu}
\end{equation}
We have six jets, $j$, where two jets rise from the $\tilde{q}_R$
decay and four jets from the decay of the two $\tilde{\mu}_L$. If the
$\tilde{\mu}_L$ decay via the 3-body decay (see Table
\ref{BRs_point_lp231a}), two jets will be $b$-jets and we will
also have two $W$s in the final state.  We also find two muons from
$\neut_1$ decay, where all charge combinations of the muons are
possible due to the Majorana nature of the $\neut_1$. We therefore
have a new source for like-sign dimuon events, which does not exist in
$\Psix$ mSUGRA scenarios with a stable $\neut_1$ LSP.  In principle,
it should be possible to reconstruct the full event,
Eq.~(\ref{squark_pair_mu}), although we have large combinatorial
backgrounds due to the many jets in the final state.

\mymed 

We show in Fig.~\ref{mu_pT_from_chi} the $p_T$-distribution of the
muons arising from $\neut_1$ decay within our example scenario,
Eq.~(\ref{example_point}). The distribution peaks at 20 GeV and
therefore we expect that most of the muons will pass standard
experimental cuts. However, the position of the peak is restricted by
the mass difference of the $\tilde{\mu}_L$ and $\neut_1$. In our
example the mass difference is 37 GeV. In general we find in
Figs.~\ref{fig:MOrder_Lamp221_a0tanb},
\ref{fig:MOrder_Lamp221_m12m0} mass differences of up to 90 GeV.

\mymed 

In a $\tilde{\nu}_i$ LSP scenario with $\lam'_{ijk}|_{\rm GUT}\not=
\lam'_{231}|_{\rm GUT}$ we get the following differences. Now 
left-handed (right-handed down-type) squarks of
generation $j$ ($k$) will couple to the $L_i Q_j \bar D_k$
operator. These squarks can now decay into a quark of generation $k$
($j$) and into a lepton of generation $i$. In addition, the masses of
these squarks will be reduced via the $\text{B}_3$ interaction. For
$i=1$, we have to replace the muons in the discussion above by
electrons. For $i=3$, we have taus instead of muons. We will get taus
with large momenta, {\it i.e.} $p_\tau=\mathcal{O}(100 \, \text{GeV})$,
from the decays of the squarks via the $\text{B}_3$ interaction. These
taus have a boost factor of $\gamma = \mathcal{O}(100)$ and are thus
long lived leading to detached vertices of
$\mathcal{O}(1 \, \text{cm})$. We finally see in
Figs.~\ref{fig:MOrder_Lamp331_m12m0},
\ref{fig:MOrder_Lamp331_a0tanb} that also in large regions of $\tilde{\nu}_\tau$ LSP parameter 
space the $\tilde{\tau}_1$ is lighter than the $\neut_1$. This might
lead to like-sign tau events from two decay chains involving a
$\neut_1$.

\subsection{Single Sparticle Production}
\label{single_spart_prod}

Here we explore single sparticle production, which is not possible if
$\Psix$ is conserved. We expect high rates due to the large
$\lam'_{ijk}$ coupling in $\tilde{\nu}_i$ LSP scenarios.

\mymed

\begin{table}
\begin{ruledtabular}
\begin{tabular}{clcc}
 & process & cross section & \\
 \hline
 & $PP \ra \tilde{\nu}_\mu + X$ & $2.2 \times 10^{6}$ fb & \\ 
 & $PP \ra \neut_1 \nu_\mu + X$ & $ 4.2 \times 10^{1}$ fb & \\
 & $PP \ra \neut_2 \nu_\mu + X$ & $ 6.2 \times 10^{0}$ fb & \\
 & $PP \ra \tilde{\chi}^-_1 \mu^+ + X$ & $ 1.3 \times 10^{1}$ fb & \\
 & $PP \ra \tilde{\mu}_L^- t + X$ & $1.3 \times 10^{4}$ fb &  
\end{tabular}
\caption{\label{single_prod_xsect} Total hadronic cross sections for single 
         sparticle production at the LHC within the $\tilde{\nu}_\mu$
         LSP scenario, Eq.~(\ref{example_point}), with
         $\lambda'_{231}|_{\rm GUT}= 0.11$. The cross sections include
         also the charge conjugated processes.}
\end{ruledtabular}
\end{table}

We show in Table~\ref{single_prod_xsect} the hadronic cross sections
for different single sparticle production processes. We again consider
the example scenario, Eq.~(\ref{example_point}), with $\lambda'_{231}|_{\rm GUT}=
0.11$. The first four cross sections are calculated with
{\tt HERWIG} and the last cross section is taken from
Ref.~\cite{Bernhardt:2008mz}. The first four processes involve a real
or virtual $\tilde{\nu}_\mu$, which is the LSP. The corresponding
processes with the $\tilde{\mu}_L$ are not possible, because one
parton in the initial state has to be a top-quark. A single
$\tilde{\mu}_L$ can therefore be produced only in association with a
SM particle, for example with a top-quark
\cite{Bernhardt:2008mz,Slepitop2}, see also
Table~\ref{single_prod_xsect}.

\mymed

We indeed observe in Table~\ref{single_prod_xsect} a large cross
section for the resonant production of single $\tilde{\nu}_\mu$s due
to the large $\lam'_{231}$ coupling, high parton luminosity (due to
small Bjorken x) and large phase space. For 10 $\text{fb}^{-1}$
integrated luminosity we will produce more than two million
$\tilde{\nu}_\mu$ LSPs. However, the $\tilde{\nu}_\mu$ can only decay
into two jets, {\it cf.} Table~\ref{BRs_point_lp231a}, where one jet
is a $b$-jet \cite{footnote_2body,footnote_4body}. This process thus
suffers from large QCD background and it will be very hard to observe
an excess over the SM background at the LHC
\cite{Hewett:1998fu}.

\mymed

The process in Table~\ref{single_prod_xsect} with the second largest
cross section is single $\tilde{\mu}_L$ production in association with
a top quark. This process suffers in general from the large SM $t\bar t+
\text{jet}$ background \cite{Bernhardt:2008mz}. However it might be 
possible to see an excess over the SM in small regions of
$\tilde{\nu}_\mu$ LSP parameter space, where the $\neut_1$ is lighter
than the $\tilde{\mu}_L$, {\it cf.} 
Figs.~\ref{fig:MOrder_Lamp221_a0tanb}, \ref{fig:MOrder_Lamp221_m12m0}.
The $\tilde{\mu}_L$ can decay in this case to $\neut_1 \mu$ and we
might employ the charge asymmetry of the muons to distinguish the
signal from the background \cite{Bernhardt:2008mz}.

\mymed

The production of a $\neut_{1}$ [$\neut_2$] in association with a
neutrino, Table~\ref{single_prod_xsect}, can lead to a muon with jets
and $\met$ in the final state, because $28\%$ [$44.8\%$] of the
$\neut_1$s [$\neut_2$s] decay into a $\tilde{\mu}_L \mu$ pair. However
the respective production cross sections are rather small, namely 42
$\text{fb}$ [6.2 $\text{fb}$].

\mymed

The production of charginos and muons, $\tilde{\chi}_1^- \mu^+$, seems
more promising. Roughly $50\%$ of the produced $\tilde{\chi}_1^-$ will
decay into $\tilde{\nu}_\mu^* \mu^-$ leading to a final state with a pair
of muons, and two jets, where one jet is a $b$-jet. But again the
cross section is small, 13 $\text{fb}$.

\mymed 

In $\tilde{\nu}_i$ LSP scenarios, where $\lam'_{ijk}|_{\rm GUT}\not=
\lam'_{231}|_{\rm GUT}$, the main difference arises if $j \not = 3$.
In this case also resonant single charged slepton, $\tilde{\ell}_{Li}$,
production, Eq.~(\ref{res_slep}), is possible via an up-type quark of
generation $j$. Therefore, if the $\neut_1$ is lighter than the
$\tilde{\ell}_{Li}$, we expect a high rate of leptons from
$\tilde{\ell}_{Li}$ decay to $\neut_1 \ell_i$. But this is only
possible in small regions of $\tilde{\nu}_i$ LSP parameter space, see
Figs.~\ref{fig:MOrder_Lamp221_a0tanb},
\ref{fig:MOrder_Lamp331_a0tanb}, \ref{fig:MOrder_Lamp221_m12m0} and
\ref{fig:MOrder_Lamp331_m12m0}.  A further bottleneck for the
observation of these leptons is the small mass difference between the
$\neut_1$ and $\tilde{\ell}_{Li}$ leading to small lepton momenta. The
mass difference will not exceed roughly 30 GeV. Large $\lam'_{ijk}$
couplings with $j\not = 3$ are also disfavoured by $D_0$--$\bar
D_0$-mixing, {\it cf.}  Sect.~\ref{indirect_bounds}.

\mymed

We conclude, that pair production of SUSY particles and their
subsequent decays lead to much more promising signatures than single
sparticle production. On the one hand, resonant single sneutrino
production, which occurs at a high rate, lead mainly to jets in the
final state and thus suffers from the large QCD background. On the
other hand, processes with one or two leptons in the final state have
small cross sections, {\it i.e.} $\lsim \mathcal{O}(10 \, fb)$.

\section{Conclusion}
\label{conclusion}

In supersymmetric models it is essential to know the nature of the
LSP, since it is involved in practically all collider signals. In the
MSSM the LSP is necessarily the lightest neutralino. However, in B$_3$
mSUGRA models this is not the case: It had been shown previously that
one can obtain a stau LSP and even a sneutrino LSP. In this paper we
have analysed in detail which B$_3$ mSUGRA parameter region leads to a
sneutrino LSP. In particular, we have found that a coupling
$\lam'_{ijk}=\mathcal{O}(10^{-1})$ at the GUT scale will lead to a
sneutrino LSP due to additional $\text{B}_3$ terms in the RGEs. We
have shown, that such a large coupling can still be consistent with
experiment, for a $\tilde\nu_{\mu,\tau}$ LSP. A $\tilde{\nu}
_e$ LSP is disfavoured due to the strong bounds on the couplings
$\lambda'_{1jk}$, see Table~\ref{RPV_couplings}.

\mymed

We have explored which conditions at the GUT scale lead
to a sneutrino LSP. We have shown that a negative trilinear
scalar coupling $A_0$ with a large magnitude enhances the negative
$\text{B}_3$ contribution to the sneutrino mass. We have found large
regions in the $\text{B}_3$ mSUGRA parameter space, where the
sneutrino is the LSP and which are consistent with the observed
anomalous magnetic moment of the muon, $a_\mu^{\rm exp}$, as well as
with BR($b\ra s\gamma$), see Figs.~\ref{fig:a0_tanb} and
\ref{fig:M0_M12}. The allowed $\tilde{\nu}_\mu$ LSP parameter space is
hereby larger than the $\tilde{\nu}_\tau$ LSP parameter space. We have
also shown that $a_\mu^{\rm exp}$ puts an upper bound of roughly 300
GeV on the sneutrino LSP mass.

\mymed

We have next investigated the phenomenology of sneutrino LSP models at
the LHC. We have considered one benchmark scenario with a $\tilde{\nu}
_\mu$ LSP which is obtained via $\lambda'_{231}|_{\rm GUT}=0.11$.
Within this scenario, we have found that direct decays of light as
well as heavy SUSY particles lead to an excess of muons in the final
state, {\it cf.} Table~\ref{BRs_point_lp231a}. We also have found that
signatures from pair production of SUSY particles are more promising
than from single sparticle production, since the latter mainly
involve hadronic final states. Promising pair production signatures
are high-$p_T$ muons of a few hundred GeV, \textit{cf.} 
Fig.~\ref{mu_pT_from_sdR_or_stop}, high-$p_T$ jets, like-sign muon
events and long-lived taus with a detached vertex of $\mathcal{O}
$(1cm).

\mymed

These signatures should be investigated by the experimental
groups in order to find supersymmetry as well as to distinguish
$\text{B}_3$ mSUGRA with a sneutrino LSP from ``normal" mSUGRA with a
stable $\neut_1$.


\begin{acknowledgments}
  We thank Benjamin Allanach for help with the as-yet unpublished
  B$_3$ version of {\tt SOFTSUSY}. We also thank Volker B\"uscher for
  helpful discussions. SG thanks the theory groups of Fermilab
  National Accelerator, Argonne National Laboratory and UC Santa Cruz
  for helpful discussions and warm hospitality. SG also thanks the
  `Deutsche Telekom Stiftung' and the `Bonn-Cologne Graduate School of
  Physics and Astronomy' for financial support.  This work was
  partially supported by BMBF grant 05 HT6PDA, by the Helmholtz
  Allianz HA-101 `Physics at the Terascale' and by the SFB Transregio
  33 `The Dark Universe'.
\end{acknowledgments}

\bibliographystyle{h-physrev}

\begin{thebibliography}{99}
\bibitem{4bodyHERWIG} 
  The version of {\tt HERWIG} \cite{Herwig} used in this paper includes
  modifications to simulate the four-body decays of a stau LSP and was
  implemented by Peter Richardson.
\bibitem{footnote_2body}
The $\tilde{\nu}_\mu$ can in principle
also decay via a RGE generated $\lambda_{121}$ ($\lambda_{323}$)
coupling into a $e \nu_e$ ($\tau \nu_\tau$) pair. However the
generated $\lambda$'s are at least ten orders of magnitude smaller than
the $\lambda'_{231}$ coupling. The relevant RGEs to generate the
$\lambda$s involve off-diagonal down-Yukawa matrix elements, which
vanish at $M_{\rm Z}$, see Sect.~\ref{tree_neut_bounds}.
\end{thebibliography}


\end{document}